%% file: HIG-17-034_temp.tex
\pdfoutput=1

\documentclass[11pt,twoside,a4paper,cmspaper,final,collab]{cms-tdr}
\def\svnVersion{1e0eb5d-D}\def\svnDate{2019/10/30}\def\cmsCernNoTag{CERN-EP-2019-029}\def\cmsCernDate{\today}\def\cmsMessage{"Published in Physical Review D as \href{http://dx.doi.org/10.1103/PhysRevD.100.112002}{\doi{10.1103/PhysRevD.100.112002}.}"}

\begin{document}\cmsNoteHeader{HIG-17-034}

\hyphenation{had-ron-i-za-tion}
\hyphenation{cal-or-i-me-ter}
\hyphenation{de-vices}
\RCS$HeadURL$
\RCS$Id$
\newlength\cmsFigWidth
\ifthenelse{\boolean{cms@external}}{\setlength\cmsFigWidth{0.85\columnwidth}}{\setlength\cmsFigWidth{0.4\textwidth}}
\ifthenelse{\boolean{cms@external}}{\providecommand{\cmsLeft}{upper\xspace}}{\providecommand{\cmsLeft}{left\xspace}}
\ifthenelse{\boolean{cms@external}}{\providecommand{\cmsRight}{lower\xspace}}{\providecommand{\cmsRight}{right\xspace}}
\ifthenelse{\boolean{cms@external}}{\providecommand{\NA}{\ensuremath{\text{ \cdots}}\xspace}}{\providecommand{\NA}{\ensuremath{\text{---}}\xspace}}
\ifthenelse{\boolean{cms@external}}{\renewcommand{\CL}{\ensuremath{\text{C.L.}}\xspace}}{}
\ifthenelse{\boolean{cms@external}}{\providecommand{\CLnp}{\ensuremath{\text{C.L}}\xspace}}{\providecommand{\CLnp}{\ensuremath{\text{CL}}\xspace}}
\newlength\cmsTabSkip\setlength\cmsTabSkip{1.6ex}
\newcommand{\mvis}{\ensuremath{m_\text{vis}}\xspace}
\newcommand{\mtautau}{\ensuremath{m_{\Pgt\Pgt}}\xspace}
\newcommand{\pth}{\ensuremath{\pt^{\Pgt\Pgt}}\xspace}
\newcommand{\ptvech}{\ensuremath{\ptvec^{\Pgt\Pgt}}\xspace}
\newcommand{\mjj}{\ensuremath{m_{JJ}}\xspace}
\newcommand{\aMCATNLO} {\textsc{MG5}\_a\MCATNLO\xspace}
\newcommand{\emu}{\ensuremath{\Pe\Pgm}\xspace}
\providecommand{\V}{\cmsSymbolFace{V}}
\providecommand{\f}{\cmsSymbolFace{f}}
\newcommand{\Hboson}{\ensuremath{\PH} boson\xspace}

\cmsNoteHeader{HIG-17-034}
\title{Constraints on anomalous	\texorpdfstring{$\PH\V\V$}{HVV} couplings from the production of Higgs bosons decaying to \texorpdfstring{$\Pgt$}{tau} lepton pairs}
\date{\today}

\abstract{
A study is presented of anomalous $\PH\V\V$ interactions of the Higgs boson, including its $CP$ properties.
The study uses Higgs boson candidates produced mainly in vector boson fusion and gluon fusion that
subsequently decay to a pair of $\Pgt$ leptons. The data were recorded by the CMS experiment at the
LHC in 2016 at a center-of-mass energy of 13\TeV and correspond to an integrated luminosity of 35.9\fbinv.
A matrix element technique is employed for the analysis of anomalous interactions.
The results are combined with those from the $\PH\to 4\ell$ decay channel presented earlier,
yielding the most stringent constraints on anomalous Higgs boson couplings
to electroweak vector bosons expressed as effective cross section fractions and phases:
the $CP$-violating parameter $f_{a3}\cos(\phi_{a3})=(0.00\pm0.27)\times10^{-3}$ and
the $CP$-conserving parameters $f_{a2}\cos(\phi_{a2})=(0.08^{+1.04}_{-0.21})\times10^{-3}$,
$f_{\Lambda1}\cos(\phi_{\Lambda1})=(0.00^{+0.53}_{-0.09})\times10^{-3}$, and
$f_{\Lambda1}^{\PZ\gamma}\cos(\phi_{\Lambda1}^{\PZ\gamma})=(0.0^{+1.1}_{-1.3})\times10^{-3}$.
The current dataset does not allow for precise constraints on $CP$ properties in the gluon fusion process.
The results are consistent with standard model expectations.
}

\hypersetup{%
pdfauthor={CMS Collaboration},%
pdftitle={Constraints on anomalous HVV couplings in the production of Higgs bosons decaying to tau lepton pairs},%
pdfsubject={CMS},%
pdfkeywords={Higgs, CP violation, anomalous couplings}}

\maketitle

\section{Introduction}
\label{sec:Introduction}

The Higgs boson ($\PH$) discovered in 2012 at the CERN LHC~\cite{Aad:2012tfa, Chatrchyan:2012xdj, Chatrchyan:2013lba}
has thus far been found to have properties consistent with expectations from the
standard model (SM)~\cite{StandardModel67_1, Englert:1964et,Higgs:1964ia,Higgs:1964pj,Guralnik:1964eu,StandardModel67_2,StandardModel67_3}.
In particular, its spin-parity quantum numbers are consistent with $J^{PC}=0^{++}$ according to measurements performed
by the CMS~\cite{Chatrchyan:2012jja,Chatrchyan:2013mxa,Khachatryan:2014kca,Khachatryan:2015mma,Khachatryan:2016tnr,Sirunyan:2017tqd,Sirunyan:2019twz}
and ATLAS~\cite{Aad:2013xqa,Aad:2015mxa,Aad:2016nal,Aaboud:2017oem,Aaboud:2017vzb,Aaboud:2018xdt} experiments.
It is still to be determined whether small anomalous couplings contribute to the $\PH\V\V$ or $\PH\f\f$ interactions,
where $\V$ stands for vector bosons and $\text{f}$ stands for fermions.
Because nonzero spin assignments of the \Hboson have been excluded~\cite{Khachatryan:2014kca,Aad:2015mxa},
we focus on the analysis of couplings of a spin-0 \Hboson.
Previous studies of anomalous $\PH\V\V$ couplings were performed by both the CMS and ATLAS
experiments using either decay-only
information~\cite{Chatrchyan:2012jja,Chatrchyan:2013mxa,Khachatryan:2014kca,Aad:2013xqa,Aad:2015mxa,Aaboud:2017oem},
including associated production
information~\cite{Khachatryan:2016tnr,Sirunyan:2017tqd,Sirunyan:2019twz,Aad:2016nal,Aaboud:2017vzb,Aaboud:2018xdt},
or including off-shell \Hboson production~\cite{Khachatryan:2015mma, Sirunyan:2019twz}.
In this paper, we report a study of $\PH\V\V$ couplings using information from production of the \Hboson decaying to $\Pgt$ leptons.
These results are combined with the previous CMS measurements using both
associated production and decay information in the $\PH\to4\ell$ channel~\cite{Sirunyan:2019twz},
resulting in stringent constraints on anomalous \Hboson couplings. Here and in the following
$\ell$ denotes an electron or muon.

\begin{figure*}[!bth]
\centering
\includegraphics[width=0.36\textwidth]{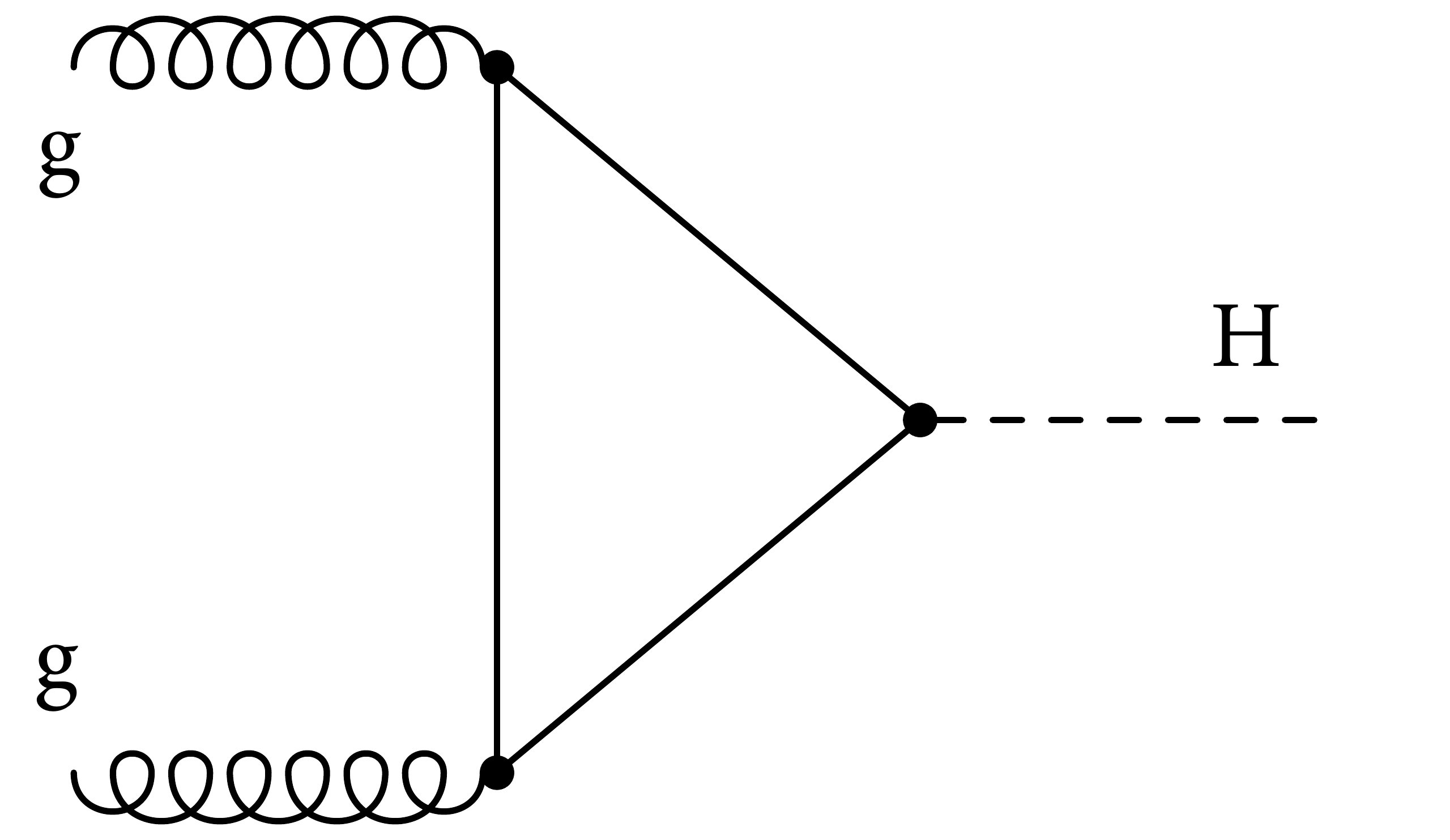}
\includegraphics[width=0.30\textwidth]{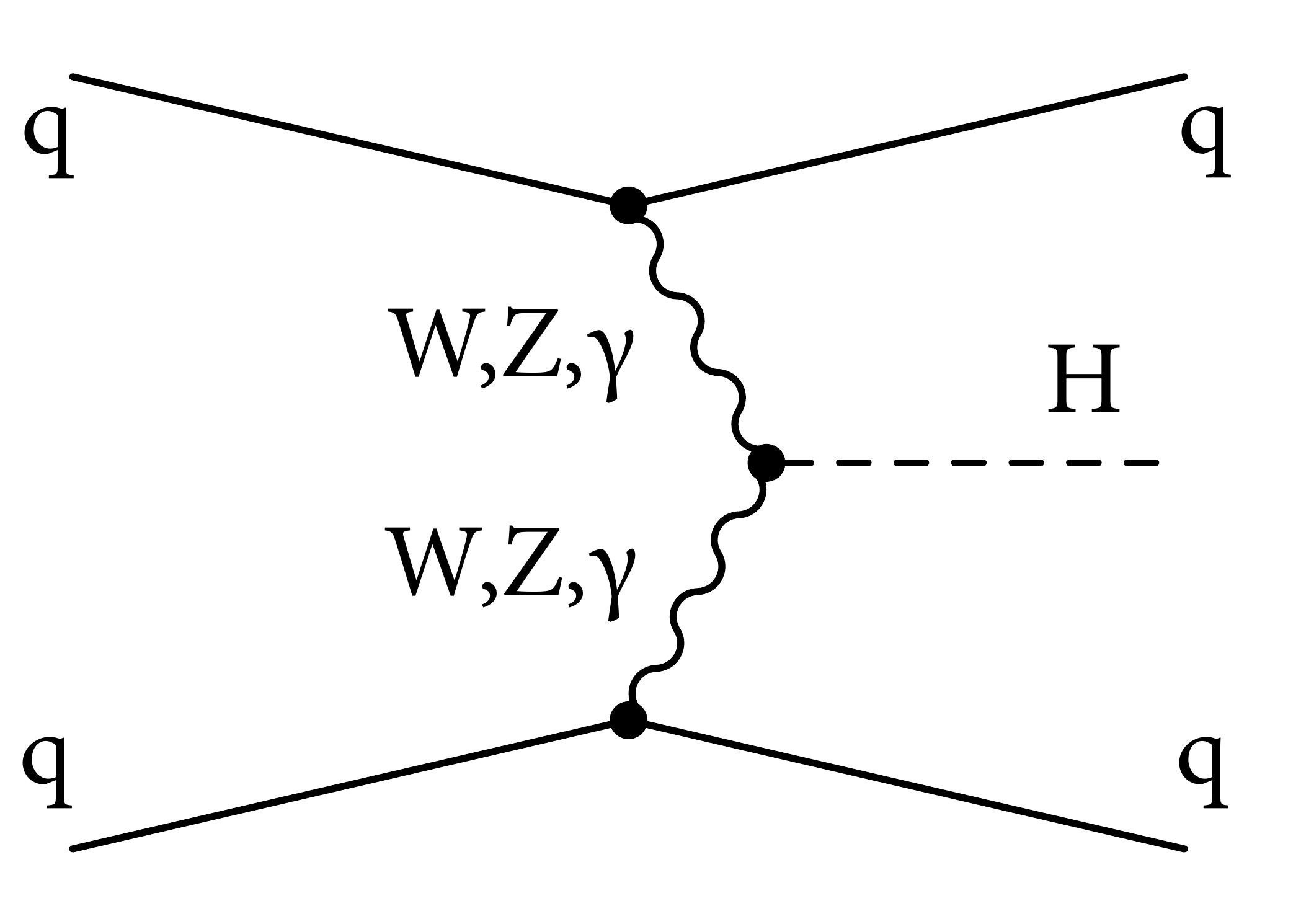}
\includegraphics[width=0.32\textwidth]{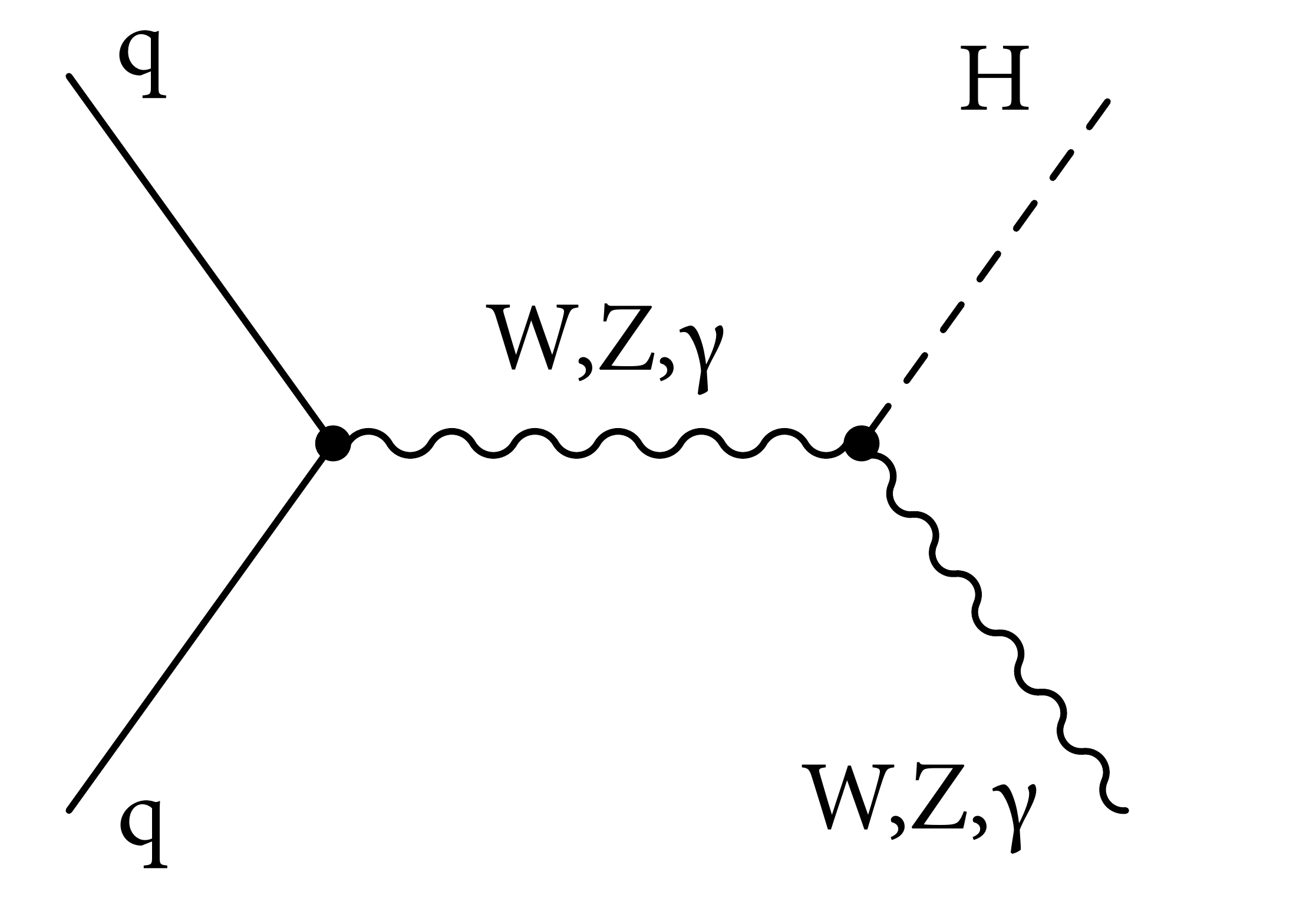}
\caption
{
Examples of leading-order Feynman diagrams for \Hboson production via the gluon fusion (left),
vector boson fusion (middle), and associated production with a vector boson (right).
The $\PH\PW\PW$ and $\PH\PZ\PZ$ couplings may appear at tree level, as the SM predicts.
Additionally, $\PH\PW\PW$, $\PH\PZ\PZ$, $\PH\PZ\gamma$, $\PH\gamma\gamma$, and $\PH\Pg\Pg$ couplings
may be generated by loops of SM or unknown particles,
as indicated in the left diagram but not shown explicitly in the middle and right diagrams.
\label{fig:diagrams}
}
\end{figure*}

The $\PH\to\Pgt\Pgt$ decay has been observed by the
CMS experiment, with over five standard deviation significance~\cite{Sirunyan:2017khh}.
The $\PH\to\Pgt\Pgt$ sample can be used to study the quantum numbers of the \Hboson and its anomalous couplings
to SM particles, including its $CP$ properties.
The dominant production mechanisms of the \Hboson considered in this paper are shown at leading order
in QCD in Fig.~\ref{fig:diagrams}. Anomalous $\PH\PW\PW$, $\PH\Z\Z$, $\PH\Z\gamma$, $\PH\gamma\gamma$, and $\PH\Pg\Pg$ couplings affect the correlations between the \Hboson, the beam line, and the two jets in vector boson fusion (VBF), in associated production with a vector boson decaying hadronically ($\V\PH$, where $\V=\PW, \PZ$), and also in gluon fusion production with additional two jets. The gluon fusion production with two additional jets appears at higher
order in QCD with an example of gluons appearing in place of the vector bosons shown in the VBF diagram in the middle of Fig.~\ref{fig:diagrams}.
A study of anomalous $\PH\ttbar$ couplings in associated production with top quarks, $\ttbar\PH$ or $\PQt\PQq\PH$,
and anomalous $\PH\Pgt\Pgt$ couplings in the decay of the \Hboson are also possible using $\Pgt\Pgt$ events~\cite{Dawson:2013bba}.
However, more data are needed to reach sensitivity to such anomalous effects,
and it has been confirmed that these anomalous couplings would not affect the measurements presented in this paper.

To increase the sensitivity to anomalous couplings in the \Hboson production, the matrix element likelihood approach
(\textsc{mela})~\cite{Chatrchyan:2012xdj,Gao:2010qx,Bolognesi:2012mm,Anderson:2013afp,Gritsan:2016hjl}
is utilized to form optimal observables. The analysis is optimized for VBF production and is not additionally optimized
for $\V\PH$ or gluon fusion production. However, all three production mechanisms are included in the analysis,
using a general anomalous coupling parametrization.
The $\PH\to\Pgt\Pgt$ channel has advantages over other \Hboson decay channels because of the
relatively high significance of the signal events in the VBF channel~\cite{Sirunyan:2017khh}.
Three mutually exclusive categories of events are reconstructed in the analysis:
the VBF category targets events with two associated jets in the VBF event topology,
the boosted category contains events with one jet or more jets if the event is not in the VBF category,
and the 0-jet category targets \Hboson events produced via gluon fusion without associated jets.
The simultaneous analysis of all three categories of events is necessary to boost the sensitivity to anomalous
$\PH\V\V$ couplings from events with partial kinematic information reconstructed in the non-VBF categories
and to normalize the relative contribution of different production mechanisms.

The analysis utilizes the same data, event selection, and categorization as Ref.~\cite{Sirunyan:2017khh} and is described in Sec.~\ref{sec:RecoSel}.
The phenomenological model and Monte Carlo (MC) simulation are described in
Sec.~\ref{sec:Phenomenology}. The matrix element techniques used to extract the kinematic information are discussed in
Sec.~\ref{sec:Kinematics}. The implementation of the likelihood fit using kinematic information in the events is presented
in Sec.~\ref{sec:Strategy}. The results are presented and discussed in Secs.~\ref{sec:Results} and~\ref{sec:Combination},
before conclusions are drawn in Sec.~\ref{sec:Summary}.

\section{CMS detector}
\label{sec:DetectorSamples}

The central feature of the CMS apparatus is a superconducting solenoid of 6\unit{m} internal diameter, providing a magnetic
field of 3.8\unit{T}. Within the solenoid volume, there are a silicon pixel and strip tracker, a lead tungstate crystal
electromagnetic calorimeter (ECAL), and a brass and scintillator hadron calorimeter, each composed of a barrel and two endcap
sections. Forward calorimeters extend the pseudorapidity, $\eta$, coverage provided by the barrel and endcap detectors.
Muons are detected in gas-ionization chambers embedded in the steel flux-return yoke outside the solenoid.

Events of interest are selected using a two-tiered trigger system~\cite{Khachatryan:2016bia}. The first level (L1),
composed of custom hardware processors, uses information from the calorimeters and muon detectors to select
events at a rate of around 100\unit{kHz} within a time interval of less than 4\mus. The second level, known as the
high-level trigger, consists of a farm of processors running a version of the full event reconstruction software
optimized for fast processing, and reduces the event rate to about 1\unit{kHz} before data storage.

A more detailed description of the CMS detector, together with a definition of the coordinate system used and
the relevant kinematic variables, can be found in Ref.~\cite{Chatrchyan:2008zzk}.

The data samples used in this analysis correspond to an integrated luminosity
of 35.9\fbinv collected in Run 2 of the LHC during 2016 at a center-of-mass energy of 13\TeV.

\section{Event reconstruction and selection}
\label{sec:RecoSel}

The analysis uses the same dataset, event reconstruction, and selection criteria as those used in the analysis
leading to the observation of the \Hboson decay to a pair of $\Pgt$ leptons~\cite{Sirunyan:2017khh}.

\subsection{Event reconstruction}
\label{subsec:Reco}

The reconstruction of observed and simulated events relies on the particle-flow (PF) algorithm~\cite{CMS-PRF-14-001},
which combines the information from the CMS subdetectors to identify and reconstruct particles emerging from $\Pp\Pp$ collisions.
Combinations of these PF candidates are used to reconstruct higher-level objects such as jets, $\Pgt$ candidates, or
missing transverse momentum, $\ptvecmiss$.
The reconstructed vertex with the largest value of summed physics object $\pt^2$ is taken to be the primary $\Pp\Pp$
interaction vertex, where $\pt$ is the transverse momentum. The physics objects are the objects constructed by a jet
finding algorithm~\cite{Cacciari:2008gp,Cacciari:2011ma} applied to all charged tracks associated with the vertex
and the corresponding associated missing transverse momentum.

Electrons are identified with a multivariate discriminant
combining several quantities describing the track quality, the shape of the energy deposits in the ECAL,
and the compatibility of the measurements from the tracker and the ECAL~\cite{Khachatryan:2015hwa}.
Muons are identified with requirements on the quality of the track reconstruction and on the number of measurements in the
tracker and the muon systems~\cite{Sirunyan:2018muon}.
To reject nonprompt or misidentified leptons, an isolation requirement $I^\ell$
is applied according to the criteria described in Ref.~\cite{Sirunyan:2017khh}.

Jets are reconstructed with an anti-\kt clustering algorithm~\cite{Cacciari:fastjet2}, as implemented in the \FASTJET
package~\cite{Cacciari:2011ma}.
It is based on the clustering of neutral and charged PF candidates within a distance parameter of 0.4. Charged PF candidates
not associated with the primary vertex of the interaction are not considered when building jets.
An offset correction is applied
to jet energies to take into account the contribution from additional
$\Pp\Pp$ interactions within the same or nearby bunch crossings.
In this analysis, jets are required to have $\pt>30$\GeV
and absolute pseudorapidity $\abs{\eta}<4.7$, and
to be separated from the selected leptons by a distance parameter
$\Delta R=\sqrt{\smash[b]{(\Delta\eta)^2+(\Delta\phi)^2}}$ of at least 0.5,
where $\phi$ is the azimuthal angle in radians.
The combined secondary vertex algorithm is used to identify jets that are
likely to originate from a bottom quark (``{\cPqb} jets'').
The algorithm exploits track-based lifetime information along with the secondary vertex of the jet to provide a
likelihood ratio discriminator for {\cPqb} jet identification.

Hadronically decaying $\Pgt$ leptons, denoted as $\tauh$, are reconstructed with the hadron-plus-strips
algorithm~\cite{Khachatryan:2015dfa, Sirunyan:2018pgf}, which is seeded with anti-\kt jets.
This algorithm reconstructs $\tauh$ candidates based on the
number of tracks and the number of ECAL strips with energy deposits within the
associated $\eta$-$\phi$ plane and reconstructs one-prong,
one-prong+$\PGpz$(s), and three-prong decay modes, identified as $M=1,2,$ and $3$, respectively.
A multivariate discriminator, including isolation
and lifetime information, is used to reduce the rate for
quark- and gluon-initiated jets to be identified as $\tauh$ candidates.
The working point used in this analysis has an efficiency of
about 60\% for genuine $\tauh$, with about 1\% misidentification
rate for quark- and gluon-initiated jets, for a $\pt$
range typical of $\tauh$ originating from a $\PZ$ boson.
Electrons and muons misidentified as $\tauh$ candidates are suppressed
using dedicated criteria based on the consistency between the
measurements in the tracker, the calorimeters, and the muon
detectors~\cite{Khachatryan:2015dfa, Sirunyan:2018pgf}.
The $\tauh$ energy scale as well as the rate and the energy
scale of electrons and muons misidentified as $\tauh$ candidates
are corrected in simulation to match those measured in
data~\cite{Sirunyan:2017khh}.

The missing transverse momentum is defined as
the negative vector sum of the transverse momenta of all PF
candidates~\cite{CMS-JME-17-001}.
The details of the corrections to $\ptvecmiss$ for the mismodeling in the simulation of $\PZ+\text{jets}$, $\PW+\text{jets}$,
and \Hboson processes are described in Ref.~\cite{Sirunyan:2017khh}.

Both the visible mass of the $\Pgt\Pgt$ system $\mvis$ and the invariant mass of the
$\Pgt\Pgt$ system $\mtautau$ are used in the analysis. The visible mass is defined as the invariant mass
of the visible decay products of the $\Pgt$ leptons. The observable $\mtautau$ is reconstructed
using the \textsc{svfit}~\cite{Bianchini:2014vza} algorithm, which combines
the \ptvecmiss and its uncertainty with the 4-vectors of both $\Pgt$ candidates
to calculate a more accurate estimate of the mass of the parent boson.
The estimate of the 4-momentum of the \Hboson provided by \textsc{svfit}
is used to calculate the kinematic observables discussed in Sec.~\ref{sec:Kinematics}.

\subsection{Event selection and categorization}
\label{subsec:Selection}

Selected events are classified according to four decay channels,
$\Pe\Pgm$, $\Pe\tauh$, $\Pgm\tauh$, and $\tauh\tauh$.
The resulting event samples are made mutually exclusive by
discarding events that have additional loosely identified
and isolated electrons or muons.

The largest irreducible source of background is Drell-Yan production of
$\PZ\to\Pgt\Pgt$, while the dominant background sources with jets misidentified as leptons are QCD multijet and
$\PW+\text{jets}$.
Other contributing background sources are $\ttbar$, single top, $\PZ\to\ell\ell$, and diboson production.

The two leptons assigned to the \Hboson decay are required to have opposite charges.
The trigger requirements, geometrical acceptances, and transverse momentum
criteria are summarized in Table~\ref{tab:inclusive_selection}.
The $\pt$ thresholds in the lepton selections are optimized to increase the sensitivity to the $\PH\to\Pgt\Pgt$ signal,
while also satisfying the trigger requirements. The pseudorapidity requirements are driven by reconstruction and trigger requirements.

\begin{table*}[htbp]
\centering
\topcaption{Kinematic selection criteria for the four decay channels. For the trigger threshold requirements,
the numbers indicate the trigger thresholds in \GeV. The lepton selection criteria   include the transverse
momentum threshold, pseudorapidity range, as well as isolation criteria.
\label{tab:inclusive_selection}
}
\begin{scotch}{llll}
  Channel           &         Trigger requirement              &    \multicolumn{2}{c}{Lepton selection}                 \\
 & $\pt$ ($\GeVns{}$) & $\pt$ ($\GeVns{}$) & $\eta$  \\
\hline
$\Pe\Pgm$        &         $\pt^{\Pe}>12\,\&\,\pt^{\Pgm}>23$    &     $\pt^{\Pe}>13$ & $\abs{\eta^\Pe}<2.5$  \\
                          &              &     $\pt^{\Pgm}>24$ & $\abs{\eta^\Pgm}<2.4$  \\[\cmsTabSkip]
         &         $\pt^{\Pe}>23\,\&\,\pt^{\Pgm}>8$    &     $\pt^{\Pe}>24$ & $\abs{\eta^\Pe}<2.5$        \\
                          &          &     $\pt^{\Pgm}>15$ & $\abs{\eta^\Pgm}<2.4$      \\
[\cmsTabSkip]
  $\Pe\tauh$        &         $\pt^{\Pe}>25$     &     $\pt^\Pe>26$  & $\abs{\eta^\Pe}<2.1$    \\
                          &       &     $\pt^{\tauh}>30$ &  $\abs{\eta^{\tauh}}<2.3$   \\
[\cmsTabSkip]
  $\mu\tauh$       &         $\pt^{\Pgm}>22$     &     $\pt^\Pgm>23$  &  $\abs{\eta^\Pgm}<2.1$         \\
                          &       &     $\pt^{\tauh}>30$ &  $\abs{\eta^{\tauh}}<2.3$   \\[\cmsTabSkip]

                   &         $\pt^{\Pgm}>19\,\&\,\pt^{\tauh}>21$     &     $20<\pt^\Pgm<23$  &  $\abs{\eta^\Pgm}<2.1$        \\
                          &       &     $\pt^{\tauh}>30$ &  $\abs{\eta^{\tauh}}<2.3$  \\
[\cmsTabSkip]
 $\tauh\tauh$    &         $\pt^{\tauh}>35\,\&\,35$              &     $\pt^{\tauh}>50\,\&\,40$ & $\abs{\eta^{\tauh}}<2.1$              \\
\end{scotch}
\end{table*}

In the $\ell\tauh$ channels, the large $\PW+\text{jets}$ background
is reduced by requiring the transverse mass, $\mT$, to be less than 50\GeV.
The transverse mass is defined as follows,
\begin{equation}
\mT \equiv \sqrt{\smash[b]{2 \pt^\ell \ptmiss [1-\cos(\Delta\phi)]}},
\end{equation}
where $\pt^\ell$ is the transverse momentum of the electron or muon
and $\Delta\phi$ is the azimuthal angle between the lepton direction and the \ptvecmiss direction.

In the $\emu$ channel, the \ttbar background is reduced by requiring
$p_\zeta - 0.85 \, p_\zeta^{\text{vis}} > -35$\GeV or $-10$\GeV depending on the category,
where $p_\zeta$ is the component of \ptvecmiss along the bisector of
the transverse momenta of the two leptons and $p_\zeta^{\text{vis}}$
is the sum of the components of the lepton transverse momenta along
the same direction~\cite{Khachatryan:2014wca}.
In addition, events with a {\cPqb}-tagged jet are discarded to further suppress
the \ttbar background in this channel.

In the same way as in Ref.~\cite{Sirunyan:2017khh}, the event samples are split into three
mutually exclusive production categories:
\begin{itemize}
\item {0-jet category}: This category targets \Hboson events produced via gluon fusion.
Events containing no jets with $\pt>30$\GeV are selected.
Simulations indicate that about 98\% of signal events in the 0-jet category arise from the gluon fusion production mechanism.\\
\item {VBF category}: This category targets \Hboson events produced via the VBF process.
Events are selected with exactly (at least) two jets with $\pt>30$\GeV in the
$\Pe\Pgm$ ($\Pe\tauh$, $\Pgm\tauh$, and $\tauh\tauh$) channels.
In the $\Pgm\tauh$, $\Pe\tauh$, and $\Pe\Pgm$ channels, the two leading jets
are required to have an invariant mass, $\mjj$, larger than 300\GeV.
The vector sum of the $\ptvecmiss$ and the $\ptvec$ of the visible decay products of the tau leptons,
defined as $\ptvech$, is required to have a magnitude greater than 50 (100)\GeV in the $\ell\tauh$ ($\tauh\tauh$) channels.
In addition, the $\pt$ threshold on the $\tauh$ candidate is raised to
40\GeV in the $\Pgm\tauh$ channel, and the two leading jets in the
$\tauh\tauh$ channel must be separated in pseudorapidity by $\abs{\Delta\eta}>2.5$.
Depending on the decay channel, up to 57\% of the signal
events in the VBF category is produced via VBF.
This fraction increases with $\mjj$. Gluon fusion production makes 40\%-50\% of the total signal,
while the $\V\PH$ contribution is less than 3\%. \\
\item {Boosted category}: This category contains all the events that do not
enter one of the previous categories, namely events with one jet and events
with several jets that fail the requirements of the VBF category. It targets events with a
\Hboson produced in gluon fusion and recoiling against an initial state radiation jet.
It contains gluon fusion events produced in association with one or more
jets (78\%-80\% of the signal events), VBF events in which one of the jets has escaped
detection or events with low $\mjj$ (11\%-13\%), as well as \Hboson events produced in
association with a $\PW$ or a $\PZ$ boson decaying hadronically (4\%-8\%).
\end{itemize}

In addition to these three signal regions for each channel, a series of control regions targeting
different background processes are included in the maximum likelihood fit used to extract the results of the analysis.
The normalization of the $\PW+\text{jets}$ background in the $\Pe\tauh$ and $\Pgm\tauh$ channels is estimated
from simulations, and adjusted to data using control regions obtained by applying all selection criteria,
with the exception that $\mT$ is required to be greater than 80\GeV instead of less than 50\GeV. 
An uncertainty on the extrapolation from the control region to the signal region is determined in the same way as described in Ref.~\cite{Sirunyan:2017khh}.
The normalization of the QCD multijet background in the $\Pe\tauh$ and $\Pgm\tauh$ channels is estimated
from events where the electron or the muon has the same charge as
the $\tauh$ candidate. The contributions from Drell--Yan, $\ttbar$, diboson, and $\PW+\text{jets}$ processes are subtracted. The factor to extrapolate from the same-sign to the opposite-sign region is determined by comparing the yield of the QCD multijet background for events with $\ell$ candidates passing inverted isolation criteria, in the same-sign and opposite-sign regions. It is constrained by adding the opposite-sign region, where the $\ell$ candidates pass inverted isolation criteria, to the global fit.

In the $\tauh\tauh$ channel, the QCD multijet background is estimated from events where the $\tauh$ candidates pass relaxed
isolation conditions, and the extrapolation factor is derived from events where the $\tauh$ candidates have charges
of the same sign. The events selected with opposite-sign $\tauh$ candidates passing relaxed isolation requirements form a control
region included in the global fit.
Finally, the normalization of the $\ttbar$ background is adjusted using a control region defined similarly to the $\Pe\Pgm$ signal region, except that the $p_\zeta$ requirement is inverted and the events are required to contain at least one jet.

\section{Phenomenology of anomalous couplings and simulation}
\label{sec:Phenomenology}

We follow the formalism used in the study of anomalous couplings in earlier analyses by
CMS~\cite{Chatrchyan:2012jja,Chatrchyan:2013mxa,Khachatryan:2014kca,Khachatryan:2015mma,Khachatryan:2016tnr,Sirunyan:2017tqd,Sirunyan:2019twz}.
The theoretical approach is described in
Refs.~\cite{Plehn:2001nj,Hankele:2006ma,Accomando:2006ga,Hagiwara:2009wt,Gao:2010qx,DeRujula:2010ys,
Bolognesi:2012mm,Ellis:2012xd,Artoisenet:2013puc,Anderson:2013afp,Dolan:2014upa,Greljo:2015sla,Gritsan:2016hjl}.
Anomalous interactions of a spin-0 \Hboson with two spin-1 gauge bosons $\V\V$,
such as $\PW\PW$, $\PZ\PZ$, $\PZ\gamma$, $\gamma\gamma$, and $\Pg\Pg$, are parametrized
by a scattering amplitude that includes three tensor structures with expansion of coefficients up to $(q^2/\Lambda^2)$
\ifthenelse{\boolean{cms@external}}{
\begin{multline}
A(\PH\V\V) \sim
\left[ a_{1}^{\V\V}
+ \frac{\kappa_1^{\V\V}q_{1}^2 + \kappa_2^{\V\V} q_{2}^{2}}{\left(\Lambda_{1}^{\V\V} \right)^{2}}
\right]
m_{\V1}^2 \epsilon_{\V1}^* \epsilon_{\V2}^*\\
+ a_{2}^{\V\V}  f_{\mu \nu}^{*(1)}f^{*(2)\mu\nu}
+ a_{3}^{\V\V}   f^{*(1)}_{\mu \nu} {\tilde f}^{*(2)\mu\nu},
\label{eq:formfact-fullampl-spin0}
\end{multline}
}{
\begin{equation}
A(\PH\V\V) \sim
\left[ a_{1}^{\V\V}
+ \frac{\kappa_1^{\V\V}q_{1}^2 + \kappa_2^{\V\V} q_{2}^{2}}{\left(\Lambda_{1}^{\V\V} \right)^{2}}
\right]
m_{\V1}^2 \epsilon_{\V1}^* \epsilon_{\V2}^*
+ a_{2}^{\V\V}  f_{\mu \nu}^{*(1)}f^{*(2)\mu\nu}
+ a_{3}^{\V\V}   f^{*(1)}_{\mu \nu} {\tilde f}^{*(2)\mu\nu},
\label{eq:formfact-fullampl-spin0}
\end{equation}
}
where $q_{i}$, $\epsilon_{Vi}$, and $m_{\V1}$ are the 4-momentum, polarization vector,
and pole mass of the gauge boson, indexed by $i=1,2$.
The gauge boson's field strength tensor and dual field strength tensor are
$f^{(i){\mu \nu}} = \epsilon_{{\V}i}^{\mu}q_{i}^{\nu} - \epsilon_{{\V}i}^\nu q_{i}^{\mu}$
and ${\tilde f}^{(i)}_{\mu \nu} = \frac{1}{2} \epsilon_{\mu\nu\rho\sigma} f^{(i)\rho\sigma}$.
The coupling coefficients $a_{i}^{\V\V}$, which multiply the three tensor structures,
and $\kappa_i^{\V\V}/(\Lambda_{1}^{\V\V})^2$,
which multiply the next term in the $q^2$ expansion for the first tensor structure,
are to be determined from data, where $\Lambda_{1}$ is the scale of beyond the SM (BSM) physics.

In Eq.~(\ref{eq:formfact-fullampl-spin0}), the only nonzero SM contributions at tree level are $a_{1}^{\PW\PW}$ and $a_{1}^{\PZ\PZ}$,
which are assumed to be equal under custodial symmetry.
All other ${\PZ\PZ}$ and ${\PW\PW}$ couplings are considered anomalous contributions,
which are either due to BSM physics or small contributions arising in the SM due to
loop effects and are not accessible with the current precision.
As the event kinematics of the \Hboson production in ${\PW\PW}$ fusion and in $\PZ\PZ$ fusion are very similar,
they are analyzed together assuming $a_i^{\PW\PW}=a_i^{\PZ\PZ}$ and
$\kappa_i^{\PZ\PZ}/\left(\Lambda_{1}^{\PZ\PZ}\right)^2=\kappa_i^{\PW\PW}/\left(\Lambda_{1}^{\PW\PW}\right)^2$.
The results can be reinterpreted for any other relationship between the $a_i^{\PW\PW}$ and $a_i^{\PZ\PZ}$
couplings~\cite{Sirunyan:2019twz}.
For convenience, we refer to these parameters as $a_i$, $\kappa_i$, and $\Lambda_1$, without the superscripts.
Among the anomalous contributions, considerations of symmetry and gauge invariance require
$\kappa_1^{\PZ\PZ}=\kappa_2^{\PZ\PZ}=-\exp({i\phi^{\PZ\PZ}_{\Lambda{1}}})$,
$\kappa_1^{\gamma\gamma}=\kappa_2^{\gamma\gamma}=0$,
$\kappa_1^{\Pg\Pg}=\kappa_2^{\Pg\Pg}=0$,
$\kappa_1^{\PZ\gamma}=0$, and $\kappa_2^{\PZ\gamma}=-\exp({i\phi^{\PZ\gamma}_{\Lambda{1}}})$,
where $\phi^{\V\V}_{\Lambda{1}}$ is the phase of the corresponding coupling.
In the case of the $\gamma\gamma$ and $\Pg\Pg$ couplings, the only contributing terms
are $a_2^{\gamma\gamma,\Pg\Pg}$ and $a_3^{\gamma\gamma,\Pg\Pg}$. Our earlier measurements in Ref.~\cite{Khachatryan:2014kca} indicated substantially tighter limits on $a_2^{\gamma\gamma,\PZ\gamma}$ and $a_3^{\gamma\gamma,\PZ\gamma}$ couplings from $\PH\to\PZ\gamma$ and $\PH\to \gamma\gamma$ decays with on-shell photons than from measurements with virtual photons, so we do not pursue measurements of these parameters in this paper.
The coupling $a_2^{\Pg\Pg}$ refers to a SM-like contribution in the gluon fusion process,
and $a_3^{\Pg\Pg}$ corresponds to a $CP$-odd anomalous contribution.
There are four other anomalous couplings targeted in this analysis:
two from the first term of Eq.~(\ref{eq:formfact-fullampl-spin0}),
$\Lambda_{1}^{\PZ\PZ}=\Lambda_1^{\PW\PW}=\Lambda_{1}$ and $\Lambda_{1}^{\PZ\gamma}$;
one coming from the second term,  $a_2^{\PZ\PZ}=a_2^{\PW\PW}=a_{2}$;
and one coming from the third term, $a_3^{\PZ\PZ}=a_3^{\PW\PW}=a_{3}$.
The $a_{3}$ coupling corresponds to the $CP$-odd amplitude, and its interference with a $CP$-even amplitude would result in $CP$ violation.

It is convenient to measure the effective cross section ratios $f_{ai}$ rather than the
anomalous couplings $a_i$ themselves, as most uncertainties cancel in the ratio.
Moreover, the effective fractions are conveniently bounded between 0 and 1, independent of the coupling
convention. The effective fractional cross sections $f_{ai}$ and phases $\phi_{ai}$ are defined as follows,
\begin{widetext}
\begin{equation}\begin{aligned}
f_{a3} &= \frac{\abs{a_3}^2 \sigma_{3}}{\abs{a_1}^2 \sigma_{1} + \abs{a_2}^2 \sigma_{2} + \abs{a_3}^2 \sigma_{3} + \tilde{\sigma}_{\Lambda{1}}/\left(\Lambda_{1}\right)^{4} +\ldots}, \qquad & \phi_{a3} = \text{arg}\left(\frac{a_{3}}{a_{1}}\right), \\
f_{a2} &= \frac{\abs{a_2}^2 \sigma_{2}}{\abs{a_1}^2 \sigma_{1} + \abs{a_2}^2 \sigma_{2} + \abs{a_3}^2 \sigma_{3} + \tilde{\sigma}_{\Lambda{1}}/\left(\Lambda_{1}\right)^{4} +\ldots}, \qquad & \phi_{a2} = \text{arg}\left(\frac{a_{2}}{a_{1}}\right), \\
f_{\Lambda1} &= \frac{\tilde{\sigma}_{\Lambda1}/\left(\Lambda_{1}\right)^{4}}{\abs{a_1}^2 \sigma_{1} + \abs{a_2}^2 \sigma_{2} + \abs{a_3}^2 \sigma_{3} + \tilde{\sigma}_{\Lambda{1}}/\left(\Lambda_{1}\right)^{4} +\ldots}, \qquad & \phi_{\Lambda1}, \\
f^{\PZ\gamma}_{\Lambda1} & = \frac{\tilde{\sigma}^{\PZ\gamma}_{\Lambda1}/\left(\Lambda^{\PZ\gamma}_{1}\right)^{4}}{\abs{a_1}^2 \sigma_{1} + \tilde{\sigma}^{\PZ\gamma}_{\Lambda1}/\left(\Lambda^{\PZ\gamma}_{1}\right)^{4} +\ldots}, \qquad & \phi^{\PZ\gamma}_{\Lambda1} ,
\label{eq:fa_definitions}
\end{aligned}\end{equation}
\end{widetext}
where $\sigma_i$ is the cross section for the process corresponding to $a_i=1$ and all other couplings are set to zero.
Since the production cross sections depend on the parton distribution functions (PDFs), the definition with respect to the decay process is more convenient.
The cross section ratios defined in the $\PH\to 2\Pe2\mu$ decay analysis~\cite{Chatrchyan:2013mxa} are adopted. Their values are
$\sigma_{1}/\sigma_{3}=6.53$, $\sigma_{1}/\sigma_{2}=2.77$,
$(\sigma_{1}/{\tilde\sigma_{\Lambda1}})\times\TeV^{4}=1.47\times 10^{4}$, and
$(\sigma_{1}/{\tilde\sigma_{\Lambda1}^{\PZ\gamma}})\times\TeV^{4}=5.80\times 10^{3}$,
as calculated using the \textsc{JHUGen}~7.0.2 event generator \cite{Gao:2010qx,Bolognesi:2012mm,Anderson:2013afp,Gritsan:2016hjl}.
The ellipsis ($\ldots$) in Eq.~(\ref{eq:fa_definitions}) indicates any other contribution not listed explicitly.
Under the assumption that the couplings in Eq.~(\ref{eq:formfact-fullampl-spin0}) are constant and real, the above formulation is equivalent to an effective Lagrangian notation. Therefore, in this paper, the real coupling constants are tested, which means only $\phi_{ai}=0$ or $\pi$ are allowed. The constraints are set on the product $f_{ai}\cos(\phi_{ai})$, which ranges from $-1$ to +1.

Anomalous effects in the $\PH\to\Pgt\Pgt$ decay and $\ttbar\PH$ production are described by the $\PH\f\f$ couplings
of the \Hboson to fermions, with generally two couplings  $\kappa_f$ and $\tilde\kappa_f$, $CP$-even and $CP$-odd, respectively.
Similarly, if the gluon coupling $\PH\Pg\Pg$ is dominated by the top quark loop, it can be described with the
$\kappa_t$ and $\tilde\kappa_t$ parameters. However, since other heavy states may contribute to the loop,
we consider the effective $\PH\Pg\Pg$ coupling using the more general parametrization
given in Eq.~(\ref{eq:formfact-fullampl-spin0}) instead of explicitly including the quark loop.
In particular, the effective cross section fraction in gluon fusion becomes
\begin{equation}
f_{a3}^{\Pg\Pg\PH} = \frac{\abs{a_3^{\Pg\Pg\PH}}^2}{\abs{a_2^{\Pg\Pg\PH}}^2 + \abs{a_3^{\Pg\Pg\PH}}^2},
\label{eq:fa3ggH_definition}
\end{equation}
where the cross sections $\sigma^{\Pg\Pg\PH}_{2}=\sigma_{3}^{\Pg\Pg\PH}$ drop out from the equation following the coupling convention in Eq.~(\ref{eq:formfact-fullampl-spin0}).

Experimentally observable effects resulting from the above anomalous couplings are discussed in the next section.
In this paper, anomalous $\PH\PW\PW$, $\PH\PZ\PZ$, and $\PH\PZ\gamma$ couplings are considered
in VBF and $\V\PH$ production, and anomalous $\PH\Pg\Pg$ couplings are considered in gluon fusion.
Since $CP$-violating effects in electroweak (VBF and $\V\PH$) and gluon fusion production modify
the same kinematic distributions, both $CP$-sensitive parameters, $f_{a3}$ and $f_{a3}^{\Pg\Pg\PH}$,
are left unconstrained simultaneously.
It has been checked that $CP$ violation in $\PH\to\Pgt\Pgt$ decays would not affect these measurements.
Under the assumption that the couplings are constant and real, the above formulation is equivalent
to an effective Lagrangian notation. Therefore, in this paper, the real coupling constants are tested
and results are presented for the product of $f_{ai}$ and $\cos(\phi_{ai})$, the latter being the sign
of the real ratio of couplings $a_i/a_1$.

Following the formalism discussed in this section, simulated samples of \Hboson events produced via anomalous
$\PH\V\V$ couplings (VBF, $\V\PH$, gluon fusion in association with two jets) are generated using \textsc{JHUGen}.
The associated production in gluon fusion with two jet is affected by anomalous interactions,
while the kinematics of the production with zero or one jet are not affected. The latter events
are generated with \POWHEG~2.0~\cite{Frixione:2007vw,Bagnaschi:2011tu,Nason:2009ai,Luisoni:2013kna},
which is used for yield normalization of events selected with two jets and for the description of event
distributions in categories of events where the correlation of the two jets is not important.
For the kinematics relevant to this analysis in VBF and $\V\PH$ production,
the effects that appear at next-to-leading order (NLO) in QCD are well approximated by
the leading-order (LO) QCD matrix elements used in \textsc{JHUGen}, combined with parton showering.
The \textsc{JHUGen} samples produced with the SM couplings are compared with the equivalent samples generated
by the \POWHEG event generator at NLO QCD, with parton showering applied in both cases,
and the kinematic distributions are found to agree.

The \PYTHIA~8.212~\cite{Sjostrand:2014zea} event generator is used to model the \Hboson decay to $\Pgt$ leptons
and the decays of the $\Pgt$ leptons.
Both scalar and pseudoscalar $\PH\to\Pgt\Pgt$ decays and their interference have been modeled
to confirm that the analysis does not depend on the decay model.
The default samples are generated with the scalar hypothesis in decay.
The PDFs used in the generators are NNPDF30~\cite{Ball:2011uy}, with their precision
matching that of the matrix elements.
All MC samples are further processed through a dedicated simulation of the CMS detector
based on \GEANTfour~\cite{Agostinelli2003250}.

To simulate processes with anomalous \Hboson couplings, for each type of anomalous coupling
we generate events with both the pure anomalous term and its interference with the SM contribution
in the production $\PH\V\V$ interaction.
This allows extraction of the various coupling components and their interference.
The \textsc{mela} package,
based on \textsc{JHUGen} matrix elements, permits the application of weights to events in any sample to model any other
$\PH\V\V$ or $\PH\f\f$ couplings with the same production mechanism.
Reweighting enables one to increase the effective simulated event count by using all samples at once
to describe any model, even if it has not been simulated.
The \textsc{mela} package also allows calculation of optimal discriminants for further analysis,
as discussed in Sec.~\ref{sec:Kinematics}.

Simulated samples for the modeling of background processes and of the \Hboson signal processes with SM couplings are the same as those used for the observation of the \Hboson decay to a pair of $\Pgt$ leptons~\cite{Sirunyan:2017khh}. All the corrections applied to samples are the same as in Ref.~\cite{Sirunyan:2017khh}.
The \aMCATNLO~\cite{Alwall:2014hca} generator is used for $\PZ+\text{jets}$ and $\PW+\text{jets}$ processes.
They are simulated at LO with the MLM jet matching and merging~\cite{Alwall:2007fs}.
The \aMCATNLO generator is also used for diboson production simulated at NLO with the FxFx jet
matching and merging~\cite{Frederix:2012ps}, whereas $\POWHEG$ versions 2.0 and 1.0 are used for $\ttbar$ and
single top quark production, respectively.
The generators are interfaced with \PYTHIA to model the parton showering
and fragmentation. The \PYTHIA parameters affecting the description
of the underlying event are set to the {CUETP8M1} tune~\cite{Khachatryan:2015pea}.

\section{Discriminant distributions}
\label{sec:Kinematics}

\begin{figure*}[!htb]
\centering
\includegraphics[width=0.45\textwidth]{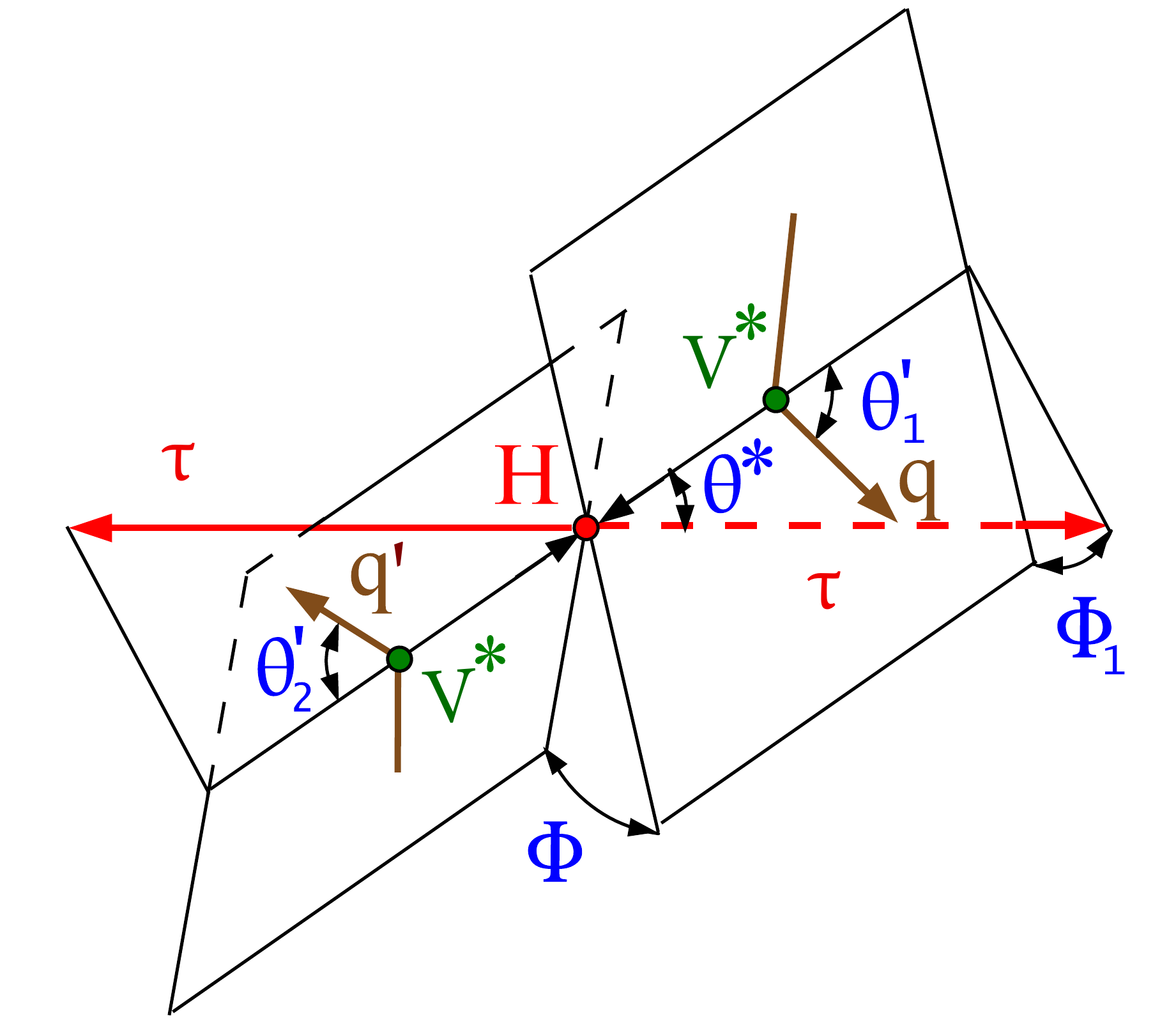}
\includegraphics[width=0.45\textwidth]{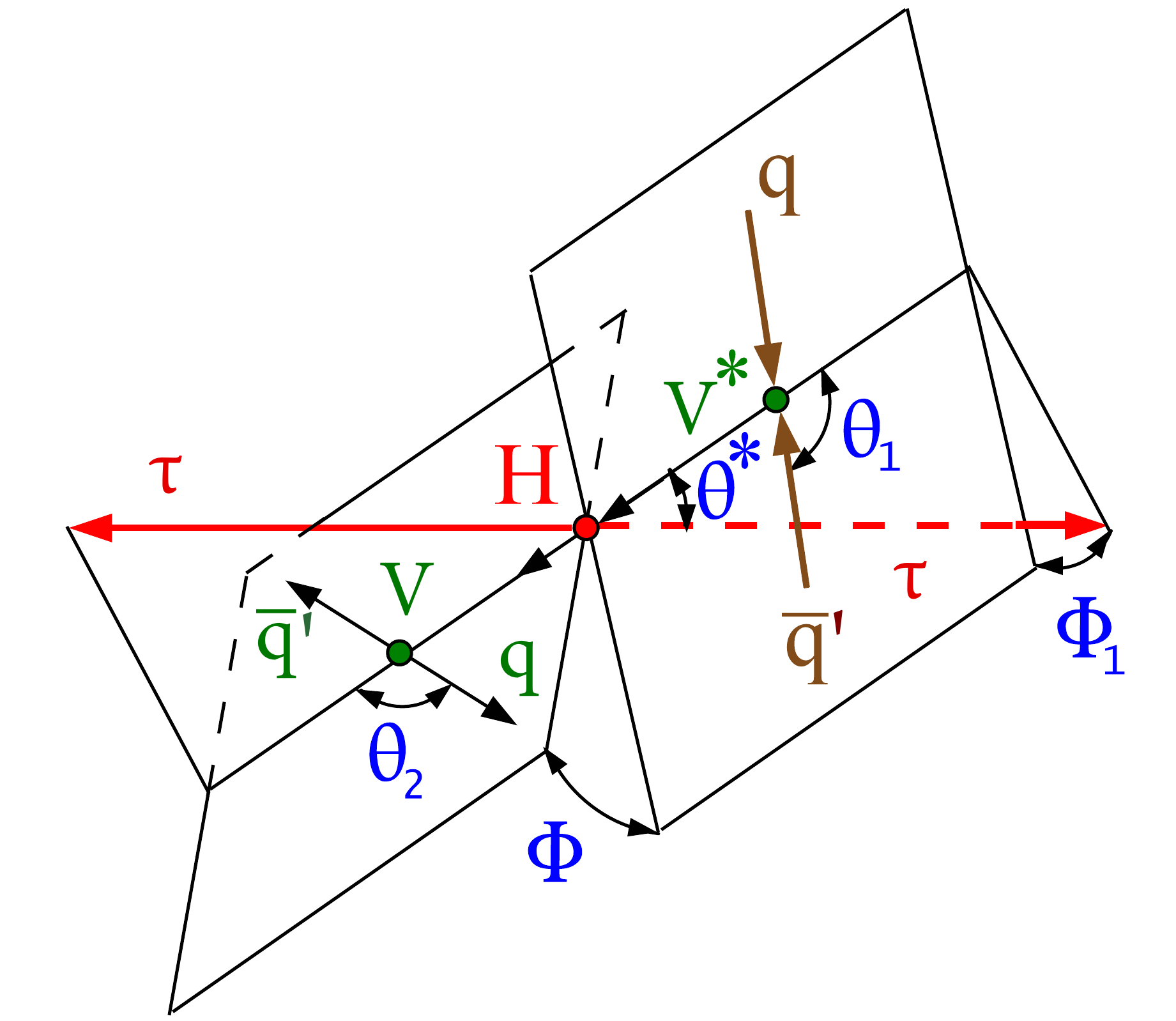}
\caption
{
Illustrations of \Hboson production in
$\Pq{\Pq^\prime}\to \Pg\Pg(\Pq{\Pq^\prime})\to \PH(\Pq{\Pq^\prime})\to \Pgt\Pgt(\Pq{\Pq^\prime})$
or VBF $\Pq{\Pq^\prime}\to \V^*\V^*(\Pq{\Pq^\prime})\to \PH(\Pq{\Pq^\prime})\to \Pgt\Pgt(\Pq{\Pq^\prime})$ (left)
and in associated production $\Pq\bar{\Pq}^\prime\to \V^*\to \V\PH \to \Pq\bar{\Pq}^\prime \Pgt\Pgt$ (right).
The $\PH\to \Pgt\Pgt$ decay is shown without further illustrating the $\Pgt$ decay chain.
Angles and invariant masses fully characterize the orientation of the production and two-body decay chain and
are defined in suitable rest frames of the $\V$ and $\PH$ bosons, except in the VBF case, where only the \Hboson
rest frame is used~\cite{Gao:2010qx,Anderson:2013afp}.
\label{fig:kinematics}
}
\end{figure*}

The full kinematic information for both production and decay of the \Hboson can be extracted from each event.
This paper focuses on the production process, illustrated in Fig.~\ref{fig:kinematics}.
The techniques discussed below are similar to those used in earlier analyses by CMS, such as in Ref.~\cite{Sirunyan:2019twz}.

Sensitivity to quantum numbers and anomalous couplings of the \Hboson is provided by the angular correlations between the two jets, the \Hboson, and the beam line direction in VBF, in $\V\PH$, and also in gluon fusion production with additional two jets.
A set of observables could be defined in VBF or $\V\PH$ production, such as
$\vec\Omega=\{\theta_1$, $\theta_2$, $\Phi$, $\theta^*$, $\Phi_1$, $q_1^{2}$, $q_2^{2} \}$
for the VBF or $\V\PH$ process with the angles illustrated in Fig.~\ref{fig:kinematics} and the $q_1^{2}$ and $q_2^{2}$
discussed in reference to Eq.~(\ref{eq:formfact-fullampl-spin0}), as described in detail in Ref.~\cite{Anderson:2013afp}.
It is, however, a challenging task to perform an optimal analysis in a multidimensional space of observables.
The \textsc{mela} is designed to reduce the number of observables to the minimum while retaining all essential information
for the purpose of a particular measurement. In this analysis, the background suppression is still provided by the
observables defined in Ref.~\cite{Sirunyan:2017khh}.

When the \Hboson and two associated jets are reconstructed, two types of discriminants can be used to optimally
search for anomalous couplings. These two discriminants
rely only on signal matrix elements and are well defined. One can apply the Neyman-Pearson lemma~\cite{Neyman289}
to prove that the two discriminants constitute a minimal and complete set of optimal observables~\cite{Anderson:2013afp,Gritsan:2016hjl}
for the measurement of the $f_{ai}$ parameter.
One type of discriminant is designed to separate the process with anomalous couplings, denoted as BSM, from the SM signal process,
\begin{equation}
\mathcal{D}_\mathrm{BSM} = \frac{\mathcal{P}_\mathrm{SM}(\vec\Omega) }{\mathcal{P}_\mathrm{SM}(\vec\Omega) + \mathcal{P}_\mathrm{BSM}(\vec\Omega)},
\label{eq:melaDbsm}
\end{equation}
where $\mathcal{P}$ is the probability for the signal VBF production process (either SM or BSM), calculated using the matrix element \textsc{mela} package
and is normalized so that the matrix elements give the same cross sections for either $f_{ai}=0$ or $1$ in the relevant phase space of each process.
Such a normalization leads to an optimal population of events in the range between 0 and 1.
The discriminants are denoted as $\mathcal{D}_\mathrm{0-}$,
$\mathcal{D}_\mathrm{0h+}$, $\mathcal{D}_{\Lambda1}$, or $\mathcal{D}_{\Lambda1}^{\PZ\gamma}$,
depending on the targeted anomalous coupling $a_3$, $a_2$, $\Lambda_{1}$, or $\Lambda_{1}^{\PZ\gamma}$, respectively.

The second type of discriminant targets the contribution from interference between the SM and BSM processes,
\begin{equation}
\mathcal{D}_\text{int} = \frac{ \mathcal{P}_\mathrm{SM-BSM}^\text{int}(\vec\Omega)}
{\mathcal{P}_\mathrm{SM}(\vec\Omega)+\mathcal{P}_\mathrm{BSM}(\vec\Omega)},
\label{eq:melaDint}
\end{equation}
where $\mathcal{P}_\mathrm{SM-BSM}^\text{int}$ is the probability distribution for interference of SM
and BSM signals in VBF production. This discriminant is used only for the $CP$-odd amplitude
analysis with $f_{a3}$ and is denoted $\mathcal{D}_{CP}$ in the rest of the paper.  In the cases of $f_{\Lambda1}$
and $f_{\Lambda1}^{\PZ\gamma}$, the interference discriminants do not carry additional information because of
their high correlation with the $\mathcal{D}_{\Lambda1}$ and $\mathcal{D}_{\Lambda1}^{\PZ\gamma}$ discriminants.
The $f_{a2}$ interference discriminant is not used in this analysis either, as it only becomes important for measurements
of smaller couplings than presently tested and because of the limited number of events available for background parametrization.

Kinematic distributions of associated particles in gluon fusion
are also sensitive to the quantum numbers of the \Hboson and to anomalous $\PH\Pg\Pg$ couplings.
A set of observables, ${\vec\Omega}$, identical to those from the VBF process also describes this process.
In this analysis, the focus is on the VBF-enhanced phase space in which the selection efficiency for the gluon
fusion process is relatively small. Furthermore, the observables defined in Eqs.~(\ref{eq:melaDbsm}) and~(\ref{eq:melaDint})
for the VBF process are found to provide smaller separation between $CP$-even and $CP$-odd
\Hboson couplings for gluon fusion production than \textsc{mela} discriminants that would be dedicated to the gluon fusion process.
Nonetheless, both parameters sensitive to $CP$ violation, $f_{a3}$ and $f_{a3}^{\Pg\Pg\PH}$,
are included in a simultaneous fit using the observables optimized for the VBF process to avoid any possible bias
in the measurement of $f_{a3}$.

While the correlations between the two jets, the \Hboson, and the beam line
provide primary information about $CP$ violation and anomalous couplings
in electroweak production (VBF and $\V\PH$), even events with reduced kinematic information can facilitate this analysis.
For example, in cases where both jets lie outside of the detector acceptance, the
$\pt$ distribution of the \Hboson is different for SM and BSM production.
This leads to different event populations across the three categories and to a different $\pt$ distribution
of the \Hboson in the boosted category. For example, the fraction of signal events is much smaller in
the 0-jet category, and the $\pt$ distribution is significantly harder in the
boosted category for pseudoscalar \Hboson production than it is for the SM case.
These effects are illustrated in Figs.~\ref{fig:0jet}, \ref{fig:boosted}, and~\ref{fig:mela}.
The same effects are, however, negligible in gluon
fusion production, where both scalar and pseudoscalar $\PH\Pg\Pg$
couplings are generated by higher-dimension operators,
which correspond to the $a_{2}^{\Pg\Pg}$ and $a_{3}^{\Pg\Pg}$ terms in Eq.~(\ref{eq:formfact-fullampl-spin0}).

Other observables, such as $\Delta\Phi_{JJ}$~\cite{Plehn:2001nj}, defined as the azimuthal difference between
the two associated jets, have been suggested for the study of $CP$ effects.
While they do provide sensitivity to $CP$ measurements, they are not as sensitive as the discriminant variables
for VBF production used in this analysis.
Nonetheless, as an alternative to the optimal VBF analysis with the \textsc{mela} discriminants,
we also performed a cross-check analysis where the $\Delta\Phi_{JJ}$ observable is used instead.
It was verified that the expected precision on $f_{a3}$ is indeed lower than in the optimal VBF analysis.
On the other hand, the sensitivity of the $\Delta\Phi_{JJ}$ observable to the $f_{a3}^{\Pg\Pg\PH}$ parameter is better
than that of the VBF discriminants, and it is close to but not as good as the optimal \textsc{mela} observables targeting
the gluon fusion topology in association with two jets. Both results are discussed in Sec.~\ref{sec:Results}.

\section{Analysis implementation}
\label{sec:Strategy}

Five anomalous $\PH\V\V$ coupling parameters defined in Sec.~\ref{sec:Phenomenology} are studied:
$f_{a3}$, $f_{a2}$, $f_{\Lambda1}$, $f_{\Lambda1}^{\PZ\gamma}$, and $f_{a3}^{\Pg\Pg\PH}$
describing anomalous couplings in VBF, $\V\PH$, and gluon fusion production.
The $CP$-sensitive parameters $f_{a3}$ and $f_{a3}^{\Pg\Pg\PH}$ are studied jointly,
while all other parameters are examined independently.
Anomalous \Hboson couplings in other production mechanisms and in the $\PH\to\Pgt\Pgt$ decay
do not affect these measurements, as the distributions studied here are insensitive to such effects.

The data, represented by a set of observables $\vec{x}$, are used to set constraints on anomalous coupling parameters.
In the case of the $CP$ study, the coupling parameters are $f_{a3}$ and $\phi_{a3}$. We also consider the scalar anomalous
couplings described by $f_{a2}$ and $\phi_{a2}$,
$f_{\Lambda1}$  and $\phi_{\Lambda1}$, and $f_{\Lambda1}^{\PZ\gamma}$  and $\phi_{\Lambda1}^{\PZ\gamma}$.
Since only real couplings are considered,
we fit for the products
$f_{a3}\cos(\phi_{a3})$ with $\cos(\phi_{a3})=\pm1$,
$f_{a2}\cos(\phi_{a2})$ with $\cos(\phi_{a2})=\pm1$,
$f_{\Lambda1}\cos(\phi_{\Lambda1})$ with $\cos(\phi_{\Lambda1})=\pm1$, and
$f_{\Lambda1}^{\PZ\gamma}\cos(\phi_{\Lambda1}^{\PZ\gamma})$ with $\cos(\phi_{\Lambda1}^{\PZ\gamma})=\pm1$.

\subsection{Observable distributions}
\label{observables}

Each event is described by its category $k$ and the corresponding observables $\vec{x}$.
In the {0-jet} and {boosted} categories, which are dominated by the gluon fusion production mechanism,
the observables are identical to those used in Ref.~\cite{Sirunyan:2017khh}, namely
$\vec{x} = \{  \mvis, M \}$ in the $\Pe\tauh$ and $\Pgm\tauh$ 0-jet categories,
$\vec{x} = \{ \mvis, \pt^\Pgm  \}$ in the $\Pe\Pgm$ 0-jet category,
$\vec{x} = \{  \mtautau  \}$  in the 0-jet $\tauh\tauh$ category, and
$\vec{x} = \{ \mtautau, \pt^\PH \}$ in the boosted categories,
where $M$ is the $\tauh$ decay mode, $\pt^\Pgm$ is the transverse momentum of the muon,
and $\pt^\PH$ is the transverse momentum of the \Hboson.
There are no dedicated observables sensitive to anomalous couplings in these categories,
as it is not possible to construct  them in the absence of a correlated jet pair.
Nonetheless, distributions of events in the above observables and categories still differ between signal models
with variation of anomalous couplings.

\begin{figure*}[!htb]
\centering
\includegraphics[width=0.45\textwidth]{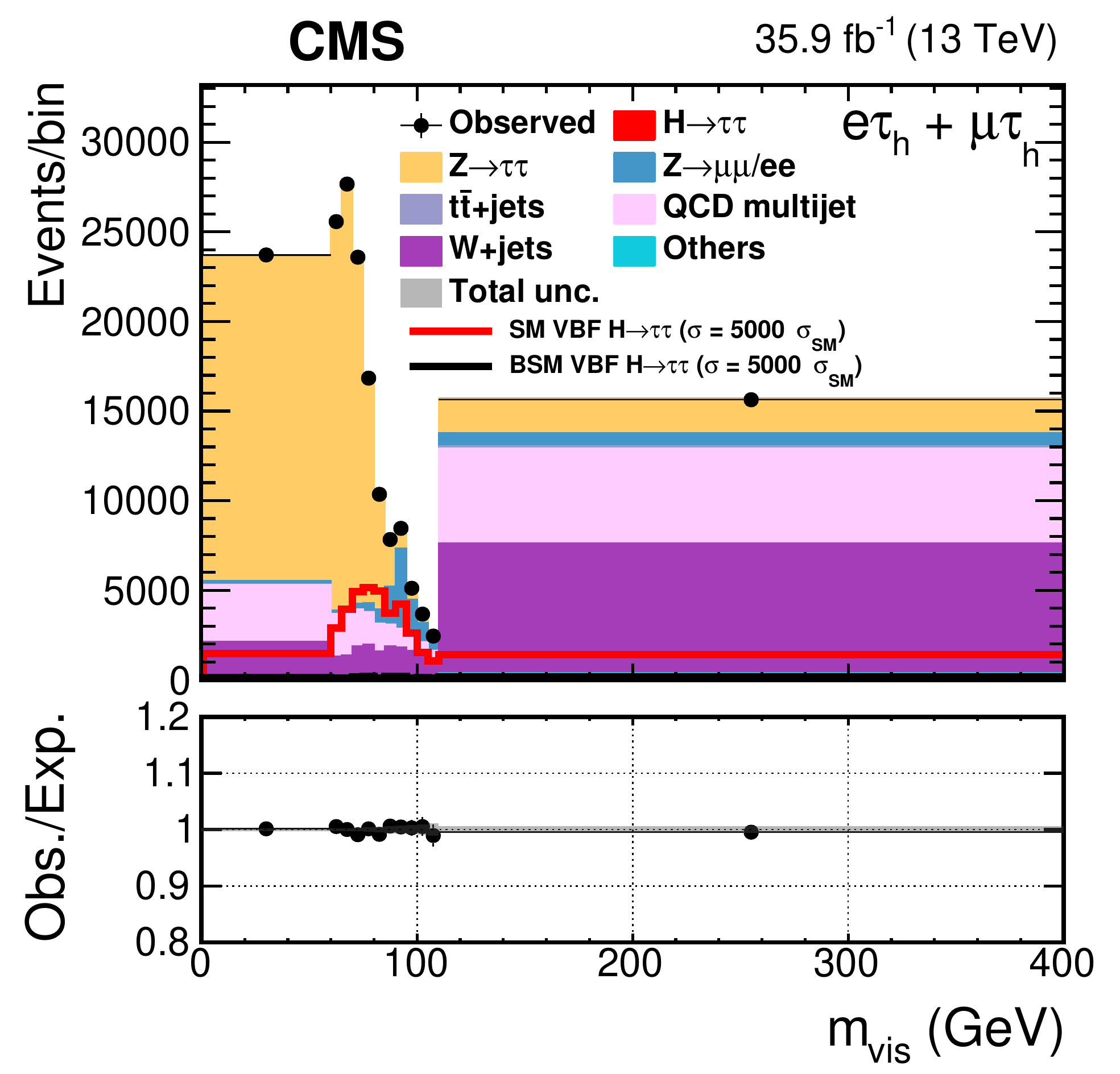}
\includegraphics[width=0.45\textwidth]{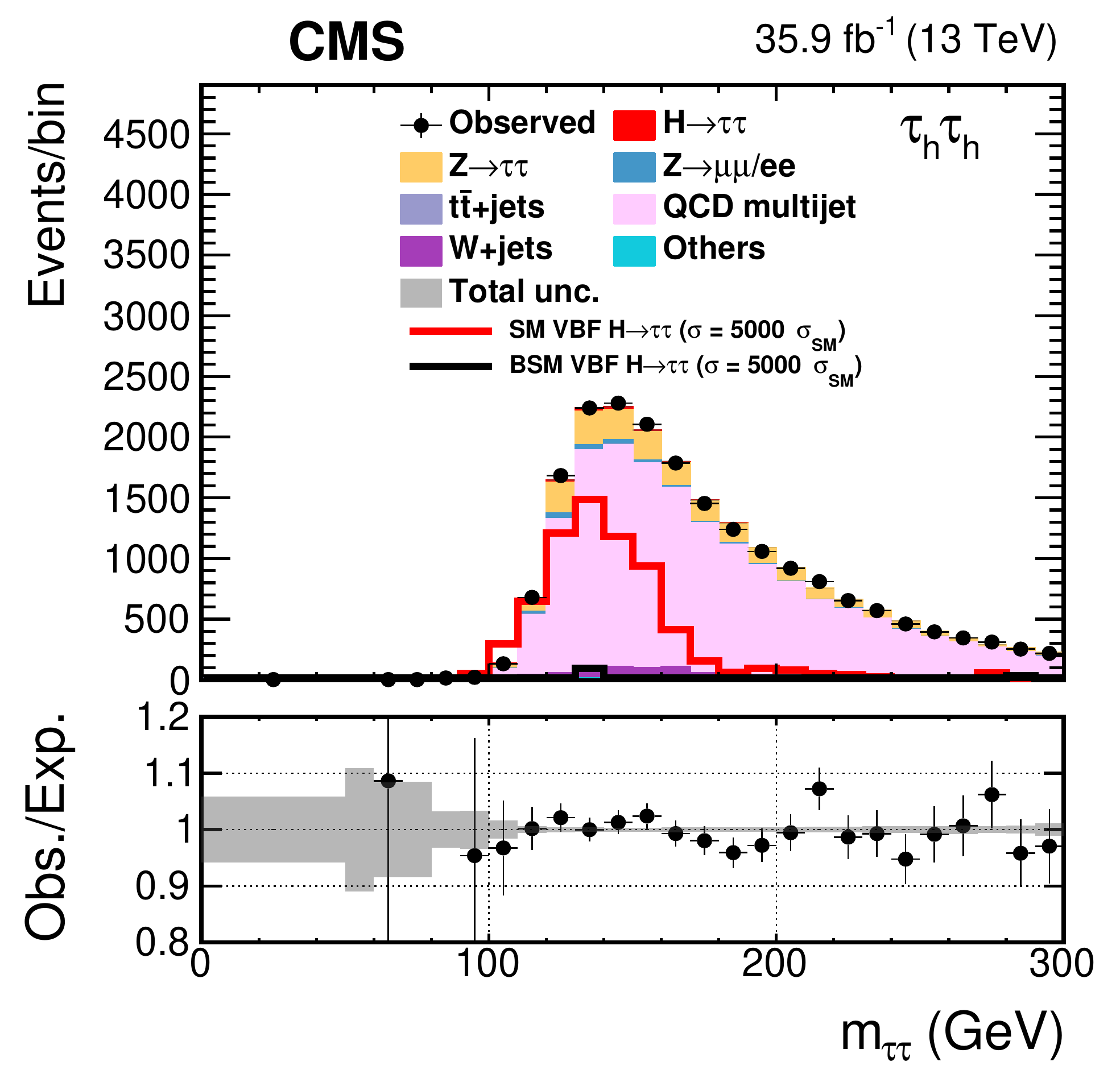}
\caption
{
The distributions of $\mvis$ and $\mtautau$ in the 0-jet category of the $\Pe\tauh+\Pgm\tauh$ (left) and $\tauh\tauh$ (right) decay channels.
The BSM hypothesis corresponds to $f_{a3}\cos(\phi_{a3})=1$.
\label{fig:0jet}
}
\end{figure*}

\begin{figure*}[!htb]
\centering
\includegraphics[width=0.45\textwidth]{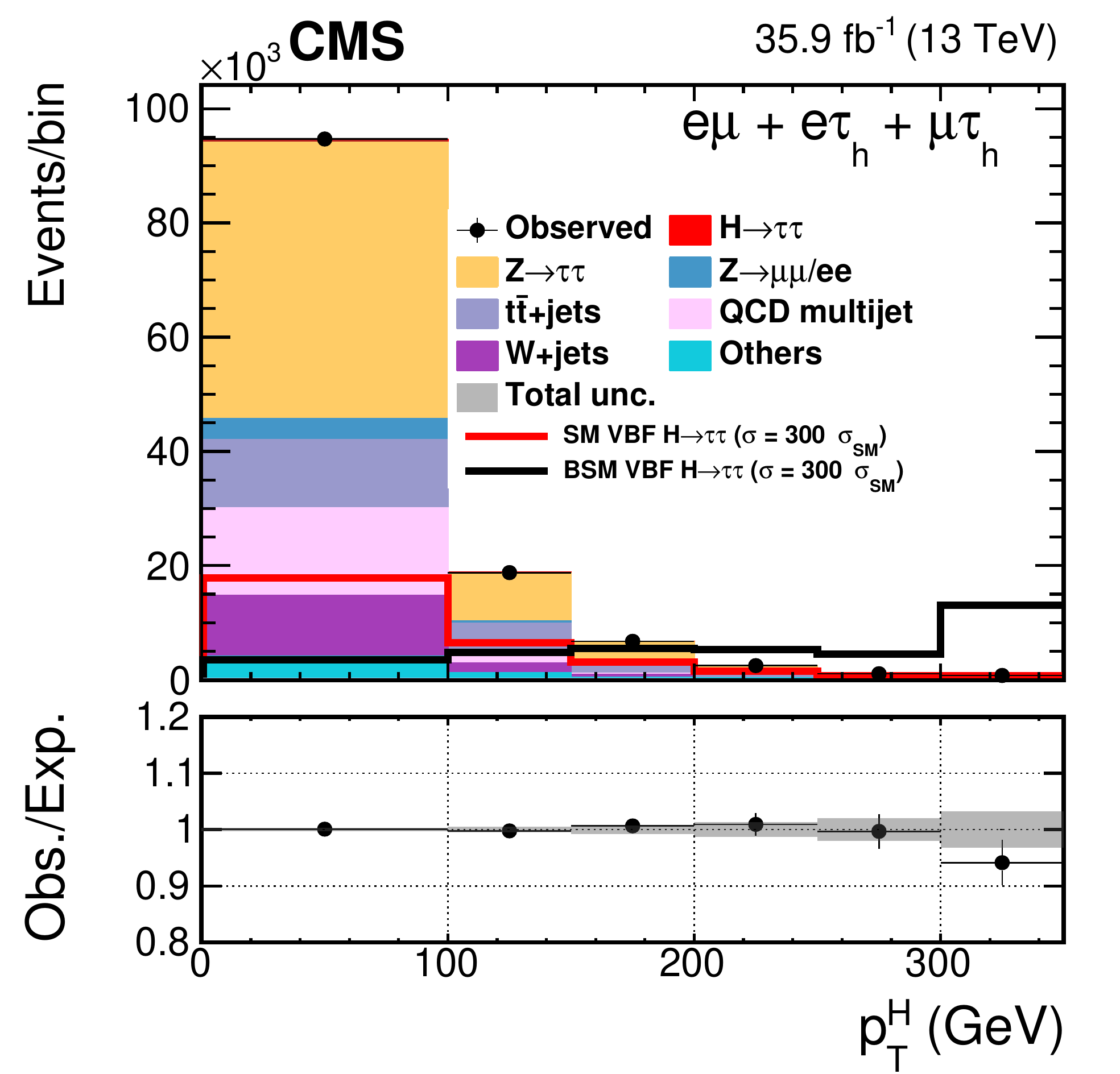}
\includegraphics[width=0.45\textwidth]{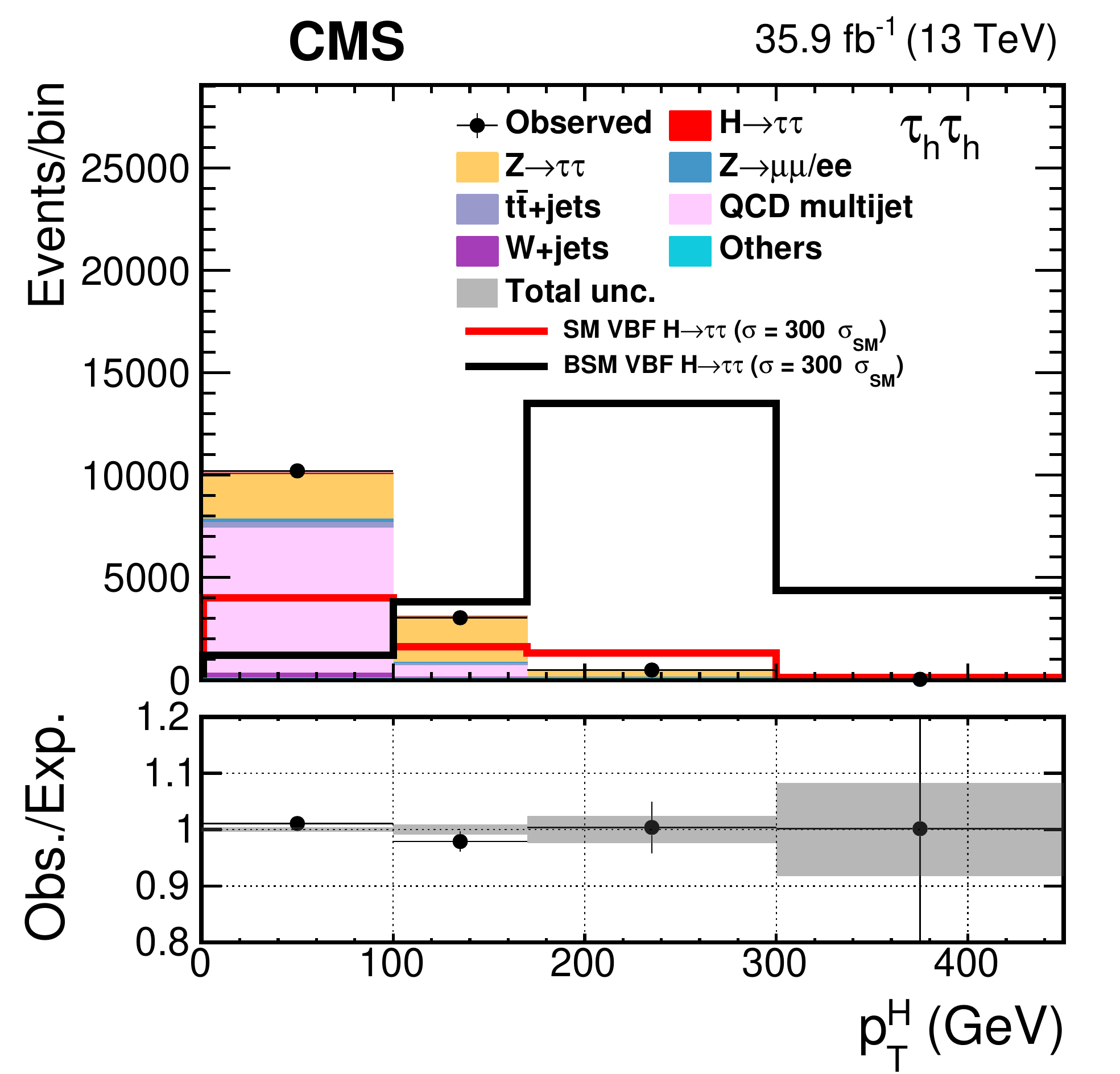}
\caption
{
The distributions of transverse momentum of the \Hboson in the boosted category of the
$\Pe\tauh+\Pgm\tauh+\Pe\Pgm$ (left) and $\tauh\tauh$ (right) decay channels.
The BSM hypothesis corresponds to $f_{a3}\cos(\phi_{a3})=1$.
\label{fig:boosted}
}
\end{figure*}

In Figs.~\ref{fig:0jet} and \ref{fig:boosted} the distributions of $\mvis$ and $\mtautau$ are displayed for selected
events in the 0-jet category, and the transverse momentum distribution of the \Hboson is shown for the boosted category.
Anomalous couplings would result in higher transverse momentum of the \Hboson and, unlike SM production,
would cause the events to preferentially populate the boosted category instead of the one with no jets in the final state. 
The observable $\mtautau$ is used in the $\tauh\tauh$ decay channel and $\mvis$ in other channels in the 0-jet category.
Two observables are used in the likelihood fit in the boosted category, $\mtautau$ and $p_\mathrm{T}^{\PH}$. The contributions from BSM and SM yields in the boosted category are different in the $\tauh\tauh$ and $\ell\tauh$ channels because of different trigger conditions and classification requirements.
In Fig.~\ref{fig:0jet}, the contribution from the $\Pe\Pgm$ channel is omitted because of its low sensitivity and
different binning in the fit.
The normalization of the predicted background distributions corresponds to the result of the likelihood
fit described in Sec.~\ref{sec:Likelihood}.
In all production modes in Figs.~\ref{fig:0jet} and \ref{fig:boosted}, the $\PH\to\Pgt\Pgt$ process is normalized
to its best-fit signal strength and couplings and is shown as an open overlaid histogram.
The background components labeled in the figures as ``others" include events from diboson and single top quark production,
as well as \Hboson decays to $\PW$ boson pairs. The uncertainty band accounts for all sources of uncertainty.
The SM prediction for the VBF $\PH\to\Pgt\Pgt$ signal, multiplied by a factor 5000 (300) in Fig.~\ref{fig:0jet} (\ref{fig:boosted}),
is shown as a red open overlaid histogram. The black open overlaid histogram represents a BSM hypothesis for the
VBF $\PH\to\Pgt\Pgt$ signal, normalized to 5000 (300) times the predicted SM cross section in Fig.~\ref{fig:0jet} (\ref{fig:boosted}).

\begin{figure*}[htbp!]
\centering
\includegraphics[width=0.45\textwidth]{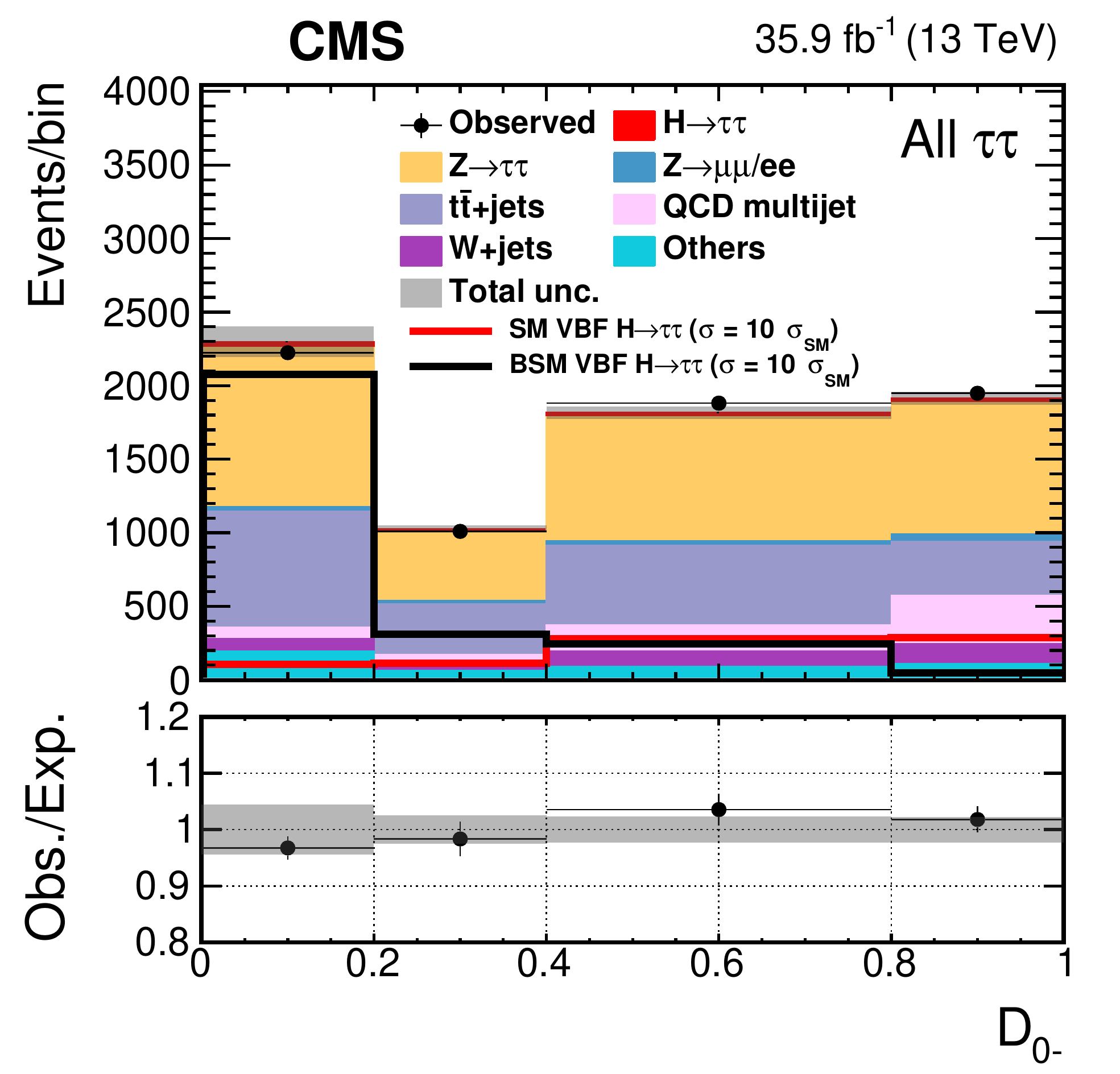}
\includegraphics[width=0.45\textwidth]{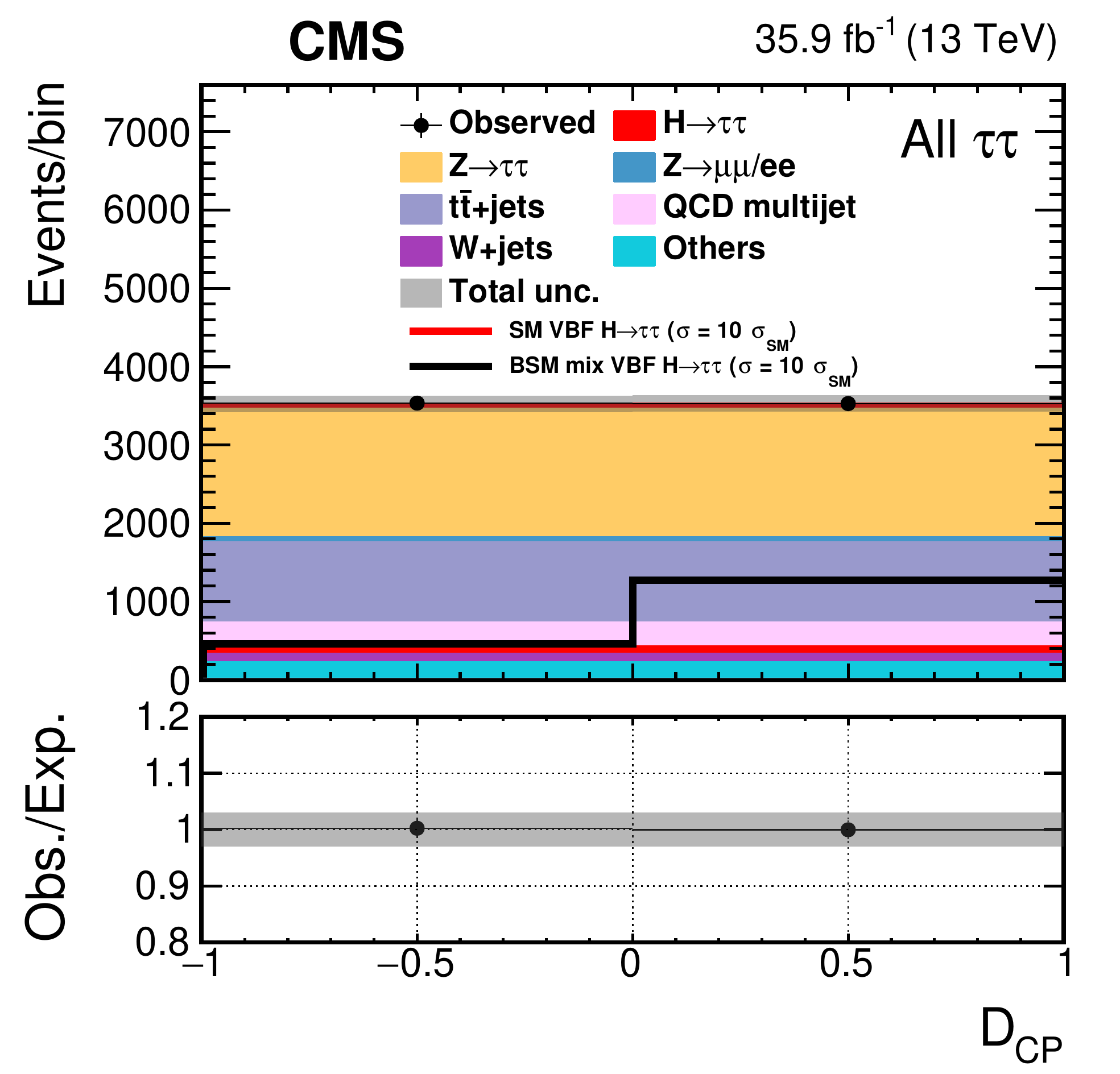}
\includegraphics[width=0.45\textwidth]{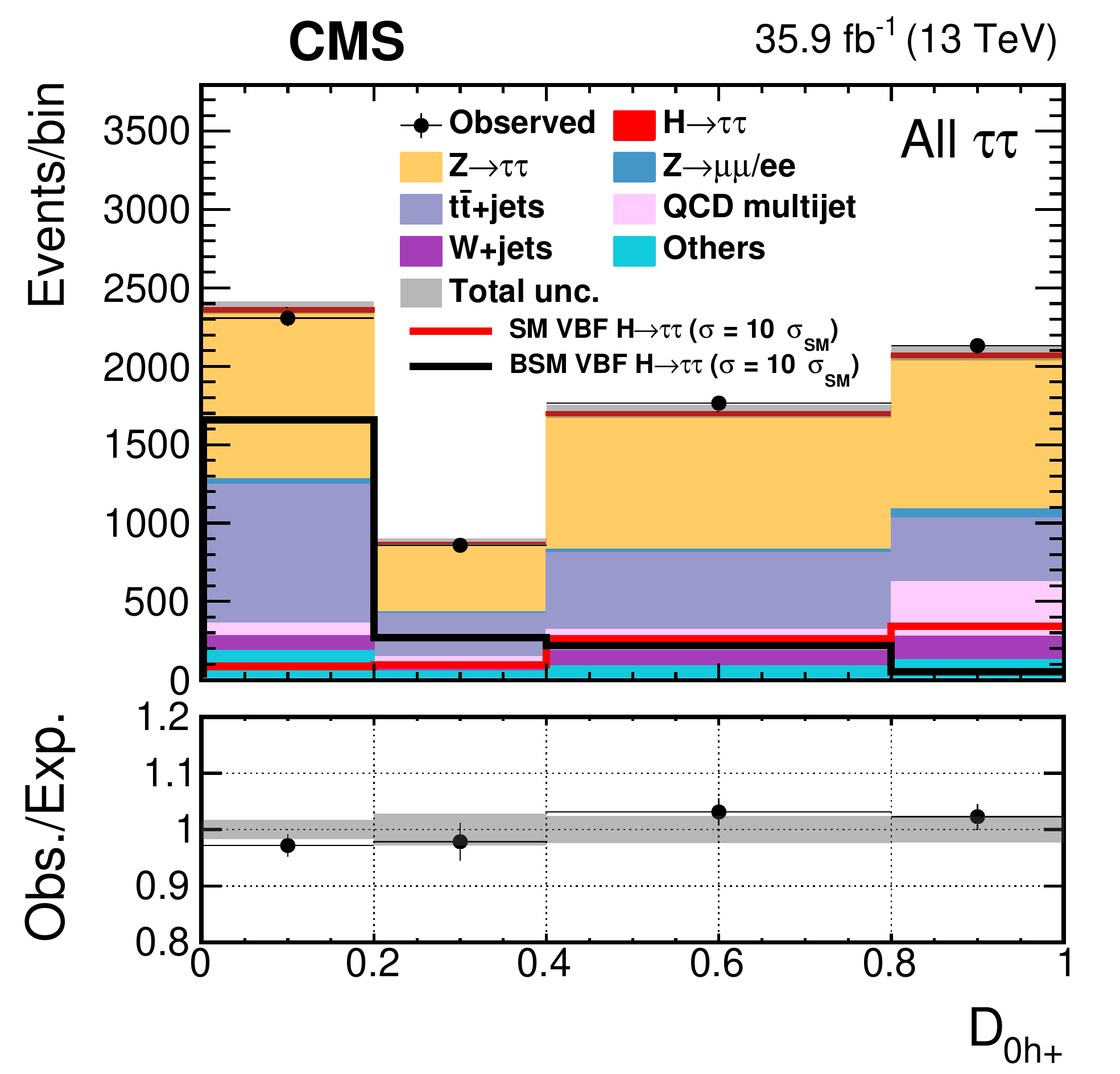}
\includegraphics[width=0.45\textwidth]{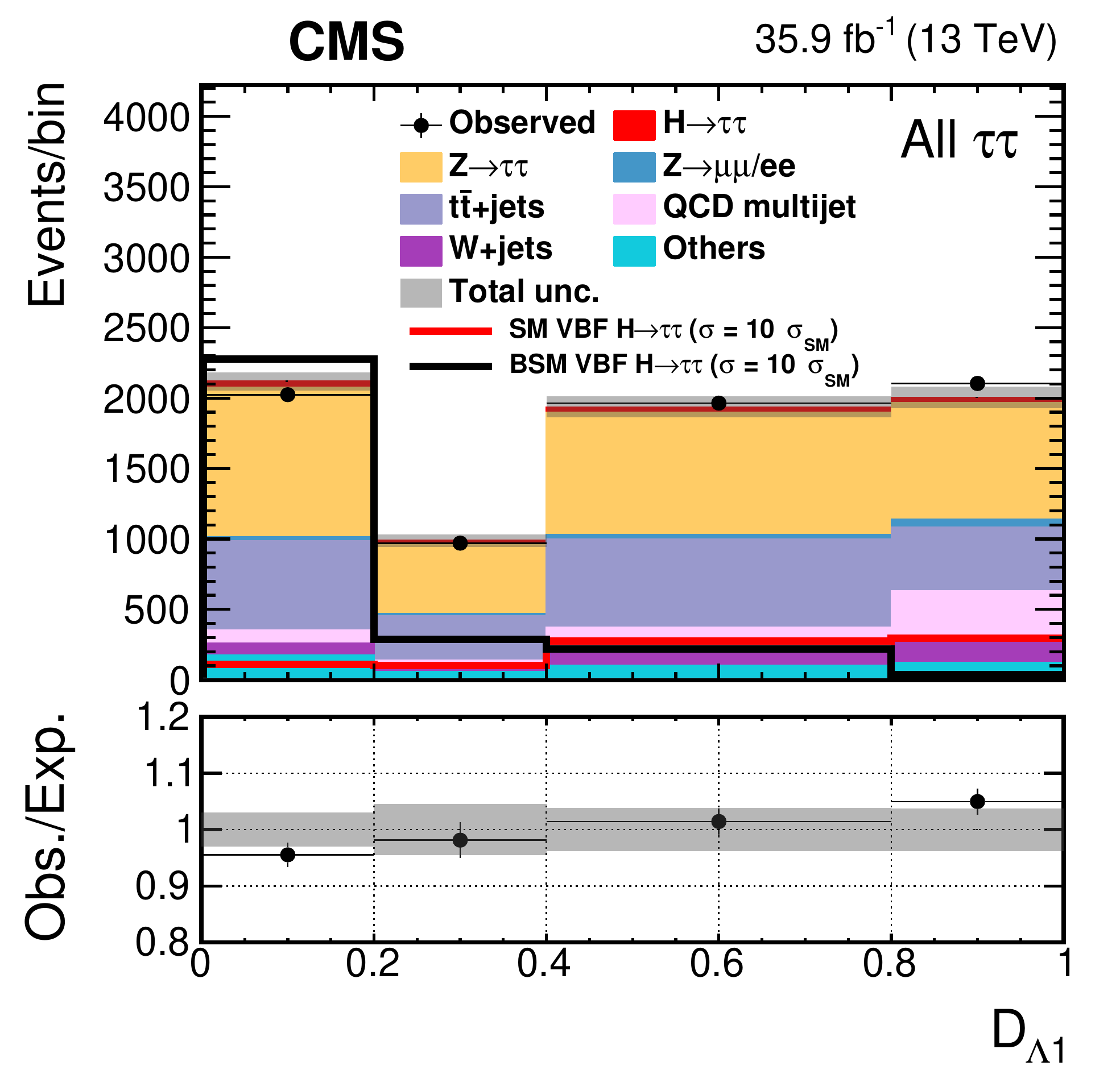}
\includegraphics[width=0.45\textwidth]{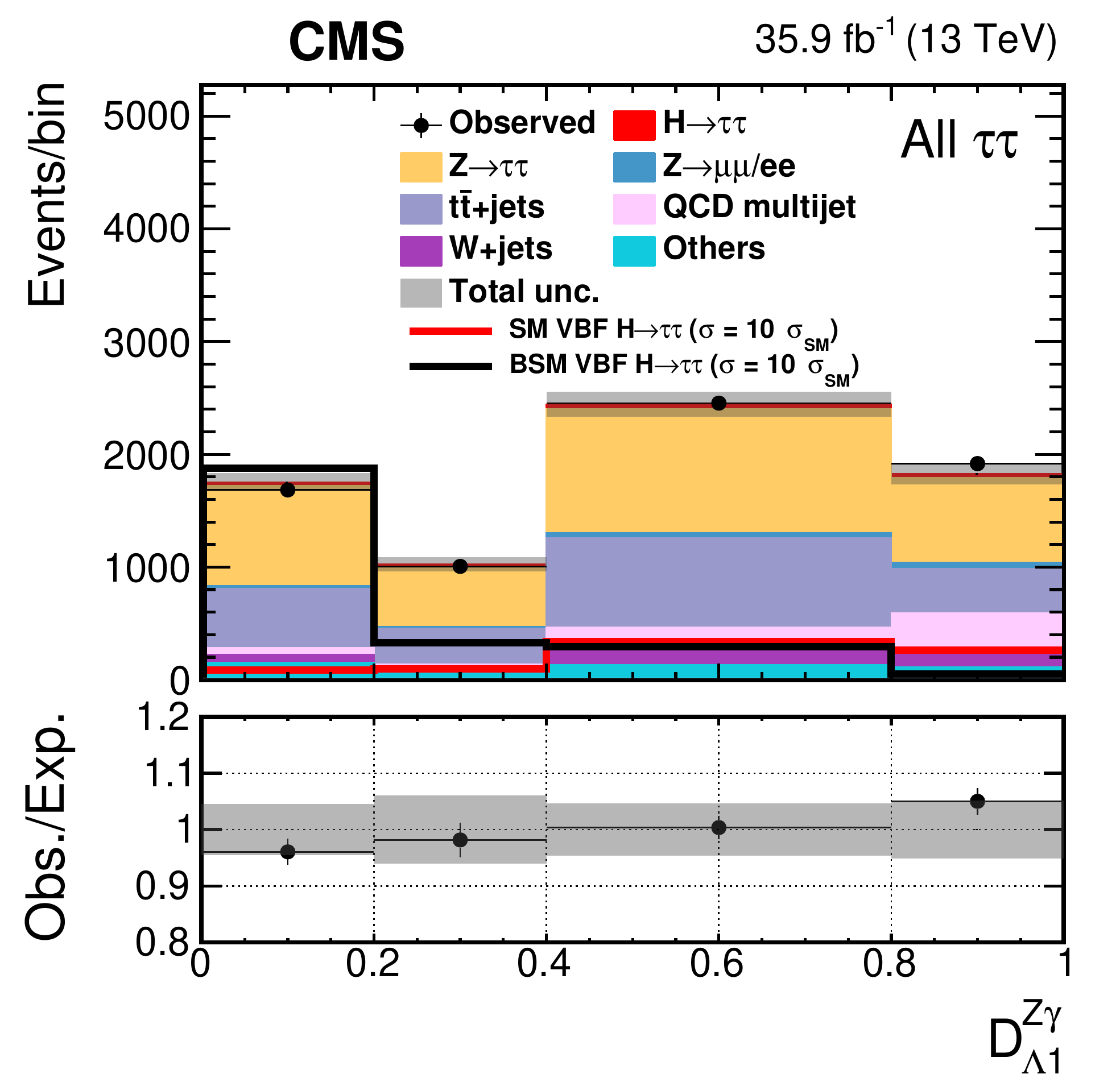}
\caption
{
The distribution of $\mathcal{D}_\mathrm{0-}$, $\mathcal{D}_{CP}$, $\mathcal{D}_\mathrm{0h+}$,
$\mathcal{D}_{\Lambda1}$, and $\mathcal{D}_{\Lambda1}^{\PZ\gamma}$ in the VBF category.
All four decay channels, $\Pe\Pgm$, $\Pe\tauh$, $\Pgm\tauh$, and $\tauh\tauh$, are summed.
The BSM hypothesis depends on the variable shown; it corresponds to $f_{a3}\cos(\phi_{a3})=1$ for the
$\mathcal{D}_\mathrm{0-}$ (upper left) distribution, the maximal mixing in VBF production
(``BSM mix" corresponding to $f_{a3}\cos(\phi_{a3})=0.013$)
for the $\mathcal{D}_{CP}$ distribution (upper right), $f_{a2}\cos(\phi_{a2})=1$ for the $\mathcal{D}_\mathrm{0h+}$
distribution (middle left), $f_{\Lambda1}\cos(\phi_{\Lambda1})=1$ for the $\mathcal{D}_{\Lambda1}$
distribution (middle right), and $f_{\Lambda1}^{\PZ\gamma}\cos(\phi_{\Lambda1}^{\PZ\gamma})=1$ for the
$\mathcal{D}_{\Lambda1}^{\PZ\gamma}$ distribution (lower). The expected $\mathcal{D}_{CP}$ distribution is always symmetric, unless a $CP$-violating effect is present in the signal.
\label{fig:mela}
}
\end{figure*}

\begin{figure*}[htbp!]
\centering
\includegraphics[width=0.9\textwidth]{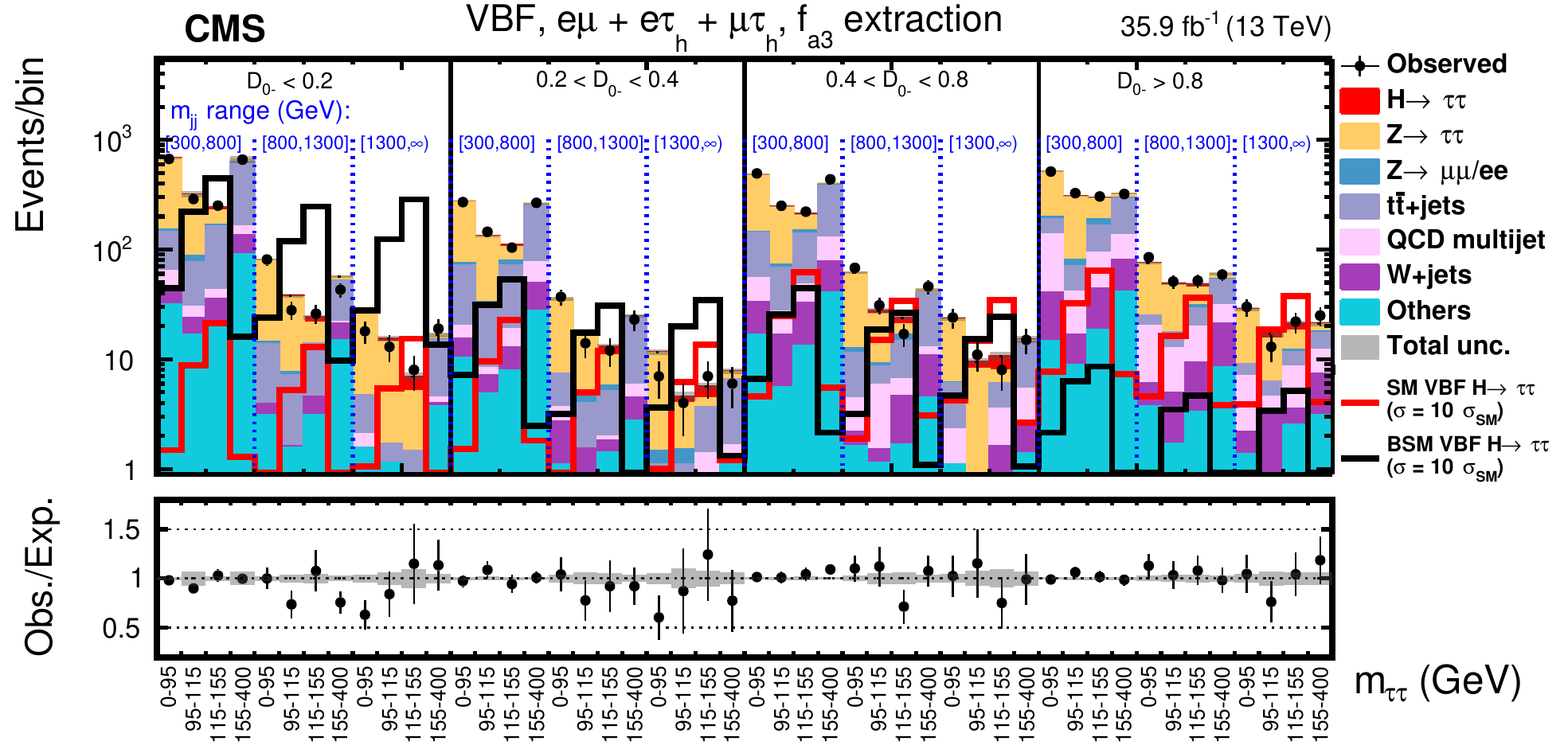}
\includegraphics[width=0.9\textwidth]{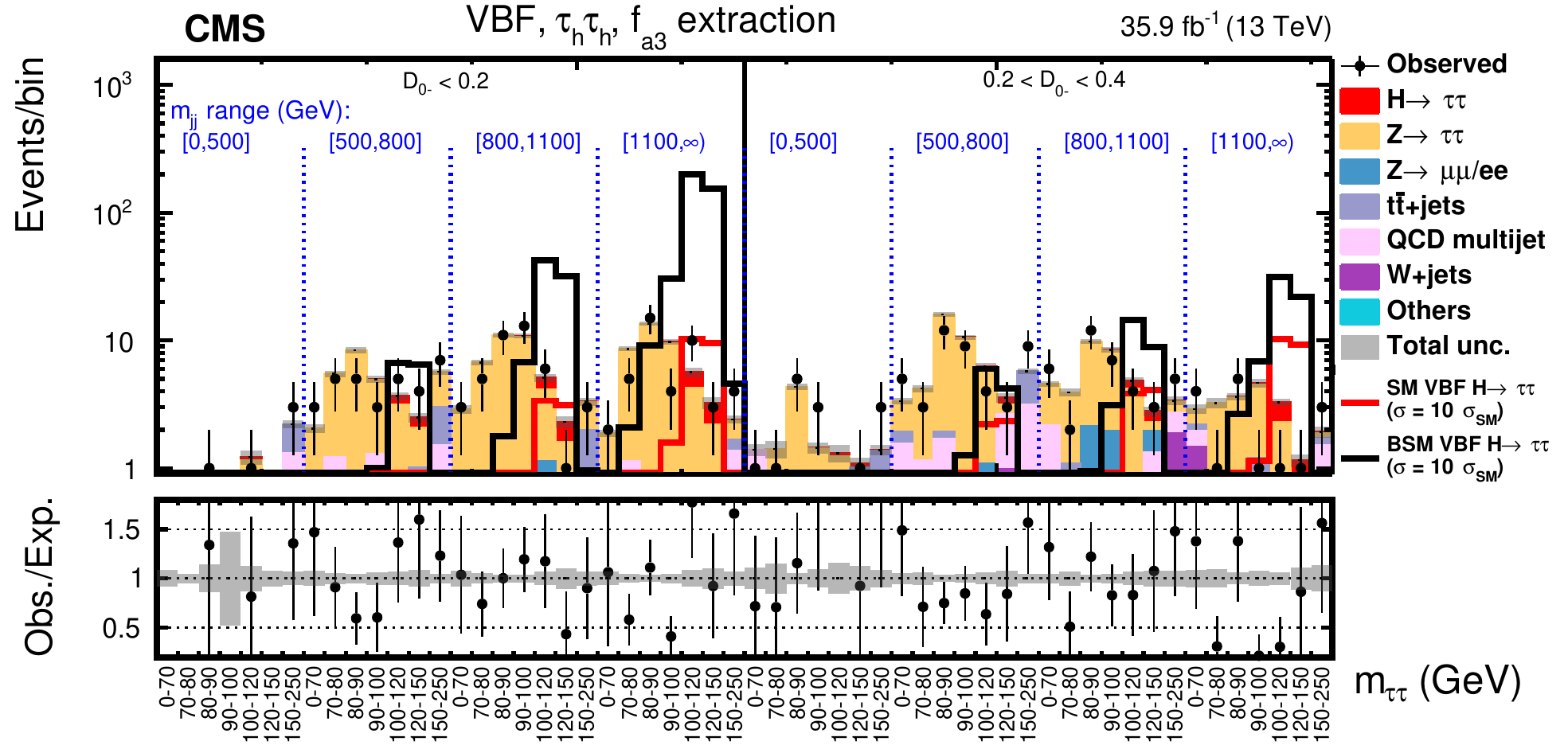}
\includegraphics[width=0.9\textwidth]{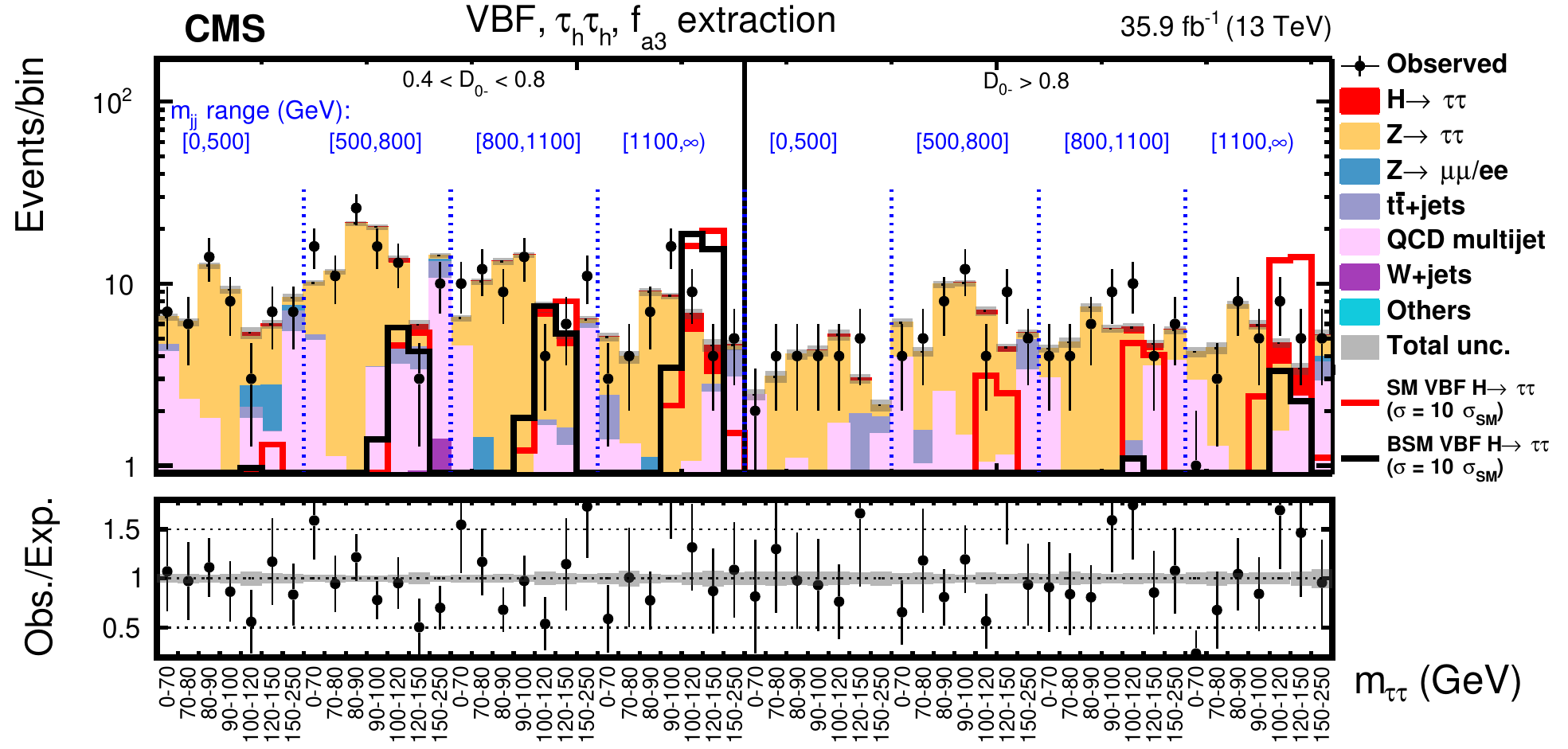}
\caption
{
Observed and expected distributions in the VBF category in bins of $\mtautau$, $\mjj$,
and $\mathcal{D}_\mathrm{0-}$ in the $f_{a3}$ analysis
for the $\Pe\Pgm+\Pe\tauh+\Pgm\tauh$ (upper) and $\tauh\tauh$ (middle and lower) decay channels.
\label{fig:unrolled_D0}
}
\end{figure*}

\begin{figure*}[htbp!]
\centering
\includegraphics[width=0.9\textwidth]{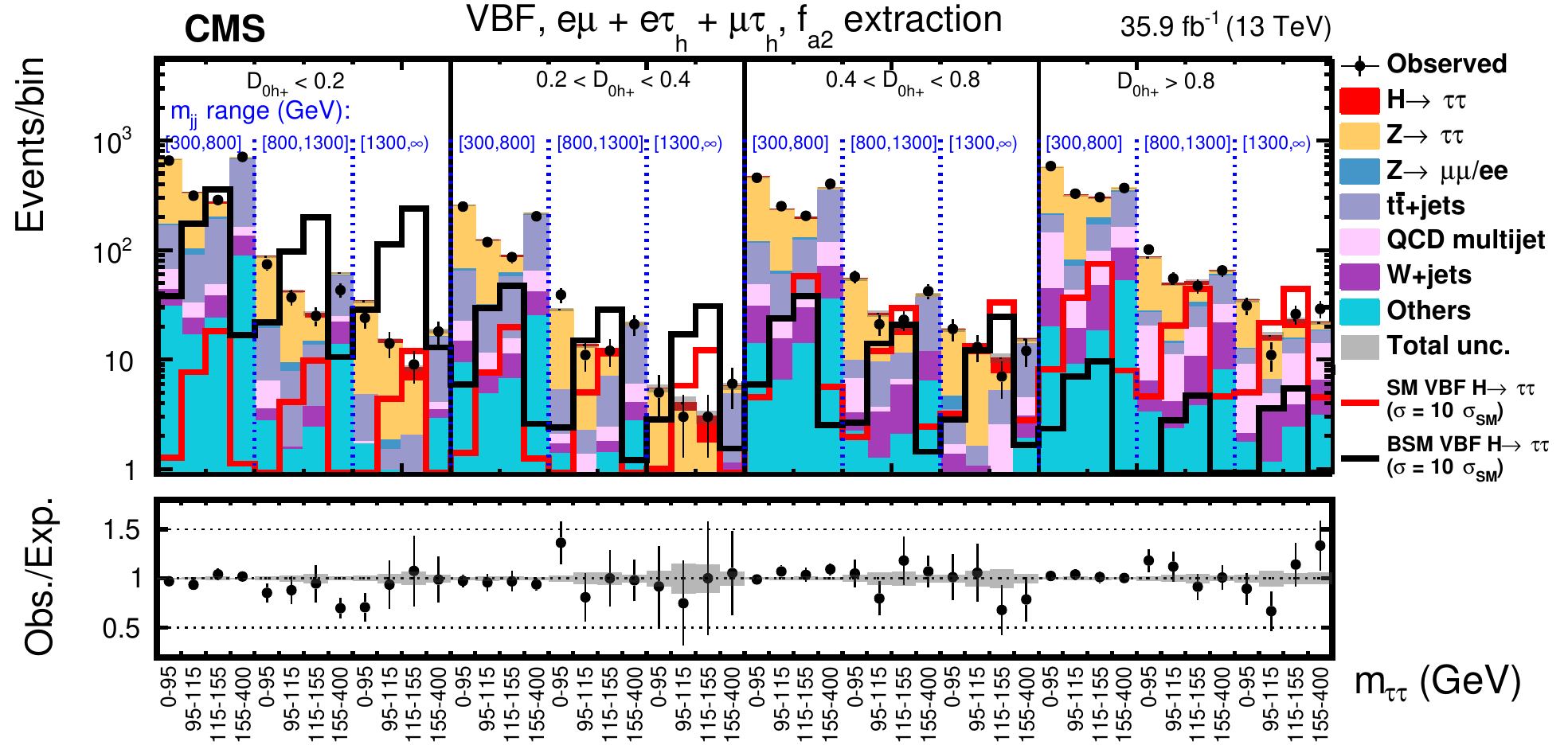}
\includegraphics[width=0.9\textwidth]{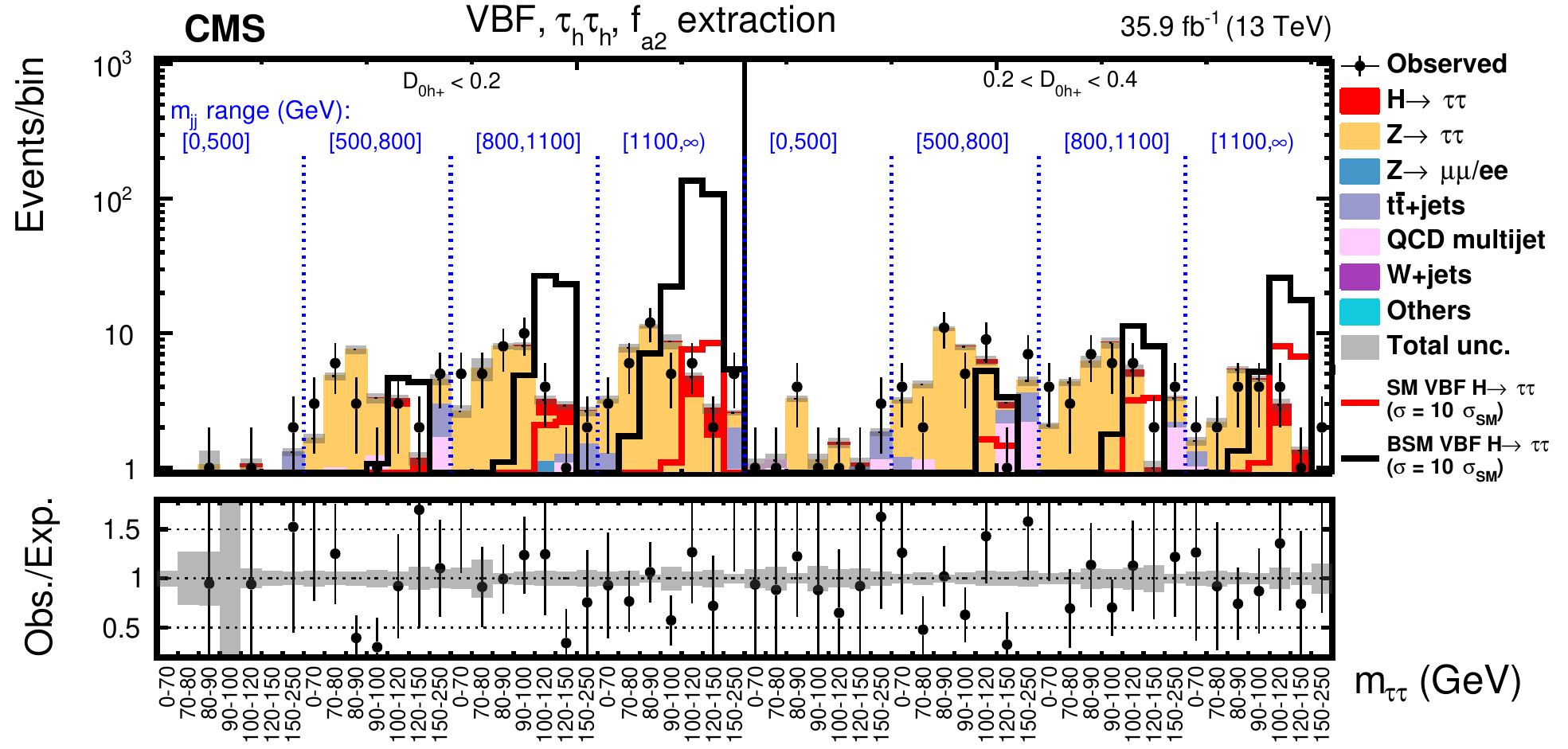}
\includegraphics[width=0.9\textwidth]{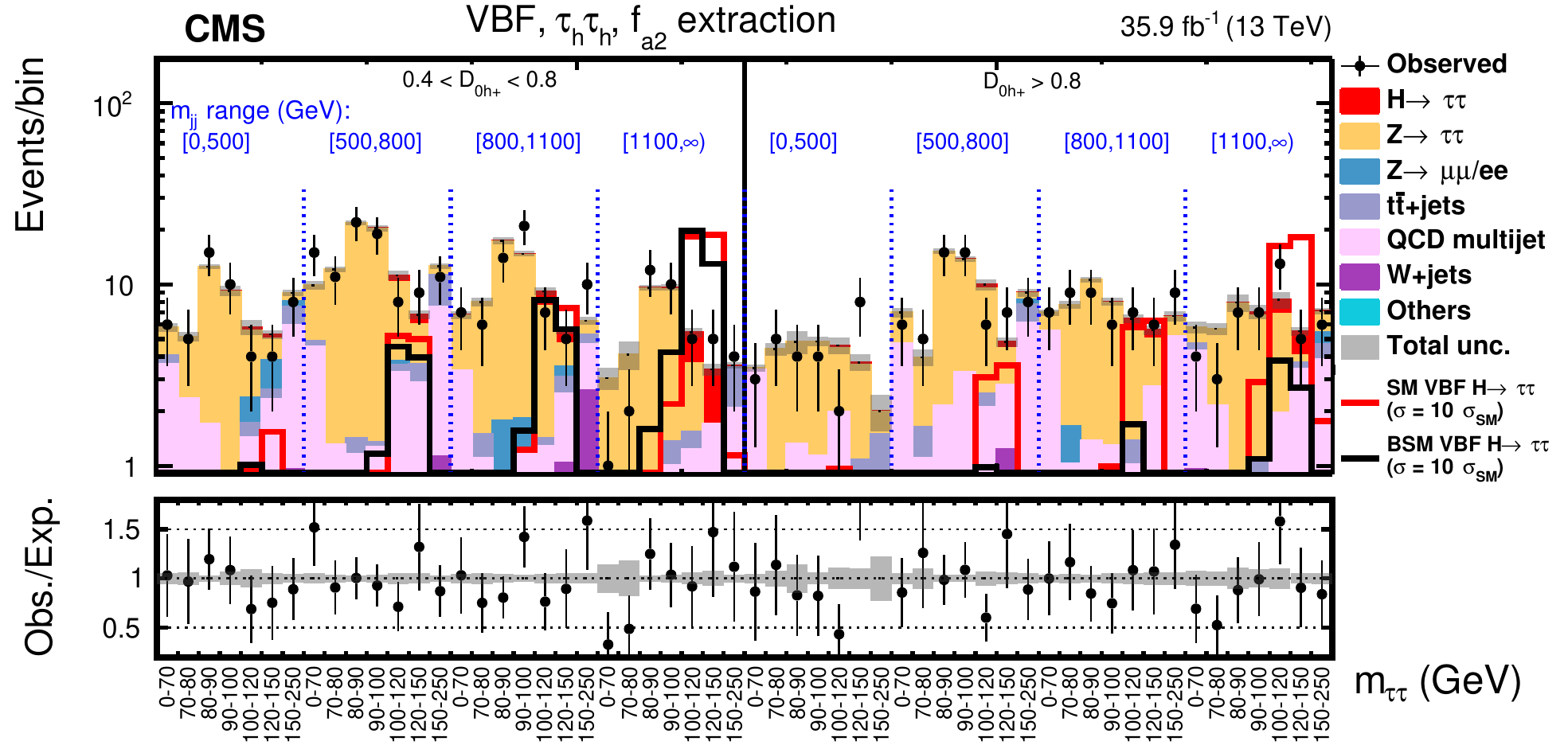}
\caption
{
Observed and expected distributions in the VBF category in bins of $\mtautau$, $\mjj$,
and $\mathcal{D}_\mathrm{0h+}$  in the $f_{a2}$ analysis
for the $\Pe\Pgm+\Pe\tauh+\Pgm\tauh$ (upper) and $\tauh\tauh$ (middle and lower) decay channels.
\label{fig:unrolled_D0h+}
}
\end{figure*}

\begin{figure*}[htbp!]
\centering
\includegraphics[width=0.9\textwidth]{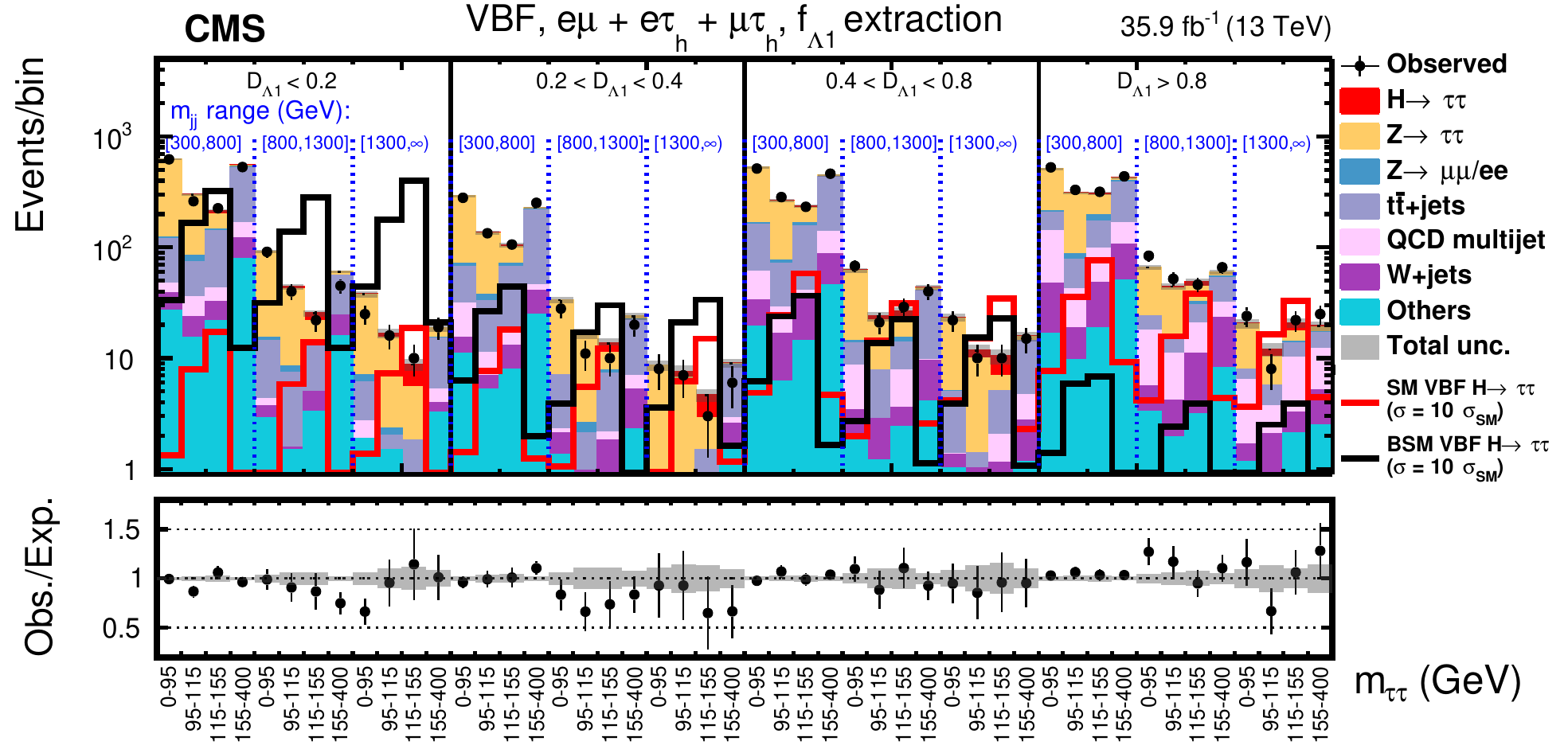}
\includegraphics[width=0.9\textwidth]{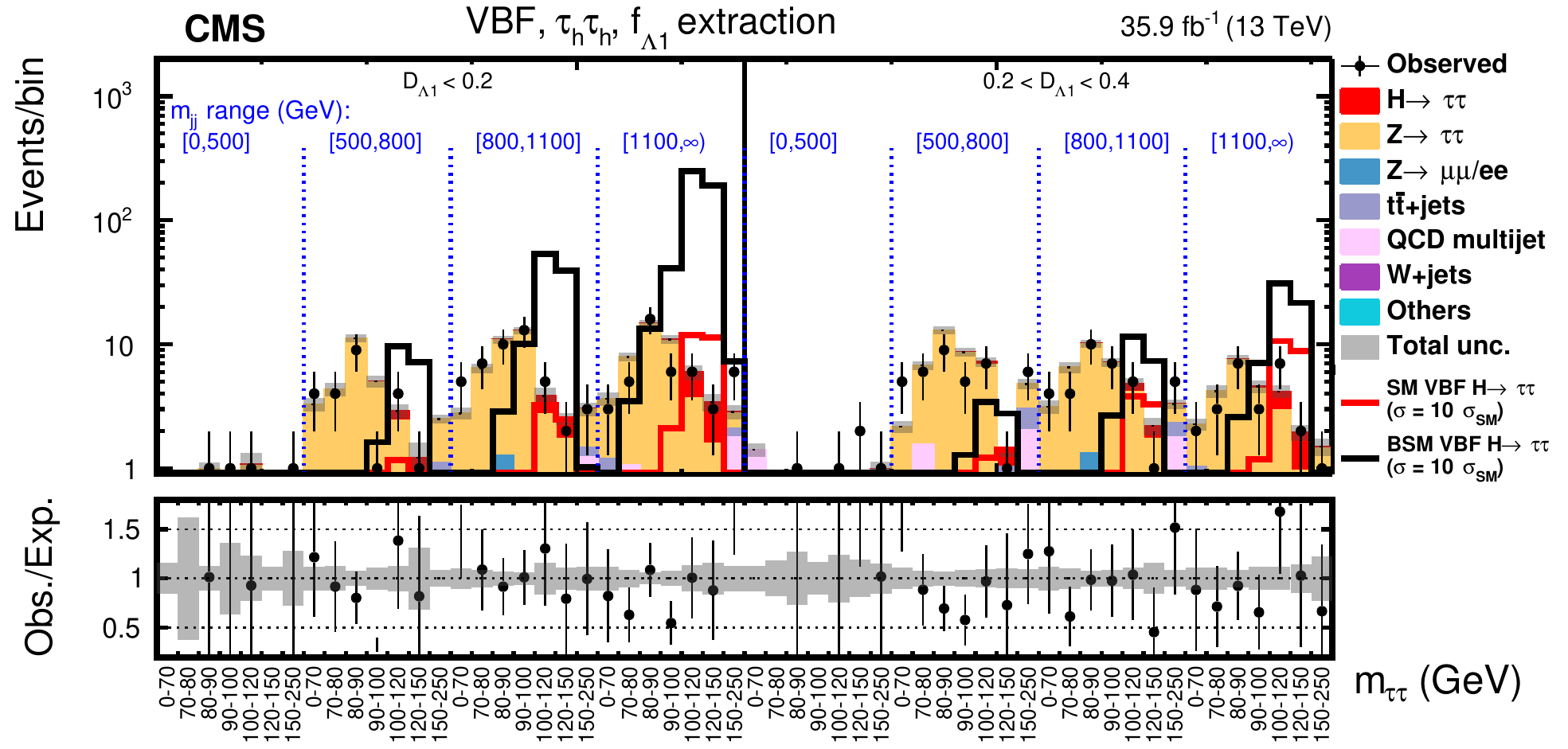}
\includegraphics[width=0.9\textwidth]{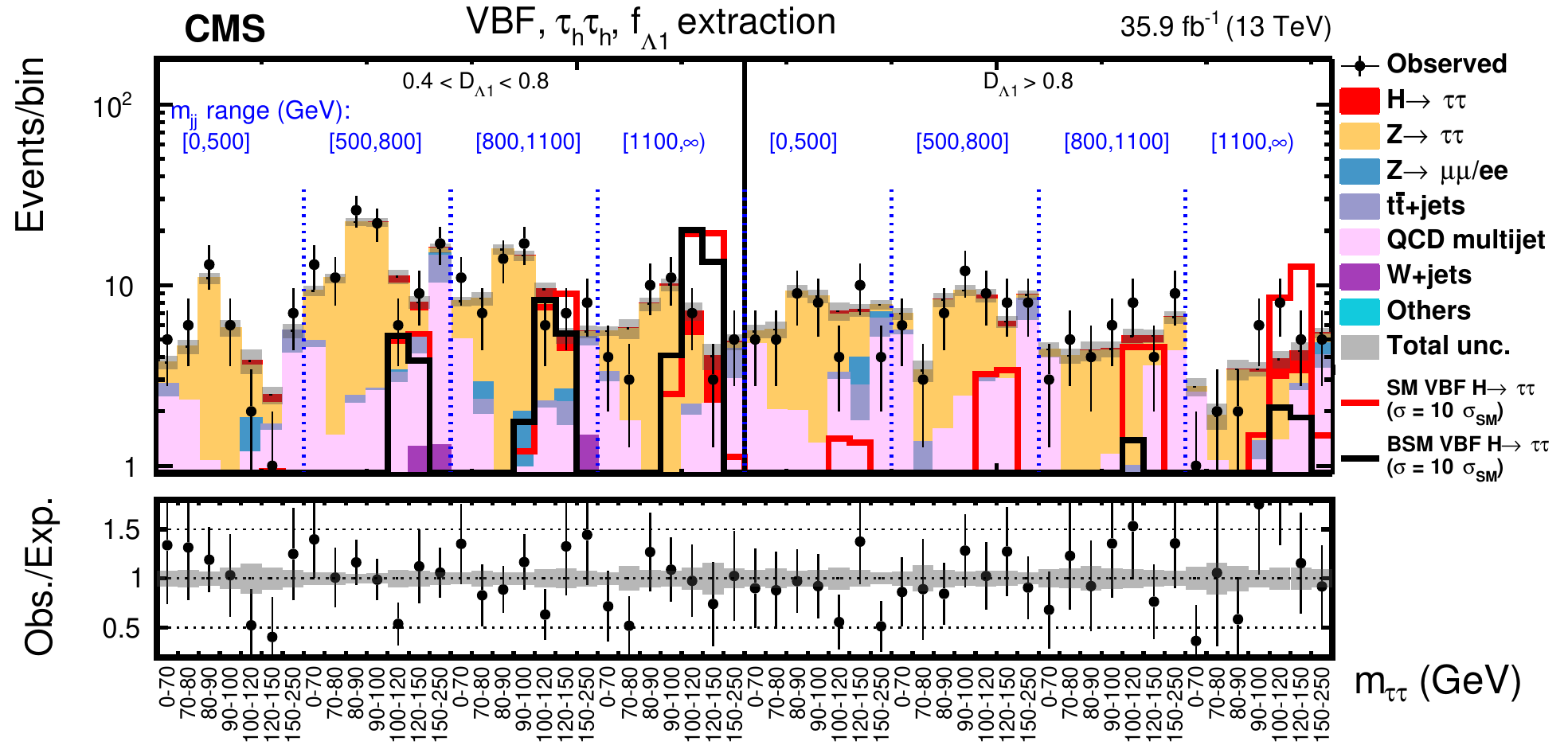}
\caption
{
Observed and expected distributions in the VBF category in bins of $\mtautau$, $\mjj$,
and $\mathcal{D}_{\Lambda1}$ in the $f_{\Lambda1}$ analysis
for the $\Pe\Pgm+\Pe\tauh+\Pgm\Pgm$ (upper) and $\tauh\tauh$ (middle and lower) decay channels.
\label{fig:unrolled_DL1}
}
\end{figure*}

\begin{figure*}[htbp!]
\centering
\includegraphics[width=0.9\textwidth]{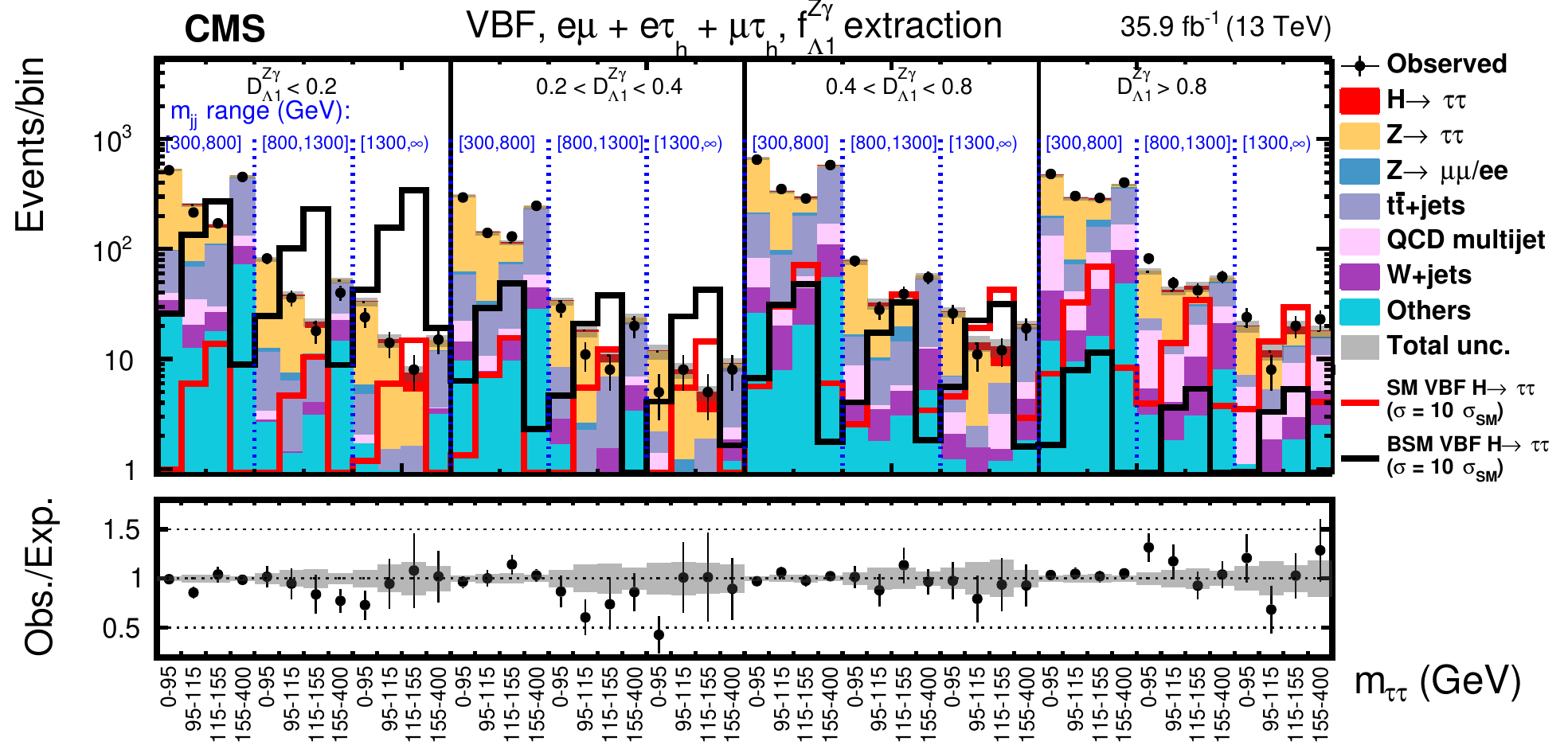}
\includegraphics[width=0.9\textwidth]{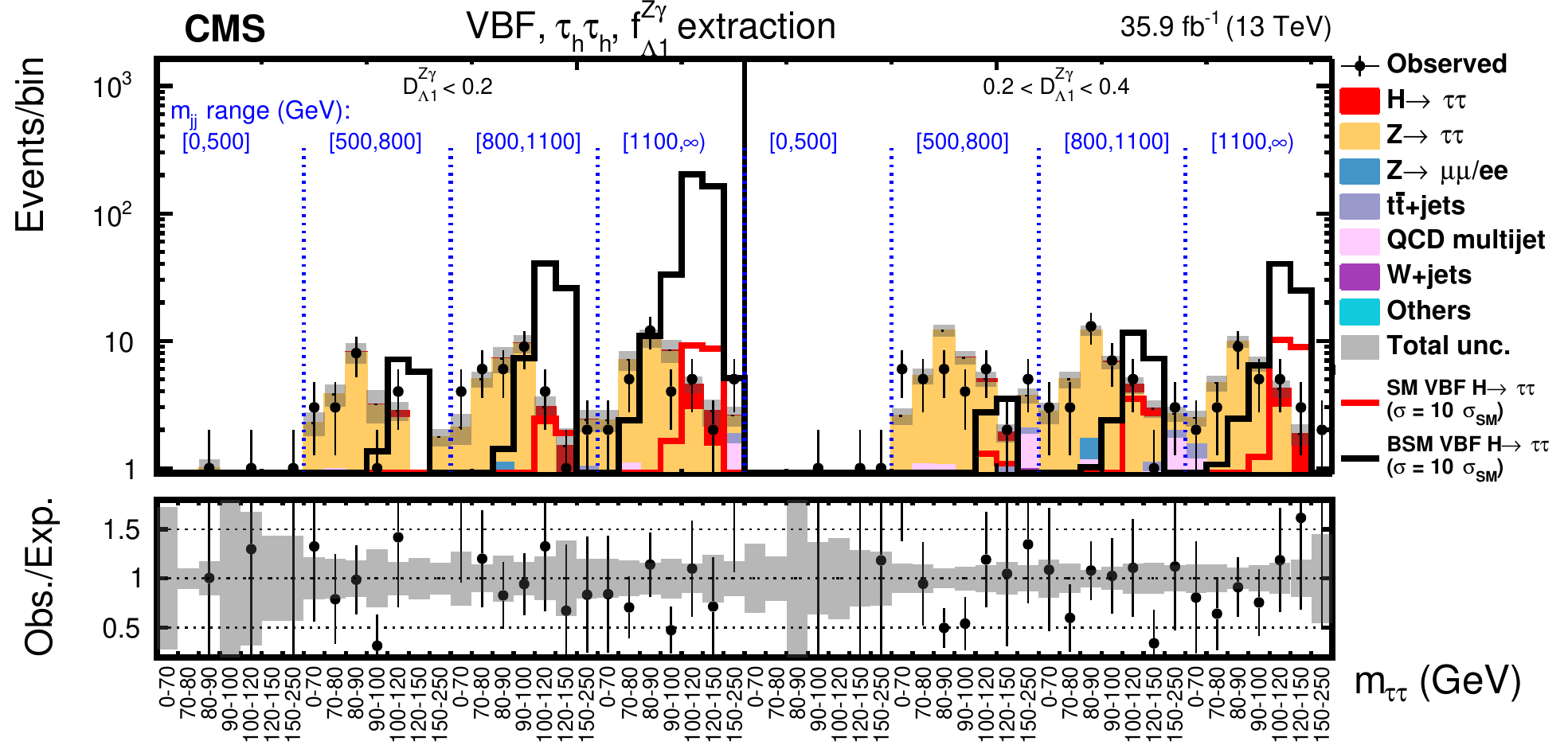}
\includegraphics[width=0.9\textwidth]{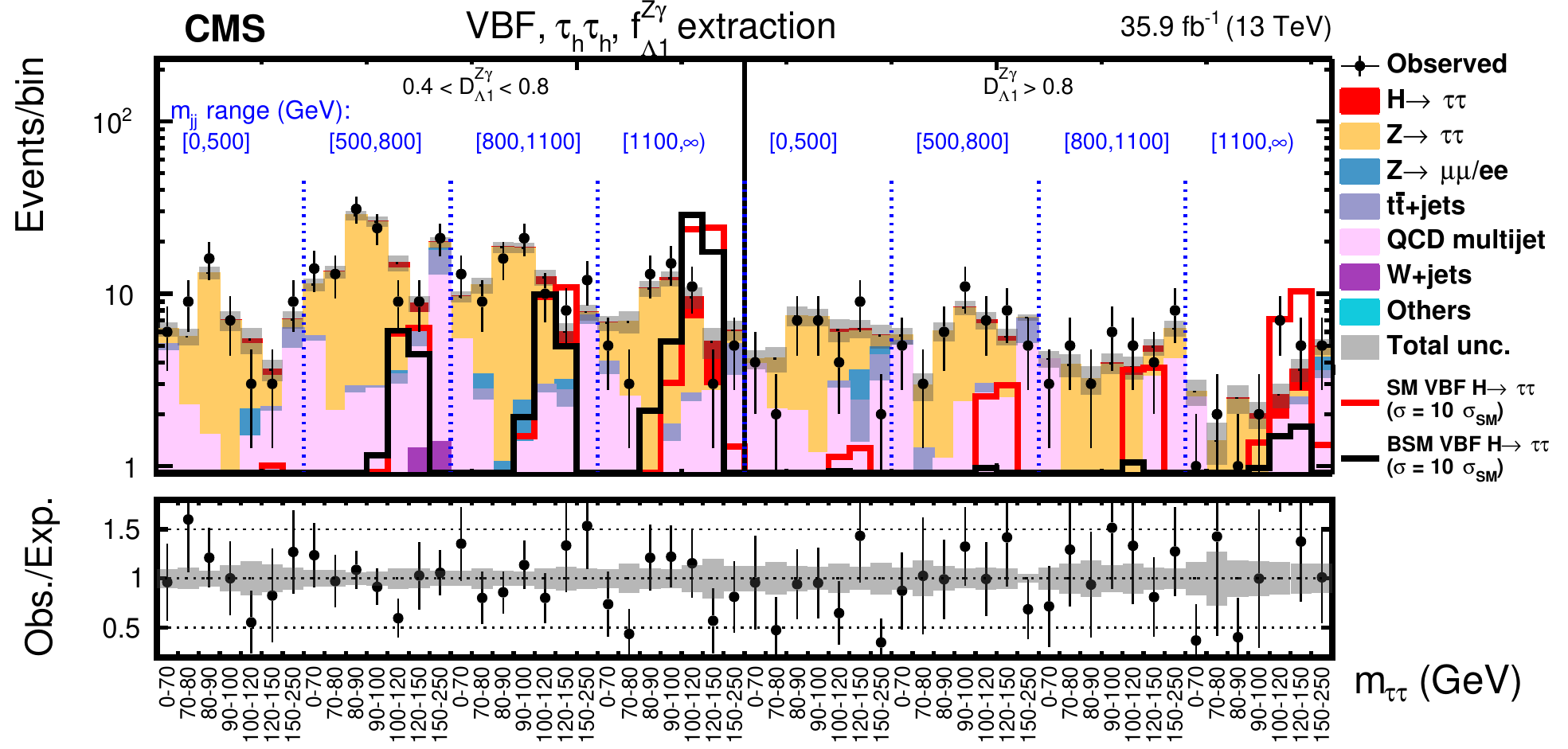}
\caption
{
Observed and expected distributions in the VBF category in bins of $\mtautau$, $\mjj$,
and $\mathcal{D}_{\Lambda1}^{\PZ\gamma}$ in the $f_{\Lambda1}^{\PZ\gamma}$ analysis
for the $\Pe\Pgm+\Pe\tauh+\tauh\tauh$ (upper) and $\tauh\tauh$ (middle and lower) decay channels.
\label{fig:unrolled_DL1Zg}
}
\end{figure*}

In Figs.~\ref{fig:mela}--\ref{fig:unrolled_DL1Zg}, the discriminant distributions in the VBF category are displayed.
In the VBF category, either three or four observables are used in the likelihood fit:
$\vec{x} = \{ \mjj, \mtautau, \mathcal{D}_\mathrm{0-}, \mathcal{D}_{CP} \}$ are used to determine the $f_{a3}$ parameter,
$\vec{x} = \{ \mjj, \mtautau, \mathcal{D}_\mathrm{0h+} \}$ are used to determine the $f_{a2}$ parameter,
$\vec{x} = \{ \mjj, \mtautau, \mathcal{D}_{\Lambda1} \}$ are used to determine the $f_{\Lambda1}$ parameter,
and $\vec{x} = \{ \mjj, \mtautau, \mathcal{D}_{\Lambda1}^{\PZ\gamma} \}$ are used to determine the $f_{\Lambda1}^{\PZ\gamma}$ parameter,
as defined in Eqs.~(\ref{eq:melaDbsm}) and~(\ref{eq:melaDint}).
In order to keep the background and signal templates sufficiently populated, a smaller number of bins is chosen for $\mjj$
and $\mtautau$ compared to Ref.~\cite{Sirunyan:2017khh}.
It was found that four bins in $\mathcal{D}_\mathrm{0-}$, $\mathcal{D}_\mathrm{0h+}$, $\mathcal{D}_{\Lambda1}$,
and $\mathcal{D}_{\Lambda1}^{\PZ\gamma}$ are sufficient for close-to-optimal performance.
At the same time, we adopt two bins in $\mathcal{D}_{CP}$ with $\mathcal{D}_{CP}<0$ and  $\mathcal{D}_{CP}>0$.
This choice does not lead to the need for additional bins in the templates, because all distributions except the $CP$-violating
interference component are symmetric in $\mathcal{D}_{CP}$, and this symmetry is enforced in the templates.
A forward-backward asymmetry in $\mathcal{D}_{CP}$ would be a clear indication of
$CP$-sensitive effects and is present only in the signal interference template.

\subsection{Likelihood parametrization}
\label{sec:Likelihood}

We perform an unbinned extended maximum likelihood fit~\cite{Barlow:1990vc} to the events split into several categories
according to the three production topologies and four tau-lepton pair final states
using the RooFit toolkit~\cite{Verkerke:2003ir,Brun:1997pa}.
The probability density functions for signal $\mathcal{P}^{j,k}_\text{sig}(\vec{x})$
and background $\mathcal{P}^{j,k}_\text{bkg}(\vec{x})$
are binned templates and are defined for each production mechanism $j$ in each category $k$.
Each event is characterized by the discrete category $k$ and up to four observables $\vec{x}$,
depending on the category. For the VBF, $\V\PH$, or gluon fusion production mechanisms,
the signal probability density function is defined as
\ifthenelse{\boolean{cms@external}}{
\begin{multline}
\mathcal{P}^{j,k}_\text{sig}\left(\vec{x} \right) =
\left(1-f_{ai}\right) \, \mathcal{T}^{j,k}_{a1}\left(\vec{x}\right)
 +  f_{ai} \, \mathcal{T}^{j,k}_{ai}\left(\vec{x}\right)\\
+  \sqrt{f_{ai}\left(1-f_{ai}\right)}\, \mathcal{T}^{j,k}_{a1,ai}\left(\vec{x}\right)\cos(\phi_{ai}) ,
\label{eq:fractions}
\end{multline}
}{
\begin{equation}
\mathcal{P}^{j,k}_\text{sig}\left(\vec{x} \right) =
\left(1-f_{ai}\right) \, \mathcal{T}^{j,k}_{a1}\left(\vec{x}\right)
 +  f_{ai} \, \mathcal{T}^{j,k}_{ai}\left(\vec{x}\right)
+  \sqrt{f_{ai}\left(1-f_{ai}\right)}\, \mathcal{T}^{j,k}_{a1,ai}\left(\vec{x}\right)\cos(\phi_{ai}) ,
\label{eq:fractions}
\end{equation}
}
where $\mathcal{T}^{j,k}_{ai}$ is the template probability of a pure anomalous coupling $a_i$ term and
$\mathcal{T}^{j,k}_{a1, ai}$ describes the interference between the anomalous coupling and SM term $a_1$,
or SM term $a_2^{\Pg\Pg\PH}$ in the case of gluon fusion.
Here $f_{ai}$ stands for either $f_{a3}$, $f_{a2}$, $f_{\Lambda1}$, $f_{\Lambda1}^{\PZ\gamma}$, or $f_{a3}^{\Pg\Pg\PH}$.
Each term in Eq.~(\ref{eq:fractions}) is extracted from a dedicated simulation.

The signal strength parameters $\mu_\V$ and $\mu_\mathrm{f}$ are introduced as two parameters of interest.
They scale the yields in the VBF+$\V\PH$ and gluon fusion production processes, respectively. They are defined such that for $f_{ai}=0$ they are equal to the ratio of the measured to the expected cross sections for the full process, including the $\PH\to\Pgt\Pgt$ decay. The likelihood is maximized with respect to the anomalous coupling $f_{ai}\cos(\phi_{ai})$
and yield ($\mu_\V$, $\mu_\mathrm{f}$) parameters and with respect to the nuisance parameters,
which include the constrained parameters describing the systematic uncertainties.
The $f_{a3}\cos(\phi_{a3})$ and $f_{a3}^{\Pg\Pg\PH}\cos(\phi_{a3}^{\Pg\Pg\PH})$ parameters are tested simultaneously, while all other $f_{ai}\cos(\phi_{ai})$ parameters are tested independently.
All parameters except the anomalous coupling parameter of interest $f_{ai}\cos(\phi_{ai})$ are profiled.
The confidence level (\CL) intervals are determined from profile likelihood scans of the respective parameters.
The allowed 68 and 95\%~\CL intervals are defined using the profile likelihood function,
$-2\,\Delta \ln{\mathcal L} = 1.00$ and 3.84, respectively,
for which exact coverage is expected in the asymptotic limit~\cite{wilks1938}.
Approximate coverage has been tested with generated samples.

\subsection{Systematic uncertainties}
\label{systematics}

A log-normal probability density function is assumed for the nuisance parameters that affect the event yields
of the various background and signal contributions, whereas systematic uncertainties that affect the distributions
are represented by nuisance parameters of which the variation results in a continuous perturbation of the
spectrum~\cite{Conway-PhyStat} and which are assumed to have a Gaussian probability density function.
The systematic uncertainties are identical to those detailed in Ref.~\cite{Sirunyan:2017khh}.
They are summarized in the following.

The rate uncertainties in the identification, isolation, and trigger efficiencies of electrons and muons amount to 2\%.
For $\tauh$, the uncertainty in the identification is 5\% per $\tauh$ candidate, and the uncertainty related to the
trigger amounts to an additional 5\% per $\tauh$ candidate~\cite{Sirunyan:2018pgf}. In the 0-jet category,
where one of the dimensions of the two-dimensional fit is the reconstructed $\tauh$ decay mode, the relative reconstruction efficiency
in a given $\tauh$ reconstructed decay mode has an uncertainty of 3\%~\cite{Sirunyan:2017khh}.
For muons and electrons
misreconstructed as $\tauh$ candidates, the $\tauh$ identification leads to rate uncertainties
of 25 and 12\%, respectively~\cite{Sirunyan:2018pgf}. 
This leads to the corresponding uncertainty in the rates of the $\PZ\to\Pgm\Pgm$ and $\PZ\to\Pe\Pe$ backgrounds misidentified as the $\Pgm\tauh$ and $\Pe\tauh$ final states, respectively.
The requirement that there are no {\cPqb}-tagged jets in $\Pe\Pgm$ decay channel events results
in a rate uncertainty as large as 5\% in the $\ttbar$ background~\cite{Sirunyan:2017ezt}.

The uncertainties in the energy scales of electrons and
$\tauh$ leptons amount to 1.0--2.5\%
and 1.2\%~\cite{Sirunyan:2017khh,Sirunyan:2018pgf} while the effect of the uncertainty in the muon energy scale is negligible. This uncertainty increases
to 3.0 and 1.5\%, respectively, for electrons
and muons misidentified as $\tauh$ candidates~\cite{Sirunyan:2017khh}. For events where
quark- or gluon-initiated jets are misidentified as $\tauh$ candidates, a linear uncertainty that increases
by 20\% per 100\GeV in transverse momentum of the $\tauh$ and amounts to 20\% for a $\tauh$ with $\pt$ of 100\GeV,
is taken into account~\cite{Sirunyan:2017khh}. This uncertainty affects simulated events with jets misidentified as $\tauh$ candidates, from various processes like the Drell-Yan, $\ttbar$, diboson, and $\PW+\text{jets}$ productions. Uncertainties in the jet and $\ptmiss$ energy scales are determined
event by event~\cite{Khachatryan:2014gga}, and propagated to the observables used in the analysis.

The uncertainty in the integrated luminosity is 2.5\%~\cite{CMS-PAS-LUM-17-001}. Per bin uncertainties
in the template probability parametrization related to the finite number of simulated events, or to the limited
number of events in data control regions, are also taken into account~\cite{Conway-PhyStat}.

The rate and acceptance uncertainties for the signal processes related to the theoretical calculations are
due to uncertainties in the PDFs, variations of the renormalization and factorization scales,
 and uncertainties in the modeling of parton showers. The magnitude of the rate uncertainty depends on the
production process and on the event category.
In particular, the inclusive uncertainty related to the PDFs amounts to 2.1\% for the
VBF production  mode~\cite{deFlorian:2016spz}, while the corresponding uncertainty for
the variation of the renormalization and factorization scales is 0.4\%~\cite{deFlorian:2016spz}. The acceptance
uncertainties related to the particular selection criteria used in this analysis are less than 1\% for all production modes.
The theoretical uncertainty in the branching fraction of
the \Hboson to $\Pgt$ leptons is 2.1\%~\cite{deFlorian:2016spz}.

An overall rate uncertainty of 3\%-10\% affects the $\PZ\to\Pgt\Pgt$ background,
depending on the category, as
estimated from a control region enriched in $\PZ\to\Pgm\Pgm$ events.
In the VBF category, this process is also affected by a shape uncertainty
that depends on $\mjj$ and $\Delta\Phi_{JJ}$, and can reach a magnitude of 20\%.
In addition to the uncertainties related to the $\PW+\text{jets}$ control regions in
the $\Pe\tauh$ and $\Pgm\tauh$
final states, the $\PW+\text{jets}$ background is affected by a rate
uncertainty ranging between 5 and 10\% to account
for the extrapolation of the constraints from the high-$\mT$ to the low-$\mT$ regions.
In the $\Pe\Pgm$ and $\tauh\tauh$ final states, the rate uncertainties in
the~$\PW+\text{jets}$ background yields are 20 and 4\%, respectively.

The uncertainty in the QCD multijet background yield in the
$\Pe\Pgm$ decay channel ranges from 10 to 20\%,
depending on the category.
In the $\Pe\tauh$ and $\Pgm\tauh$ decay channels,
uncertainties derived from the control regions are considered
for the QCD multijet background, together with an additional 20\%
uncertainty that accounts for the extrapolation
from the relaxed-isolation control region to the isolated signal region.
In the $\tauh\tauh$ decay channel, the
uncertainty in the QCD multijet background yield is a combination of the uncertainties obtained from fitting
the dedicated control regions with $\tauh$ candidates passing relaxed isolation criteria, of the extrapolation to
the signal region ranging from 3 to 15\%, and of residual differences between
prediction and data in signal-free regions with various loose isolation criteria.

The uncertainty from the fit
in the $\ttbar$ control region results in an uncertainty of about 5\% on the $\ttbar$ cross section in the
signal region. The combined systematic uncertainty in the background yield arising from diboson and
single top quark production processes is taken to be 5\%~\cite{Sirunyan:2017zjc,Sirunyan:2016cdg}.

The additional $\mathcal{D}_\mathrm{0-}$, $\mathcal{D}_\mathrm{0h+}$, $\mathcal{D}_{\Lambda1}$,
and $\mathcal{D}_{\Lambda1}^{\PZ\gamma}$ observables do not change the procedure for estimating the systematic uncertainty,
as any mismodeling due to detector effects is estimated with the same procedure as for any other distribution.
None of the systematic uncertainties introduces asymmetry in the $\mathcal{D}_{CP}$ distributions
which remain symmetric, except for the antisymmetric signal interference contribution.

\section{Results}
\label{sec:Results}

\begin{figure*}[!htb]
\centering
\includegraphics[width=0.45\textwidth]{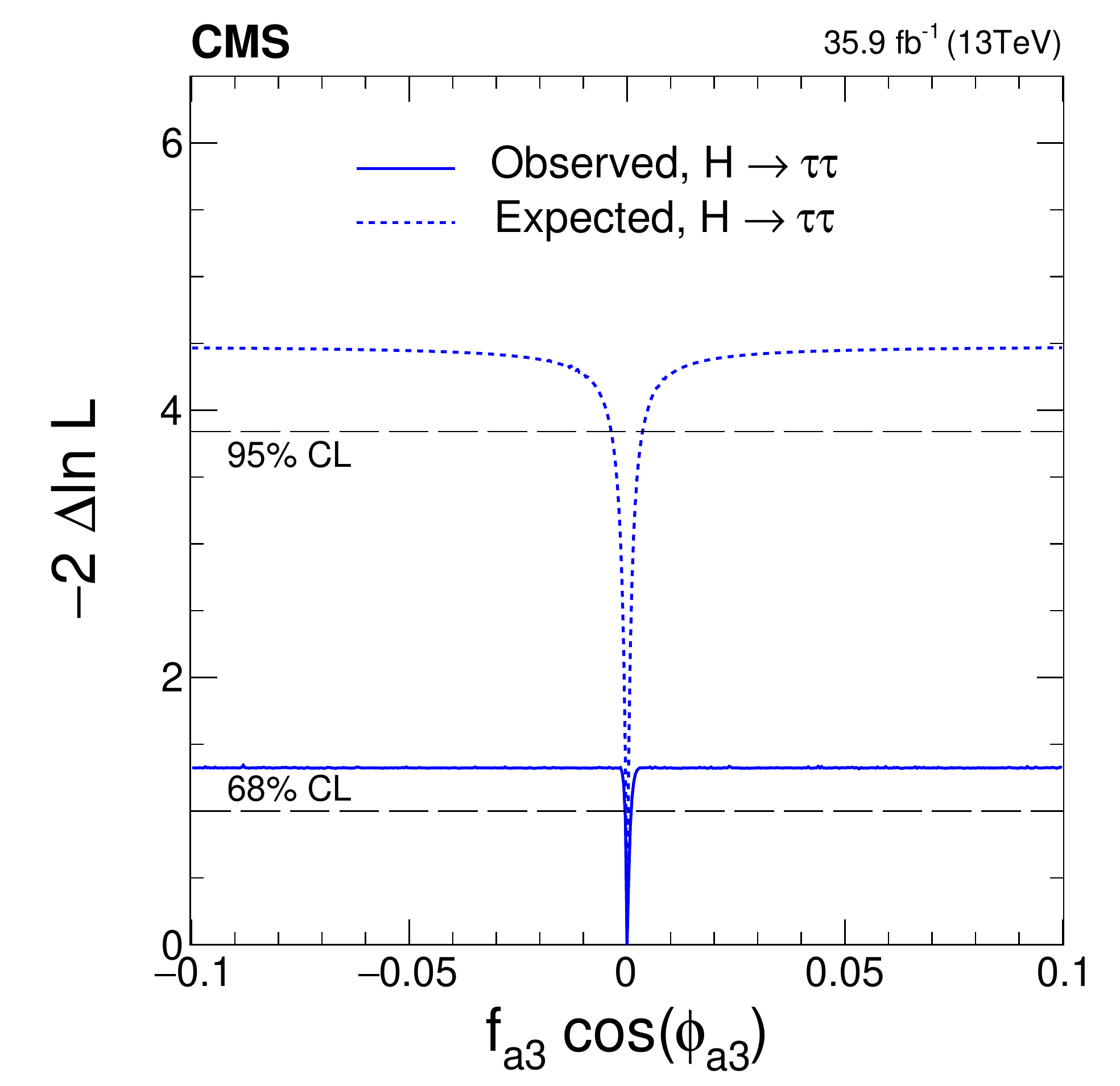}
\includegraphics[width=0.45\textwidth]{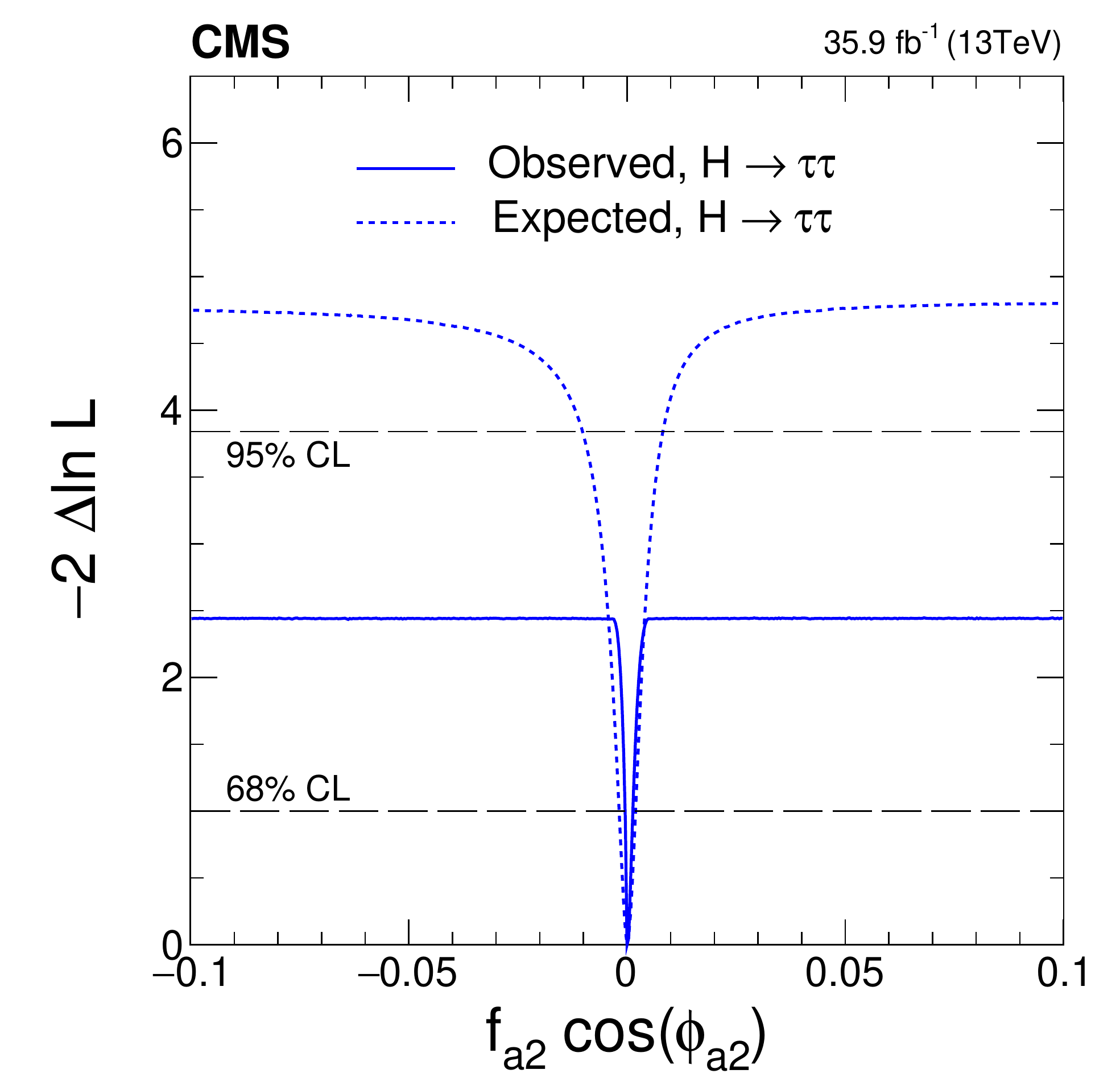}
\includegraphics[width=0.45\textwidth]{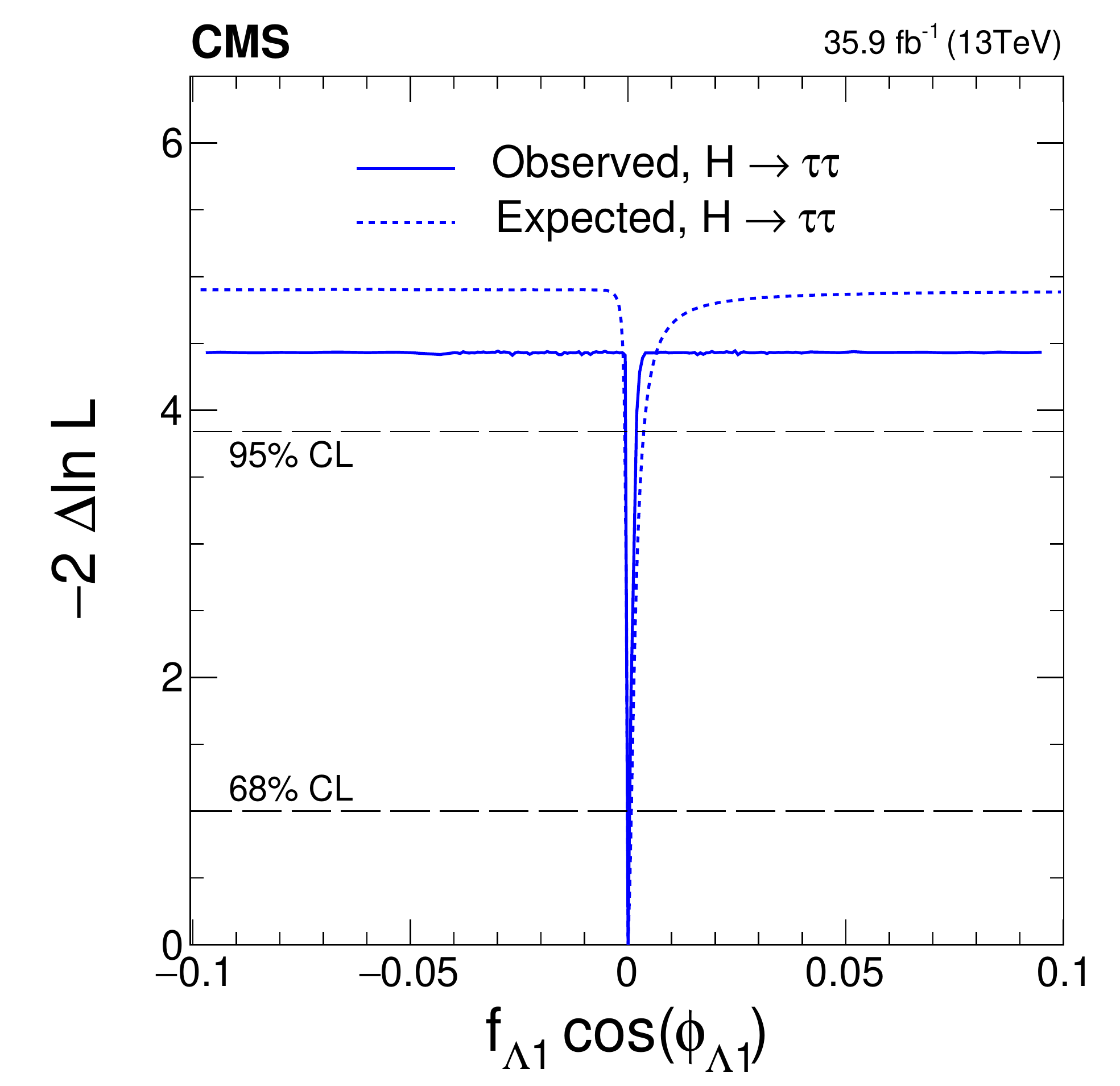}
\includegraphics[width=0.45\textwidth]{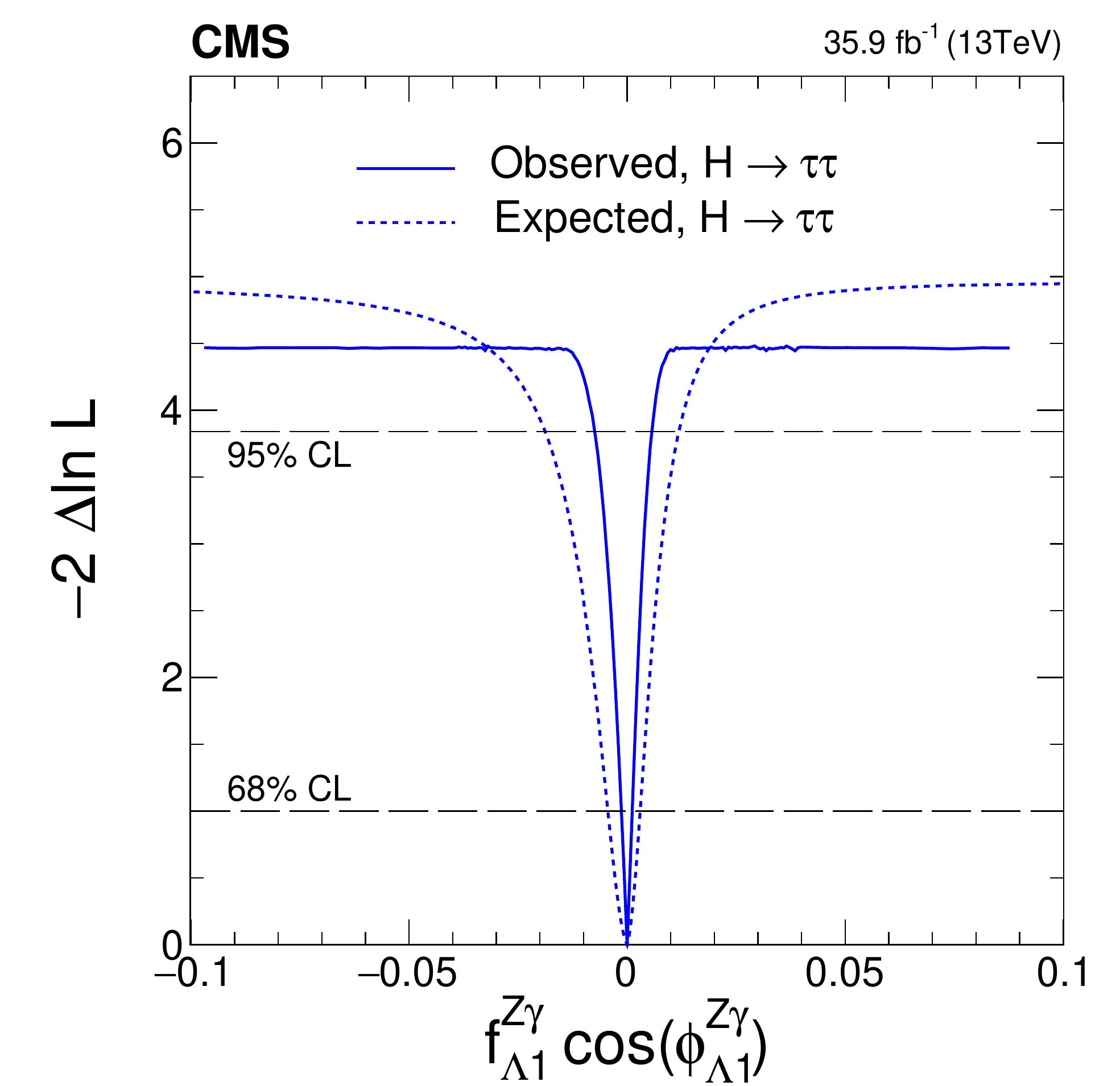}
\caption
{
Observed (solid) and expected (dashed) likelihood scans of
$f_{a3}\cos(\phi_{a3})$ (top left),
$f_{a2}\cos(\phi_{a2})$ (top right),
$f_{\Lambda1}\cos(\phi_{\Lambda1})$ (bottom left), and
$f_{\Lambda1}^{\PZ\gamma}\cos(\phi_{\Lambda1}^{\PZ\gamma})$ (bottom right).
\label{fig:fa3}
}
\end{figure*}

The four sets of $f_{ai}$ and $\phi_{ai}$ parameters describing anomalous $\PH\V\V$ couplings, as defined in Eqs.~(\ref{eq:formfact-fullampl-spin0})
and~(\ref{eq:fa_definitions}), are tested against the data according to the probability density defined in Eq.~(\ref{eq:fractions}).
The results of the likelihood scans are shown in Fig.~\ref{fig:fa3} and listed in Table~\ref{tab:summary_spin0}.
In each fit, the values of the other anomalous coupling parameters are set to zero.
In the case of the $CP$ fit, the $f_{a3}$ parameter is measured simultaneously with $f_{a3}^{\Pg\Pg\PH}$,
as defined in Eq.~(\ref{eq:fa3ggH_definition}).
All other parameters, including the signal strength parameters $\mu_\V$ and $\mu_\mathrm{f}$, are profiled.
The results are presented for the product of $f_{ai}$ and $\cos(\phi_{ai})$,
the latter being the sign of the real $a_i/a_1$ ratio of couplings.
In this approach, the $f_{ai}$ parameter is constrained to be in the physical range $f_{ai}\ge 0$.
Therefore, in the SM it is likely for the best-fit value to be at the physical boundary $f_{ai} = 0$ for both signs of the $a_i/a_1$ ratio.

\begin{table*}[h]
\centering
\topcaption{
Allowed 68\%~\CL (central values with uncertainties) and 95\%~\CL (in square brackets)
intervals on anomalous coupling parameters using the $\PH\to\Pgt\Pgt$ decay.
The observed 95\%~\CL constraints on $f_{a3}\cos(\phi_{a3})$ and $f_{a2}\cos(\phi_{a2})$ allow the full range $[-1,1]$.
}
\renewcommand{\arraystretch}{1.25}
\begin{scotch}{ccccc}
 Parameter                                   &  \multicolumn{2}{c}{Observed$/(10^{-3})$} &  \multicolumn{2}{c}{Expected$/(10^{-3})$}    \\
                                                     & 68\%~\CL &  95\%~\CL& 68\%~\CL &  95\%~\CL \\
\hline
$f_{a3}\cos(\phi_{a3})$ & $0.00^{+0.93}_{-0.43}$ & \NA    &   $0.00\pm0.28$ & $[-3.6,3.6]$  \\
$f_{a2}\cos(\phi_{a2})$  & $0.0^{+1.2}_{-0.4}$ & \NA    &   $0.0^{+2.0}_{-1.8}$ & $[-10.0,8.0]$  \\
$f_{\Lambda1}\cos(\phi_{\Lambda1})$ & $0.00^{+0.39}_{-0.10}$ & $[-0.4,1.8]$    &   $0.00^{+0.75}_{-0.16}$ & $[-0.8,3.6]$  \\
$f_{\Lambda1}^{\PZ\gamma}\cos(\phi_{\Lambda1}^{\PZ\gamma})$  & $0.0^{+1.2}_{-1.3}$ & $[-7.4,5.6]$    &   $0.0^{+3.0}_{-4.5}$ & $[-19,12]$  \\
\end{scotch}
\label{tab:summary_spin0}
\end{table*}

The constraints on $f_{ai}\cos(\phi_{ai})$ appear relatively tight compared to similar constraints utilizing the \Hboson decay information,
e.g., in Ref.~\cite{Sirunyan:2019twz}.
This is because the cross section in VBF and $\V\PH$ production increases quickly with $f_{ai}$.
The definition of $f_{ai}$ in Eq.~(\ref{eq:fa_definitions}) uses the cross section ratios defined in the $\PH\to 2\Pe2\mu$
decay as the common convention across various measurements.
Because the cross section increases with respect to $f_{ai}$ at different rates for production and decay,
relatively small values of $f_{ai}$ correspond to a substantial anomalous contribution to the production cross section.
This leads to the plateau in the $-2\ln({\cal L}/{\cal L}_\mathrm{max})$ distributions for larger values of $f_{ai}\cos(\phi_{ai})$ in Fig.~\ref{fig:fa3}.
If we had used the cross section ratios for VBF production in the $f_{ai}$ definition in Eq.~(\ref{eq:fa_definitions}),
the appearance of the plateau and the narrow exclusion range would change.
For example, the 68\%~\CL upper constraint on $f_{a3}\cos(\phi_{a3}) < 0.00093$ is dominated by the VBF production information. If we were to use
the VBF cross section ratio $\sigma_{1}^\mathrm{VBF} /\sigma_{3}^\mathrm{VBF} =0.089$ in the $f_{a3}^\mathrm{VBF}$ definition
in Eq.~(\ref{eq:fa_definitions}), this would correspond to the upper constraint $f_{a3}^\mathrm{VBF}\cos(\phi_{a3}) < 0.064$ at 68\%~\CLnp.

The observed maximum value of $-2\ln({\cal L}/{\cal L}_\mathrm{max})$ is somewhat different from expectation
and between the four analyses, mostly due to statistical fluctuations in the distribution of events across the dedicated
discriminants and other observables, leading to different significances of the observed signal driven by VBF and $\V\PH$ production.
In particular, the best-fit values for $(\mu_\V, \mu_\mathrm{f})$ in the four analyses, under the assumption that $f_{ai}=0$, are
$(0.55\pm0.48,1.03^{+0.45}_{-0.40})$ at $f_{a3}=0$,
$(0.72^{+0.48}_{-0.46},0.89^{+0.43}_{-0.37})$  at $f_{a2}=0$,
$(0.92^{+0.44}_{-0.45},0.82^{+0.46}_{-0.38})$  at $f_{\Lambda1}=0$, and
$(0.94^{+0.48}_{-0.46},0.79\pm0.40)$  at $f_{\Lambda1}^{\PZ\gamma}=0$.
This results in a somewhat lower yield of VBF and $\V\PH$ events observed in the first two cases,
leading to lower confidence levels in constraints on $f_{a3}\cos(\phi_{a3})$ and $f_{a2}\cos(\phi_{a2})$.

In the $f_{a3}$ analysis, a simultaneous measurement of $f_{a3}$ and $f_{a3}^{\Pg\Pg\PH}$ is performed.
These are the parameters sensitive to $CP$ in the VBF and gluon fusion processes, respectively.
Both the observed and expected exclusions from the null hypothesis for any BSM gluon fusion scenario with either \textsc{mela} or the $\Delta\Phi_{JJ}$ observable are below one standard deviation.

\section{Combination of results with other channels}
\label{sec:Combination}

The precision of the coupling measurements can be improved by combining the results in the
$\PH\to\Pgt\Pgt$ channel, presented here, with those of other \Hboson decay channels.
A combination is possible only with those channels where anomalous couplings in the
$\V\PH$, VBF, and gluon fusion processes are taken into account in the fit in a consistent way.
If it is not done, the kinematics of the associated jets and of the \Hboson
would not be modeled correctly for BSM values of the $f_{ai}$ or $f_{a3}^{\Pg\Pg\PH}$ parameters.

In the example of the $CP$ fit, in the stand-alone fit with the $\PH\to\Pgt\Pgt$ channel, the parameters
of interest are $f_{a3}\cos(\phi_{a3})$, $f_{a3}^{\Pg\Pg\PH}\cos(\phi_{a3}^{\Pg\Pg\PH})$, $\mu_{\V}^{\PH\Pgt\Pgt}$, and $\mu_\mathrm{f}^{\PH\Pgt\Pgt}$.
When reporting one parameter, all other parameters are profiled.
In a combined fit of the $\PH\to\Pgt\Pgt$ and  $\PH\to\V\V$ channels, such as in Ref.~\cite{Sirunyan:2019twz},
in principle there are four signal strength parameters in the two channels
($\mu_{\V}^{\PH\Pgt\Pgt}$,  $\mu_\mathrm{f}^{\PH\Pgt\Pgt}$, $\mu_{\V}^{\PH \V\V}$,  $\mu_\mathrm{f}^{\PH \V\V}$).
However, this can be reduced to three parameters because the ratio between the
VBF+$\V\PH$ and gluon fusion cross sections is expected to be the same in each of the two channels, that is
$\mu_{\V}^{\PH\Pgt\Pgt} / \mu_\mathrm{f}^{\PH\Pgt\Pgt} = \mu_{\V}^{\PH \V\V} / \mu_\mathrm{f}^{\PH \V\V}$.
Therefore, the three signal strength parameters are chosen as $\mu_{\V}$, $\mu_\mathrm{f}$, and  $\eta_{\tau}$,
where the last one is the relative strength of the \Hboson coupling to the $\Pgt$ leptons.
We should note that, as discussed earlier, the ${\PH\PW\PW}$ couplings are analyzed together
with the $\PH\PZ\PZ$ couplings assuming $a_i^{\PZ\PZ}=a_i^{\PW\PW}$.
The results can be reinterpreted for a different assumption of the $a_i^{\PZ\PZ}/a_i^{\PW\PW}$ ratio~\cite{Sirunyan:2019twz}.
In the combined likelihood fit, all common systematic uncertainties are correlated
between the channels, both theoretical uncertainties, such as those due to the PDFs,
and experimental uncertainties, such as jet energy calibration.

\begin{table*}[th]
\centering
\topcaption{
Allowed 68\%~\CL (central values with uncertainties) and 95\%~\CL (in square brackets)
intervals on anomalous coupling parameters using a combination of the $\PH\to\Pgt\Pgt$ and
$\PH\to 4\ell$~\cite{Sirunyan:2019twz} decay channels.
}
\renewcommand{\arraystretch}{1.25}
\begin{scotch}{ccccc}
 Parameter                                   &  \multicolumn{2}{c}{Observed$/(10^{-3})$} &  \multicolumn{2}{c}{Expected$/(10^{-3})$}    \\
                                                     & 68\%~\CL &  95\%~\CL& 68\%~\CL &  95\%~\CL \\
\hline
$f_{a3}\cos(\phi_{a3})$ & $0.00\pm 0.27$ & $[-92, 14]$ &   $0.00\pm0.23$ & $[-1.2, 1.2]$  \\
$f_{a2}\cos(\phi_{a2})$  & $0.08^{+1.04}_{-0.21}$ & $[-1.1, 3.4]$ & $0.0^{+1.3}_{-1.1}$ & $[-4.0, 4.2]$  \\
$f_{\Lambda1}\cos(\phi_{\Lambda1})$ & $0.00^{+0.53}_{-0.09}$ & $[-0.4, 1.8]$   &  $0.00^{+0.48}_{-0.12}$ & $[-0.5, 1.7]$   \\
$f_{\Lambda1}^{\PZ\gamma}\cos(\phi_{\Lambda1}^{\PZ\gamma})$  &  $0.0^{+1.1}_{-1.3}$ & $[-6.5, 5.7]$ & $0.0^{+2.6}_{-3.6}$ & $[-11, 8.0]$  \\
\end{scotch}
\label{tab:summary_combine}
\end{table*}

\begin{figure*}[!htb]
\centering
\includegraphics[width=0.45\textwidth]{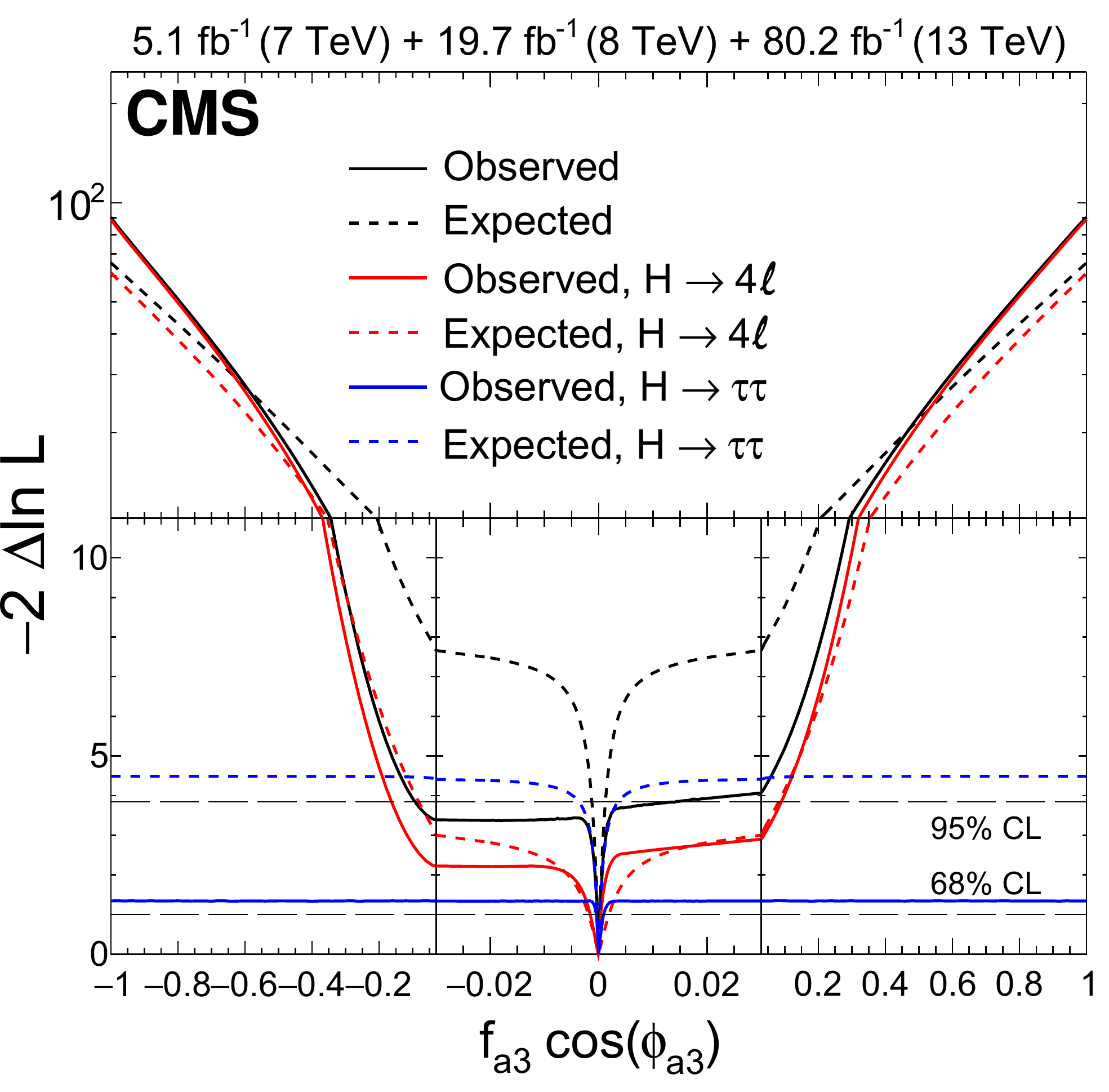}
\includegraphics[width=0.45\textwidth]{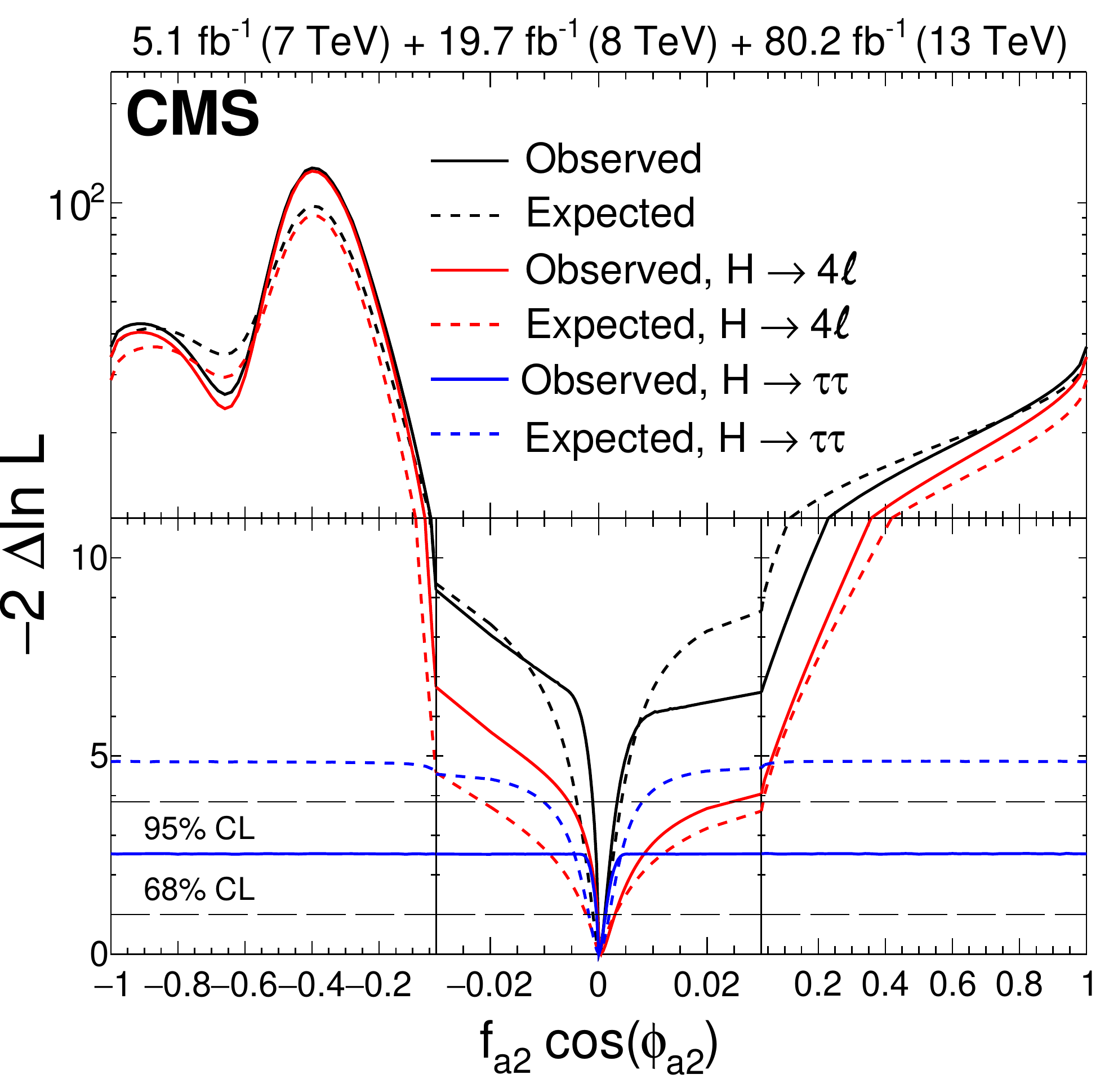}
\includegraphics[width=0.45\textwidth]{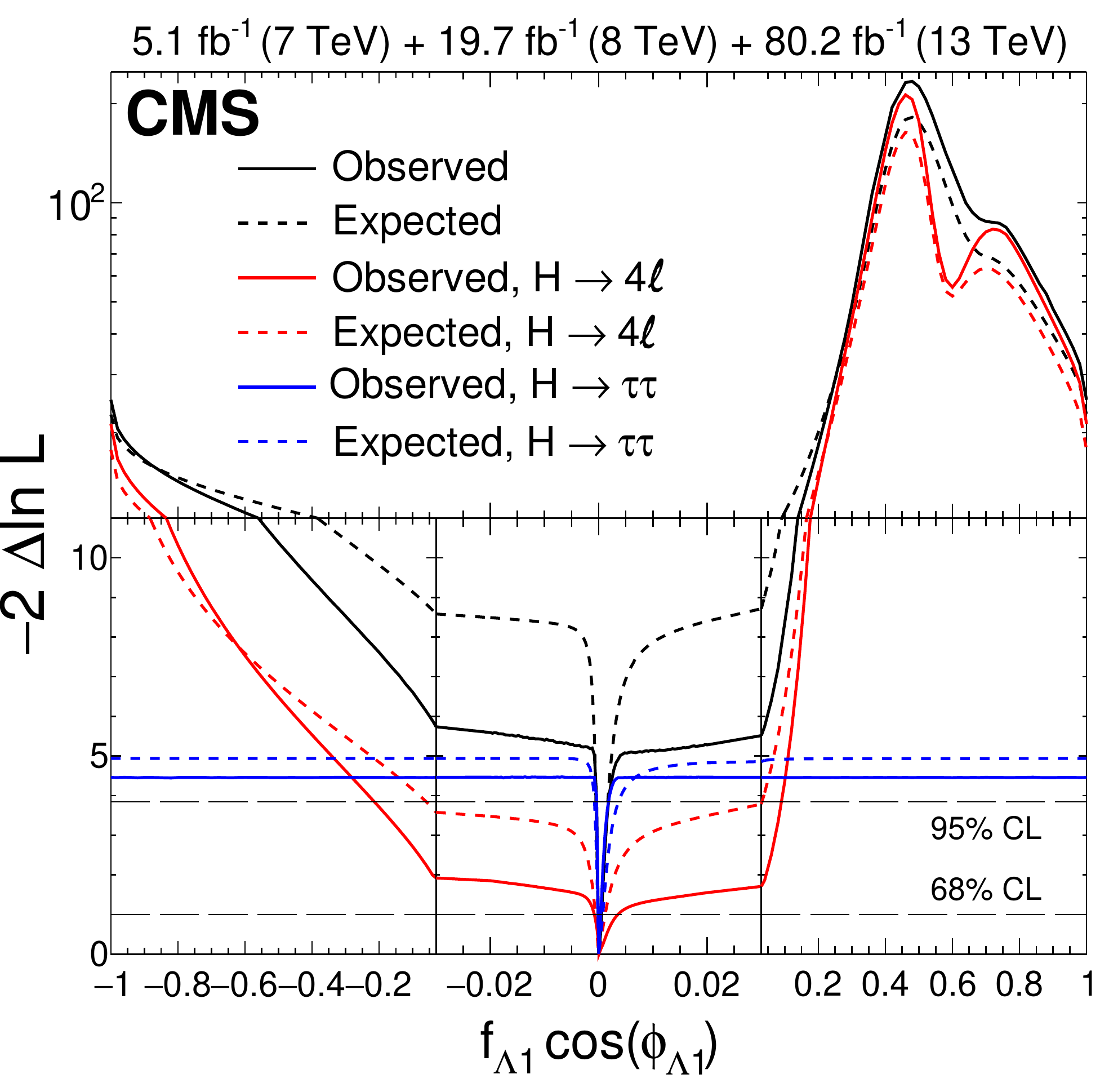}
\includegraphics[width=0.45\textwidth]{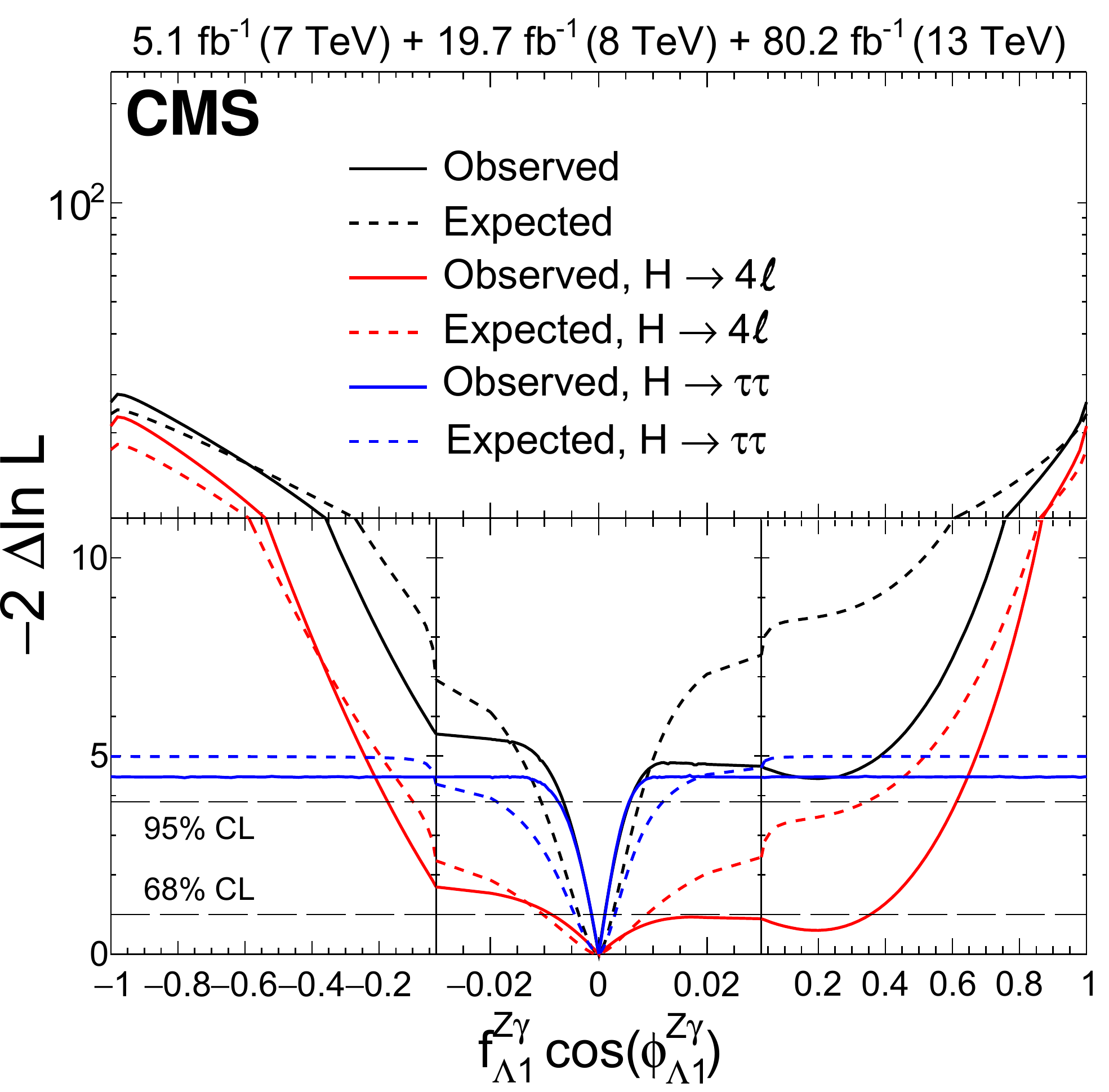}
\caption
{
Combination of results using the $\PH\to\Pgt\Pgt$ decay (presented in this paper) and the $\PH\to4\ell$ decay~\cite{Sirunyan:2019twz}.
The observed (solid) and expected (dashed) likelihood scans of
$f_{a3}\cos(\phi_{a3})$ (top left), $f_{a2}\cos(\phi_{a2})$ (top right), $f_{\Lambda1}\cos(\phi_{\Lambda1})$ (bottom left), and
$f_{\Lambda1}^{\PZ\gamma}\cos(\phi_{\Lambda1}^{\PZ\gamma})$ (bottom right) are shown.
For better visibility of all features, the  $x$ and $y$ axes are presented with variable scales. On the linear-scale $x$ axis, a zoom is applied in the range $-$0.03 to 0.03. The $y$ axis is shown in linear (logarithmic) scale for values of $-2\,\Delta \ln{\mathcal L}$ below (above) 11.
\label{fig:combination}
}
\end{figure*}

The results using the $\PH\to\Pgt\Pgt$ decay are combined with those presented in Ref.~\cite{Sirunyan:2019twz}
using the on-shell $\PH\to4\ell$ decay.
The latter employs results from Run 1 (from 2011 and 2012) and Run 2 (from 2015, 2016, and 2017) with data
corresponding to integrated luminosities of  5.1, 19.7, and 80.2\fbinv at center-of-mass energies
7, 8, and 13\TeV, respectively. In this analysis, information about $\PH\V\V$ anomalous couplings both
in VBF+$\V\PH$ production and in $\PH\to\V\V\to 4\ell$ decay is used.
In all cases, the signal strength parameters are profiled, and the parameters common to the two analyses
are correlated.
The combined 68\%~\CL and 95\%~\CL intervals are presented in Table~\ref{tab:summary_combine}
and the likelihood scans are shown in Fig.~\ref{fig:combination}.
While the constraints at large values of $f_{ai}$ are predominantly driven by the decay information in the $\PH\to\V\V$ analysis, the constraints in the narrow range of $f_{ai}$ near 0 are dominated by the production information where the $\PH\to\Pgt\Pgt$ channel dominates over the $\PH\to4\ell$. This results in the most stringent limits on anomalous $\PH\V\V$ couplings.
Reverting the transformation in Eq.~(\ref{eq:fa_definitions})~\cite{Sirunyan:2019twz},
the $f_{ai}\cos(\phi_{ai})$ results can be interpreted for the coupling parameters used in
Eq.~(\ref{eq:formfact-fullampl-spin0}), as shown in Table~\ref{tab:summary_aia1}.

\begin{table*}[th]
\centering
\topcaption{
Summary of the allowed 95\%~\CL intervals for the anomalous $\PH\V\V$ couplings using the results in Table~\ref{tab:summary_combine}.
The coupling ratios are assumed to be real and include the factor $\cos(\phi_{\Lambda1})$ or $\cos(\phi_{\Lambda1}^{\PZ\gamma})=\pm1$.
}
\renewcommand{\arraystretch}{1.25}
\begin{scotch}{ccll}
 Parameter     & \multicolumn{1}{c}{Observed} & \multicolumn{1}{c}{Expected}  \\
\hline
\vspace{-0.45cm} \\
 $a_3/a_1$ & $[-0.81, 0.31]$ & $[-0.090, 0.090]$ \\
 $a_2/a_1$ & $[-0.055, 0.097]$ & $[-0.11, 0.11]$ \\
$(\Lambda_1\sqrt{\abs{a_1}}) \cos(\phi_{\Lambda1})$  ~(GeV)
                  &  $[-\infty, -650]\cup[440, \infty]$  &  $[-\infty, -610]\cup[450, \infty]$ \\
$(\Lambda_1^{\PZ\gamma}\sqrt{\abs{a_1}}) \cos(\phi_{\Lambda1}^{\PZ\gamma}) $  ~(GeV)
                  &  $[-\infty, -400]\cup[420, \infty]$      &$[-\infty, -360]\cup[390, \infty]$ \\
\end{scotch}
\label{tab:summary_aia1}
\end{table*}

\section{Conclusions}
\label{sec:Summary}

A study is presented of anomalous $\PH\V\V$ interactions of the \Hboson with vector bosons $\V$, including $CP$-violation,
using its associated production with two hadronic jets in vector boson fusion, in the $\V\PH$ process,
and in gluon fusion, and subsequently decaying to a pair of $\Pgt$ leptons.
Constraints on the $CP$-violating parameter $f_{a3}\cos(\phi_{a3})$ and on the $CP$-conserving parameters $f_{a2}\cos(\phi_{a2})$, $f_{\Lambda1}\cos(\phi_{\Lambda1})$, and $f_{\Lambda1}^{\PZ\gamma}\cos(\phi_{\Lambda1}^{\PZ\gamma})$, defined in Eqs.~(\ref{eq:formfact-fullampl-spin0}) and~(\ref{eq:fa_definitions}), are set using matrix element techniques.
The observed and expected limits on the parameters are summarized in Table~\ref{tab:summary_spin0}.
The 68\% confidence level constraints are generally tighter than those from previous measurements using either production or decay information.
Further constraints are obtained in the combination of the $\PH\to\Pgt\Pgt$ and $\PH\to 4\ell$ decay~\cite{Sirunyan:2019twz} channels
and are summarized in Table~\ref{tab:summary_combine}.
This combination places the most stringent constraints on anomalous \Hboson couplings:
$f_{a3}\cos(\phi_{a3})=(0.00\pm0.27)\times10^{-3}$, $f_{a2}\cos(\phi_{a2})=(0.08^{+1.04}_{-0.21})\times10^{-3}$,
$f_{\Lambda1}\cos(\phi_{\Lambda1})=(0.00^{+0.53}_{-0.09})\times10^{-3}$, and
$f_{\Lambda1}^{\PZ\gamma}\cos(\phi_{\Lambda1}^{\PZ\gamma})=(0.0^{+1.1}_{-1.3})\times10^{-3}$.
A simultaneous measurement of $f_{a3}\cos(\phi_{a3})$ and $f_{a3}^{\Pg\Pg\PH}\cos(\phi_{a3}^{\Pg\Pg\PH})$ parameters is performed,
where the latter parameter, defined in Eqs.~(\ref{eq:formfact-fullampl-spin0}) and~(\ref{eq:fa3ggH_definition}),
is sensitive to $CP$ violation effects in the gluon fusion process.
The current dataset does not allow for precise constraints on $CP$ properties in the gluon fusion process.
The results are consistent with expectations for the standard model \Hboson.

\begin{acknowledgments}

\hyphenation{Bundes-ministerium Forschungs-gemeinschaft Forschungs-zentren Rachada-pisek}
We thank Markus Schulze for optimizing the \textsc{JHUGen} Monte Carlo simulation program and matrix element library for this analysis.

We congratulate our colleagues in the CERN accelerator departments for the excellent performance of the LHC and thank the technical and administrative staffs at CERN and at other CMS institutes for their contributions to the success of the CMS effort. In addition, we gratefully acknowledge the computing centers and personnel of the Worldwide LHC Computing Grid for delivering so effectively the computing infrastructure essential to our analyses. Finally, we acknowledge the enduring support for the construction and operation of the LHC and the CMS detector provided by the following funding agencies: BMBWF and FWF (Austria); FNRS and FWO (Belgium); CNPq, CAPES, FAPERJ, FAPERGS, and FAPESP (Brazil); MES (Bulgaria); CERN; CAS, MoST, and NSFC (China); COLCIENCIAS (Colombia); MSES and CSF (Croatia); RPF (Cyprus); SENESCYT (Ecuador); MoER, ERC IUT, and ERDF (Estonia); Academy of Finland, MEC, and HIP (Finland); CEA and CNRS/IN2P3 (France); BMBF, DFG, and HGF (Germany); GSRT (Greece); NKFIA (Hungary); DAE and DST (India); IPM (Iran); SFI (Ireland); INFN (Italy); MSIP and NRF (Republic of Korea); MES (Latvia); LAS (Lithuania); MOE and UM (Malaysia); BUAP, CINVESTAV, CONACYT, LNS, SEP, and UASLP-FAI (Mexico); MOS (Montenegro); MBIE (New Zealand); PAEC (Pakistan); MSHE and NSC (Poland); FCT (Portugal); JINR (Dubna); MON, RosAtom, RAS, RFBR, and NRC KI (Russia); MESTD (Serbia); SEIDI, CPAN, PCTI, and FEDER (Spain); MOSTR (Sri Lanka); Swiss Funding Agencies (Switzerland); MST (Taipei); ThEPCenter, IPST, STAR, and NSTDA (Thailand); TUBITAK and TAEK (Turkey); NASU and SFFR (Ukraine); STFC (United Kingdom); and DOE and NSF (USA).

\hyphenation{Rachada-pisek} Individuals have received support from the Marie-Curie program and the European Research Council and Horizon 2020 Grant, Contract Nos.\ 675440 and 765710 (European Union); the Leventis Foundation; the A.P.\ Sloan Foundation; the Alexander von Humboldt Foundation; the Belgian Federal Science Policy Office; the Fonds pour la Formation \`a la Recherche dans l'Industrie et dans l'Agriculture (FRIA-Belgium); the Agentschap voor Innovatie door Wetenschap en Technologie (IWT-Belgium); the F.R.S.-FNRS and FWO (Belgium) under the ``Excellence of Science - EOS" be.h Project No.\ 30820817; the Beijing Municipal Science \& Technology Commission, Grant No. Z181100004218003; the Ministry of Education, Youth and Sports (MEYS) of the Czech Republic; the Lend\"ulet (``Momentum") Program and the J\'anos Bolyai Research Scholarship of the Hungarian Academy of Sciences, the New National Excellence Program \'UNKP, the NKFIA research Grants No. 123842, No. 123959, No. 124845, No. 124850, No. 125105, No. 128713, No. 128786, and No. 129058 (Hungary); the Council of Science and Industrial Research, India; the HOMING PLUS program of the Foundation for Polish Science, cofinanced from European Union, Regional Development Fund, the Mobility Plus program of the Ministry of Science and Higher Education, the National Science Center (Poland), Contracts No. Harmonia 2014/14/M/ST2/00428, No. Opus 2014/13/B/ST2/02543, No. 2014/15/B/ST2/03998, and No. 2015/19/B/ST2/02861, Sonata-bis 2012/07/E/ST2/01406; the National Priorities Research Program by Qatar National Research Fund; the Programa Estatal de Fomento de la Investigaci{\'o}n Cient{\'i}fica y T{\'e}cnica de Excelencia Mar\'{\i}a de Maeztu, Grant No. MDM-2015-0509 and the Programa Severo Ochoa del Principado de Asturias; the Thalis and Aristeia programs cofinanced by EU-ESF and the Greek NSRF; the Rachadapisek Sompot Fund for Postdoctoral Fellowship, Chulalongkorn University and the Chulalongkorn Academic into Its 2nd Century Project Advancement Project (Thailand); the Welch Foundation, Contract No. C-1845; and the Weston Havens Foundation (USA).

\end{acknowledgments}

\bibliography{auto_generated}
\cleardoublepage \appendix\section{The CMS Collaboration \label{app:collab}}\begin{sloppypar}\hyphenpenalty=5000\widowpenalty=500\clubpenalty=5000\input{HIG-17-034-authorlist.tex}\end{sloppypar}
\end{document}

%% file: HIG-17-034-authorlist.tex
\vskip\cmsinstskip
\textbf{Yerevan Physics Institute, Yerevan, Armenia}\\*[0pt]
A.M.~Sirunyan, A.~Tumasyan
\vskip\cmsinstskip
\textbf{Institut f\"{u}r Hochenergiephysik, Wien, Austria}\\*[0pt]
W.~Adam, F.~Ambrogi, E.~Asilar, T.~Bergauer, J.~Brandstetter, M.~Dragicevic, J.~Er\"{o}, A.~Escalante~Del~Valle, M.~Flechl, R.~Fr\"{u}hwirth\cmsAuthorMark{1}, V.M.~Ghete, J.~Hrubec, M.~Jeitler\cmsAuthorMark{1}, N.~Krammer, I.~Kr\"{a}tschmer, D.~Liko, T.~Madlener, I.~Mikulec, N.~Rad, H.~Rohringer, J.~Schieck\cmsAuthorMark{1}, R.~Sch\"{o}fbeck, M.~Spanring, D.~Spitzbart, W.~Waltenberger, J.~Wittmann, C.-E.~Wulz\cmsAuthorMark{1}, M.~Zarucki
% ...
\vskip\cmsinstskip
% ...
1:  Also at Vienna University of Technology, Vienna, Austria\\
% ...
